\documentclass[twocolumn,showpacs,pra]{revtex4-1}

\usepackage{amsmath,amssymb,graphicx}

\begin{document}

\title{Coherence and dimensionality of intense spatio-spectral twin beams}

\author{Jan Pe\v{r}ina Jr.\email{jan.perina.jr@upol.cz}}
\affiliation{RCPTM, Joint Laboratory of Optics of Palack\'{y}
University and Institute of Physics of Academy of Sciences of the
Czech Republic, Faculty of Science, Palack\'{y} University, 17.
listopadu 12, 771~46 Olomouc, Czech Republic}

\begin{abstract}
Spatio-spectral properties of twin beams at their transition from
low to high intensities are analyzed in parametric and paraxial
approximations using the decomposition into paired spatial and
spectral modes. Intensity auto- and cross-correlation functions
are determined and compared in the spectral and temporal domains
as well as the transverse wave-vector and crystal output planes.
Whereas the spectral, temporal and transverse wave-vector
coherence increases with the increasing pump intensity, coherence
in the crystal output plane is practically independent on the pump
intensity owing to the mode structure in this plane. The
corresponding auto- and cross-correlation functions approach each
other for larger pump intensities. Entanglement dimensionality of
a twin beam is determined comparing several approaches.
\end{abstract}

\pacs{42.65.Lm,42.65.Yj,42.50.Dv}

\keywords{twin beam, entangled photon pair, coherence, number of
modes}

\maketitle

\section{Introduction}

The nonlinear process of parametric down-conversion
\cite{Boyd2003} is the most frequently used process for the
generation of light with nonclassical properties. It provides
entangled photon pairs in its spontaneous regime
\cite{Mandel1995}. Photons comprising a photon pair can be
entangled in different degrees of freedom including polarization,
frequency, emission direction or orbital angular momentum.
Entanglement in all these degrees of freedom has been found useful
both for testing the rules of quantum mechanics
\cite{Bouwmeester2000} and applications \cite{Carrasco2004}.

On the other hand, parametric down-conversion generates the
so-called twin beams when stimulated emission is important. Such
twin beams are composed of a signal and an idler fields with large
intensities that are mutually strongly correlated. These
correlations occur both in the spectra and emission directions as
a consequence of the properties that give different kinds of
entanglement at single-photon level. Moreover, the intensity
correlations are so strong that they violate the standard
shot-noise limit (sub-shot-noise correlations)
\cite{Kolobov1989,Kolobov1989a}. This nonclassical property
originates in the genuine pairwise character of parametric
down-conversion at its quantum level. An experimental evidence of
sub-shot-noise intensity correlations has been given in
\cite{Jedrkiewicz2004,Bondani2007,Blanchet2008,Brida2009a}. Such
states are useful also for quantum imaging. Even imaging based
upon sub-shot-noise intensity correlations has been recently
demonstrated \cite{Brida2010,Boyer2008}.

Spontaneous parametric down-conversion with its production of
entangled photon pairs has been studied by far the most
frequently. In theory, the first-order solution of the appropriate
Schr\"{o}dinger equation provides a two-photon amplitude that
determines all properties of entangled photon pairs
\cite{Rubin1994,PerinaJr1999a}. This simple formalism is also
easily applicable to more complex nonlinear structures generating
photon pairs (waveguides, 
fibers, 
layered structures, 
Bragg-reflector waveguides, 
periodically-poled structures,  
for details, see, e.g., \cite{PerinaJr2014}).
At present, photon pairs with
more-less arbitrary properties can be efficiently generated due to
many kinds of available sources.

On the other hand, investigations of twin beams have been
concentrated to more intense twin beams \cite{Jedrkiewicz2012}
because of the lack of detection techniques capable to detect
intensities at the transition from the single-photon level to the
intense ('classical') one \cite{Machulka2014}. This has required
intense pump lasers that have provided quasi-monochromatic beams
(usually picosecond pulses). As a consequence, the developed
theoretical models usually invoke the quasi-monochromatic
approximation. When combined with the pump-field quasi-plane-wave
approximation, a twin beam in the model has been decomposed into
many more-less independent pairs of signal and idler modes
localized both in the spectrum and transverse wave-vector plane
\cite{Gatti2003,Brambilla2010,Caspani2010,Dayan2007}. The dynamics
of individual pairs of modes has then been treated by the solution
of linear Heisenberg equations. Under more general conditions,
numerical solution of the Maxwell equations using a statistical
ensemble of initial conditions has occurred useful
\cite{Brambilla2004}.

The Schmidt decomposition \cite{Law2000,Law2004} of two-photon
amplitudes introduced for pure states at single-photon level has
become popular in the last couple of years due to its ability to
quantify entanglement in larger Hilbert spaces and to reveal the
genuine dual structure of a bipartite entangled state
\cite{Christ2011,Avella2014}. Such modes can even be selected from
a beam \cite{Shapiro1997,Bennink2002,Bobrov2013,Brecht2014} or
changed on demand. However, the revealed paired modes have been
found suitable also for making the bridge between the perturbation
theory of weak paired fields and that of the intense beams. The
corresponding theory has been based upon the solution of
Heisenberg equations for individual and independent paired modes,
similarly as the theory developed earlier for intense twin beams.
Contrary to this theory based on localized modes, it uses the
Schmidt modes spread over the whole spectrum or transverse
wave-vector plane \cite{Annamalai2011}. This makes the theory
suitable for describing coherence of the twin beams and especially
its growth with the increasing pump intensity. It has already been
applied to describe spectral properties and amplitude squeezing of
weaker as well as more intense twin beams
\cite{Wasilewski2006,Lvovsky2007,Christ2013}. Also angular
properties of the twin beams have been addressed
\cite{Sharapovova2015}. Even nonlinear waveguides with
back-scattering have been investigated using this approach
\cite{PerinaJr2013}.

However, the spectral and spatial properties of a twin beam have
only been considered separately in this approach up to now. This
represents a serious simplification as the spectral and spatial
modes are inevitably mutually coupled in the nonlinear
interaction. The consideration of only the spectral (or spatial)
modes allows to understand the behavior of twin beams only to
certain extent as the approach is not able to describe correctly a
common dynamics of both spectral and spatial degrees of freedom.
The 'one-dimensional Schmidt-mode models' developed up to now are
in fact simple 'fenomenological' models that are conveniently
applied for interpreting the experimental results obtained under
specific conditions [strong spatial (spectral) filtering for the
spectral (radial transverse direction) model]. On the other hand,
real down-conversion occurs in a nonlinear crystal with all
possible spatio-spectral modes participating in the interaction.
As bulk nonlinear crystals are commonly used for the generation of
intense twin beams, the number of participating modes is large.
Moreover, the number of modes giving an important (intensity)
contribution to the generated twin beam is also large. That is
why, a general spatio-spectral model of twin-beam generation is
necessary.

In this contribution, we develop the Schmidt-mode approach for
describing such a general spatio-spectral twin beam. Using the
paired spatio-spectral modes, we address coherence of the twin
beams determining their auto- and cross-correlation functions in
the spectral and temporal domains, transverse wave-vector plane,
and crystal output plane. Properties of the twin beams manifested
under different experimental conditions are compared. Special
attention is paid to the dependence of coherence on the pump
intensity. Coherence properties of the twin beams together with
their photon-number statistics are used to compare several
suitable quantifiers of their dimensionality
\cite{Stobinska2012,Chekhova2015}. This provides a complete
picture of an intense spatio-spectral twin beam.

The developed model can easily be generalized to more complex
nonlinear structures with nonlinearity homogeneous in the
transverse plane including poled nonlinear crystals
\cite{Hum2007,Svozilik2009} and nonlinear layered structures
\cite{PerinaJr2014}. It can even be applied for the description of
intense twin beams originating in the nonlinear process of
four-wave mixing, both in nonlinear crystals with $ \chi^{(3)} $
nonlinearity and atomic ensembles. The four-wave mixing in atomic
ensembles \cite{Kolchin2006}, though being partly noisy
\cite{Glorieux2010}, is very attractive due to the high effective
nonlinear coupling constants. Noiseless spatially-resolved
amplification \cite{Corzo2012} as well as entangled images
\cite{Boyer2008} in this scheme using $ {}^{85}Rb $ vapors have
been experimentally demonstrated.

Recently, the first experimental investigations of ultra-intense
twin beams have been reported both for multi-mode
\cite{Machulka2014,Allevi2014,Allevi2014a,Allevi2015,Allevi2015a,Spasibko2012}
as well as single-mode (bright squeezed-vacuum states) fields
\cite{Perez2014}. The effects of pump-field depletion have been
observed. Also back-flow of energy from the twin beam into the
pump field may occur. These processes affect spectra as well as
transverse profiles of the twin beams
\cite{Allevi2014a,Allevi2015,Allevi2015a}. Here, we restrict our
attention to the case of un-depleted pump fields (parametric
approximation). However and importantly, the generalization of the
theory to account for pump-field depletion and back-flow of energy
is possible due to the fact that the model incorporates all
spatio-spectral degrees of freedom. This generalization will be
reported elsewhere.

The paper is organized as follows. A spatio-spectral model of
parametric down-conversion based upon the Schmidt paired modes is
developed in Sec.~II. Quantities characterizing spectral and
temporal properties of twin beams are defined in Sec.~III. Several
suitable quantifiers of dimensionality of a twin beam are
introduced in Sec.~IV. Spectral and temporal properties of twin
beams as functions of the pump intensity are discussed in Sec.~V.
Properties of twin beams in the transverse wave-vector plane and
the crystal output plane are analyzed in Sec.~VI. Conclusions are
drawn in Sec.~VII. In Appendix~A, twin beams are described in
their transverse wave-vector plane (far field) and the crystal
output plane (near field).

\section{Theory of a spatio-spectral twin beam}

Optical parametric down-conversion and its evolution along a
nonlinear medium characterized by tensor $ d $ of second-order
nonlinear coefficients is described by the momentum operator $
\hat{G}_{\rm int} $ written in the interaction representation as
follows \cite{Perina1991,PerinaJr2000}:
\begin{eqnarray}   
 \hat{G}_{\rm int}(z) &=& 4 \epsilon_0 \int dxdy \int_{-\infty}^{\infty} dt \nonumber \\
 & & \hspace{-3mm} \left[
  d : E^{(+)}_{\rm p}({\bf r},t) \hat{E}^{(-)}_{\rm s}({\bf r},t)
  \hat{E}^{(-)}_{\rm i}({\bf r},t) + {\rm H.c.} \right];
\label{1}
\end{eqnarray}
$ {\bf r} = (x,y,z) $. In Eq.~(\ref{1}), $ E^{(+)}_{\rm p} $
describes the positive-frequency part of a classical pump
electric-field amplitude and $ \hat{E}^{(-)}_{\rm s} $ [$
\hat{E}^{(-)}_{\rm i} $] stands for the negative-frequency part of
a signal- [idler-] field operator amplitude. Symbol : is shorthand
for tensor shortening with respect to its three indices, $
\epsilon_0 $ is permittivity of vacuum and $ {\rm H.c.} $ replaces
the Hermitian conjugated term.

The electric-field amplitudes $ E^{(+)}_a({\bf r},t) $ [$
E^{(+)}_a({\bf r},t) = E^{(-)*}_a({\bf r},t) $] of the interacting
fields can be decomposed into harmonic plane waves with wave
vectors $ {\bf k}_a $ and frequencies $ \omega_a $:
\begin{eqnarray}   
 E^{(+)}_a({\bf r},t) &=& \frac{1}{\sqrt{2\pi}^3} \int
  d{\bf k}_a \, E^{(+)}_a({\bf k}_a) \exp(i{\bf k}_a{\bf r} -i\omega_a
  t), \nonumber \\
 & & \hspace{1cm} a={\rm p,s,i}.
\label{2}
\end{eqnarray}
We assume that the interacting fields can be described in paraxial
approximation which provides the relation $ {\bf k} =
(k_x,k_y,k_z) = ( k_x,k_y,k-[k_x^2+k_y^2]/2k) $, $ k = \sqrt{k_x^2
+k_y^2+k_z^2} $, valid for fields propagating close to the $ z $
direction.

We consider a strong pump field with the Gaussian spectrum and
Gaussian transverse profile. It is described in paraxial
approximation as follows:
\begin{eqnarray}   
 E^{(+)}_{\rm p}({\bf r},t) &=& \frac{1}{\sqrt{2\pi}^3} \int d{\bf k}_{\rm p}^\perp
  \int_{0}^{\infty} d\omega_{\rm p}
  \, E_{\rm p}^\perp({\bf k}_{\rm p}^\perp) E_{\rm p}^\parallel (\omega_{\rm p}) \nonumber \\
 & &  \hspace{-1cm} \mbox{} \times \exp(ik_{{\rm p},x} x) \exp(ik_{{\rm p},y} y) \exp(ik_{\rm p} z)
  \nonumber \\
 & &  \hspace{-1cm} \mbox{} \times \exp\left(-i\frac{k_{{\rm p},x}^2+k_{{\rm p},y}^2}{2k_{\rm p}}z\right)
  \exp(-i\omega_{\rm p} t) ;
\label{3}
\end{eqnarray}
$ {\bf k}_{\rm p}^\perp \equiv (k_{{\rm p},x},k_{{\rm p},y}) $ and
$ k_{\rm p} = k_{\rm p}(\omega_{\rm p}) $. Temporal spectrum $
E_{\rm p}^\parallel (\omega_{\rm p}) $ of the Gaussian pump pulse
is expressed as:
\begin{eqnarray}   
 E_{\rm p}^\parallel (\omega_{\rm p}) &=& \xi_{\rm p}
  \sqrt{\frac{\tau_{\rm p}}{\sqrt{2\pi}} } \exp\left[
  - \frac{\tau_{\rm p}^2 (\omega_{\rm p}-\omega_{\rm p}^0)^2}{4} \right].
\label{4}
\end{eqnarray}
According to Eq.~(\ref{4}), the pump pulse has amplitude $
\xi_{\rm p} $, duration $ \tau_{\rm p} $ and carrying frequency $
\omega^0_{\rm p} $. Provided that the pulsed pump field source has
power $ P_{\rm p} $ and repetition rate $ f $, amplitude $
\xi_{\rm p} $ is given by the relation
\begin{equation} 
 \xi_{\rm p} = \sqrt{\frac{P_{\rm p} \omega_{\rm p}}{\epsilon_0 c^2 k_{\rm p} f} },
\label{5}
\end{equation}
where $ c $ is the speed of light in vacuum. The pump field
radially symmetric in its transverse plane is characterized by the
Gaussian spatial spectrum
\begin{eqnarray}   
 E_{\rm p}^\perp ({\bf k}_{\rm p}^\perp) &=& \frac{w_{\rm p}}{\sqrt{2\pi}} \exp\left[
  - \frac{w_{\rm p}^2( k_{{\rm p},x}^2 + k_{{\rm p},y}^2)}{4}
  \right],
\label{6}
\end{eqnarray}
where $ w_{\rm p} $ gives the beam radius.

The signal and idler electric-field operator amplitudes $
\hat{E}^{(-)}_{\rm s} $ and $ \hat{E}^{(-)}_{\rm i} $ are
decomposed similarly as the pump field in paraxial approximation:
\begin{eqnarray}   
 \hat{E}^{(-)}_a({\bf r},t) &=& \frac{1}{\sqrt{2\pi}^3} \int d{\bf k}_a^\perp \int_{0}^{\infty}
  d\omega_a \, \hat{E}_a^{(-)}({\bf k}_a^\perp,\omega_a) \nonumber \\
 & &  \hspace{-1.3cm} \mbox{} \times \exp(-ik_{a,x} x) \exp(-ik_{a,y} y) \exp(-ik_a z)
  \nonumber \\
 & &  \hspace{-1.3cm} \mbox{} \times \exp\left(i\frac{k_{a,x}^2+k_{a,y}^2}{2k_a}z\right)
  \exp(i\omega_a t) , \hspace{3mm} a={\rm s,i}.
\label{7}
\end{eqnarray}
The spectral operator amplitudes $ \hat{E}_a^{(-)}({\bf
k}_a^\perp,\omega_a) $ can be expressed using creation operators $
\hat{a}^\dagger({\bf k}_a^\perp,\omega_a) $ that add a photon into
the mode with transverse wave vector $ {\bf k}_a^\perp $ and
frequency $ \omega_a $:
\begin{equation}  
 \hat{E}_a^{(-)}({\bf k}_a^\perp,\omega_a) = -i \sqrt{
  \frac{\hbar\omega_a^2}{2\epsilon_0 c^2 k_a} } \,
  \hat{a}_a^{\dagger}({\bf k}_a^\perp,\omega_a) ;
\label{8}
\end{equation}
$ \hbar $ is the reduced Planck constant. We note that $ n_a = c
k_a/\omega_a $ gives the index of refraction of field $ a $.  The
creation and annihilation operators fulfil the usual boson
commutation relations appropriate for the quantization of photon
flux \cite{Huttner1990,Vogel2001},
\begin{eqnarray} 
 [ \hat{a}_a({\bf k}_a^\perp,\omega_a), \hat{a}_{a'}^{\dagger}({\bf
  k'}_{a'}^\perp,\omega'_{a'})] &=& \delta_{aa'} \delta({\bf
  k}_a^\perp- {\bf k'}_{a'}^\perp) \nonumber \\
 & & \mbox{} \times \delta(\omega_a - \omega_{a'}),
\label{9}
\end{eqnarray}
where $ \delta $ means the Dirac $ \delta $ function and $
\delta_{aa'} $ stands for the Kronecker symbol.

Substituting expressions (\ref{3}) and (\ref{7}) into
Eq.~(\ref{1}) for momentum operator $ \hat{G}_{\rm int} $ we
arrive at the formula
\begin{eqnarray}   
 \hat{G}_{\rm int}(z) &=& -\frac{2\hbar d_{\rm eff}}{\sqrt{2\pi}^3 c^2}
  \int d{\bf k}_{\rm s}^\perp  \int d{\bf k}_{\rm i}^\perp
  \int_{0}^{\infty} d\omega_{\rm s} \int_{0}^{\infty} d\omega_{\rm i}  \nonumber \\
 & &  \hspace{-1cm} \int_{0}^{\infty} d\omega_{\rm p} \, \delta (\omega_{\rm p}
  -\omega_{\rm s}-\omega_{\rm i}) E_{\rm p}^\parallel(\omega_{\rm p})
  \frac{\omega_{\rm s} \omega_{\rm i}}{\sqrt{k_{\rm s} k_{\rm i}}} T_L({\bf k}_{\rm s}^\perp,{\bf
  k}_{\rm i}^\perp) \nonumber \\
 & &  \hspace{-1cm} \mbox{} \times \exp \left( i[k_{\rm p}(\omega_{\rm s}+\omega_{\rm i})
   -k_{\rm s}(\omega_{\rm s})-k_{\rm i}(\omega_{\rm i})]z \right) \nonumber \\
 & &  \hspace{-1cm} \mbox{} \times \hat{a}_{\rm s}^{\dagger}({\bf k}_{\rm s}^\perp,\omega_{\rm s},z)
  \hat{a}_{\rm i}^{\dagger}({\bf k}_{\rm i}^\perp,\omega_{\rm i},z)  + {\rm H.c.},
\label{10}
\end{eqnarray}
where $ d_{\rm eff} $ is an effective nonlinear coefficient.
Function $ T_L $ describes correlations between the signal and
idler fields in the transverse wave-vector plane:
\begin{eqnarray}  
 T_L({\bf k}_{\rm s}^\perp,{\bf k}_{\rm i}^\perp) &=& \int d{\bf k}_{\rm p}^\perp
  \delta ({\bf k}_{\rm p}^\perp-{\bf k}_{\rm s}^\perp-{\bf k}_{\rm i}^\perp)
  E_{\rm p}^\perp({\bf k}_{\rm p}^\perp) \nonumber \\
 & &  \hspace{-2.7cm} \mbox{} \times \frac{1}{L} \int_{0}^{L} dz
  \exp\left( -i\left[ \frac{|{\bf k}_{\rm p}^\perp|^2}{2k_{\rm p}}
  -\frac{|{\bf k}_{\rm s}^\perp|^2}{2k_{\rm s}} - \frac{|{\bf k}_{\rm i}^\perp|^2}{2k_{\rm i}}
  \right] z \right);
\label{11}
\end{eqnarray}
$ |{\bf k}_a^\perp|^2 = k_{a,x}^2 + k_{a,y}^2 $. We note that an
average value of the phase mismatch determined along the crystal
of length $ L $ occurs in formula (\ref{11}).

These correlations are conveniently expressed using dual
ortho-normal transverse modes of the signal and idler fields.
These modes are revealed by the Schmidt decomposition of the
normalized function $ T_L^n $, $ T_L = t^\perp T_L^n $ and $
t^{\perp 2} = \int d{\bf k}_{\rm s}^\perp \int d{\bf k}_{\rm
i}^\perp |T_L({\bf k}_{\rm s}^\perp,{\bf k}_{\rm i}^\perp)|^2 $.
As we are interested in the radially symmetric geometry, the use
of radial variables $ k_a^\perp $ and $ \varphi_a $ is convenient
[$ {\bf k}_a^\perp = ( k_a^\perp \cos(\varphi_a),k_a^\perp
\sin(\varphi_a)) $]. We note that the considered radial symmetry
is broken for narrow pump beams owing to the crystal anisotropy
\cite{Fedorov2008,Fedorov2007,PerinaJr2015}. In the radially
symmetric geometry, the normalized function $ T_L^n $ can be
rewritten into the form:
\begin{eqnarray}  
 T_L^n({\bf k}_{\rm s}^\perp,{\bf k}_{\rm i}^\perp) \hspace{-.5mm} &=& \hspace{-.5mm}
  \frac{1}{2\pi} \sum_{m=-\infty}^{\infty} T_{L,m}(k_{\rm s}^\perp,k_{\rm i}^\perp) \exp\left[
  i m (\varphi_{\rm s}-\varphi_{\rm i}) \right]. \nonumber \\
 & &
\label{12}
\end{eqnarray}
Functions $ T_{L,m} $ introduced in Eq.~(\ref{12}) and defined as
\begin{eqnarray}  
 T_{L,m}(k_{\rm s}^\perp,k_{\rm i}^\perp) &=& \int_{0}^{2\pi}
  d(\varphi_{\rm s}-\varphi_{\rm i}) \, T_L^n({\bf k}_{\rm s}^\perp,{\bf k}_{\rm i}^\perp)
  \nonumber \\
 & & \mbox{} \times \exp\left[-i m (\varphi_{\rm s}-\varphi_{\rm i}) \right]
\label{13}
\end{eqnarray}
can be decomposed as follows:
\begin{eqnarray} 
 \sqrt{k_{\rm s}^\perp k_{\rm i}^\perp} T_{L,m}(k_{\rm s}^\perp,k_{\rm i}^\perp) &=& \sum_{l=0}^{\infty}
  \lambda_{ml}^\perp u_{{\rm s},ml}(k_{\rm s}^\perp) u_{{\rm i},ml}(k_{\rm i}^\perp).
  \nonumber \\
 & &
\label{14}
\end{eqnarray}
Eigenfunctions $ u_{{\rm s},ml} $ and $ u_{{\rm i},ml} $ form the
ortho-normal dual bases and $ \lambda_{ml}^\perp $ denote the
corresponding eigenvalues. Substituting Eq.~(\ref{14}) into
Eq.~(\ref{12}) we reveal the Schmidt decomposition of the
normalized function $ T_L^n $:
\begin{eqnarray} 
 \sqrt{k_{\rm s}^\perp k_{\rm i}^\perp} T_{L}^n({k}_{\rm s}^\perp,\varphi_{\rm s},{k}_{\rm i}^\perp,\varphi_{\rm i})
  &=& \sum_{m=-\infty}^{\infty}
  \sum_{l=0}^{\infty} \lambda_{ml}^\perp
  t_{{\rm s},ml}(k_{\rm s}^\perp,\varphi_{\rm s})\nonumber \\
 & & \mbox{} \times
  t_{{\rm i},ml}(k_{\rm i}^\perp,\varphi_{\rm i}).
\label{15}
\end{eqnarray}
The transverse mode functions $ t_{{\rm s},ml} $ and $ t_{{\rm
i},ml} $ occurring in Eq.~(\ref{15}) take the form:
\begin{eqnarray}  
 t_{{\rm s},ml}(k_{\rm s}^\perp,\varphi_{\rm s}) &=& \frac{u_{{\rm s},ml}(k_{\rm s}^\perp)\exp(im\varphi_{\rm s})}{
  \sqrt{2\pi}} ,
  \nonumber \\
 t_{{\rm i},ml}(k_{\rm i}^\perp,\varphi_{\rm i}) &=&  \frac{u_{{\rm i},ml}(k_{\rm i}^\perp)\exp(-im\varphi_{\rm i})}{
  \sqrt{2\pi}} .
\label{16}
\end{eqnarray}

Introduction of field operators $ \hat{a}_{a,ml}(\omega_a,z) $
related to transverse mode functions $ t_{a,ml} $,
\begin{eqnarray}  
 \hat{a}_{a,ml}(\omega_a,z) &=& \int_{0}^{\infty} dk_a^\perp \int_{0}^{2\pi} d\varphi_a
  \, t_{a,ml}^*(k_a^\perp,\varphi_a) \nonumber \\
 & & \mbox{} \times  \hat{a}_{a}(k_a^\perp,\varphi_a,\omega_a,z),
  \hspace{5mm} a={\rm s,i},
\end{eqnarray}
allows to rewrite the interaction momentum operator $ \hat{G}_{\rm
int} $ in Eq.~(\ref{10}) as follows:
\begin{eqnarray}   
 \hat{G}_{\rm int}(z) &=& -\frac{2\hbar d_{\rm eff}t^\perp}{\sqrt{2\pi}^3 c^2}
  \sum_{m,l} \lambda_{ml}^\perp \int_{0}^{\infty} d\omega_{\rm s} \int_{0}^{\infty}
  d\omega_{\rm i}
  \, \frac{\omega_{\rm s} \omega_{\rm i}}{\sqrt{k_{\rm s} k_{\rm i}}} \nonumber \\
 & & \hspace{-12mm} \mbox{} \times  E_{\rm p}^\parallel(\omega_{\rm s}+\omega_{\rm i})
  \exp \left( i[k_{\rm p}(\omega_{\rm s}+\omega_{\rm i})-k_{\rm s}(\omega_{\rm s})-k_{\rm i}(\omega_{\rm i})]z \right)
  \nonumber \\
 & & \hspace{-12mm} \mbox{} \times \hat{a}_{{\rm s},ml}^{\dagger}(\omega_{\rm s},z)
   \hat{a}_{{\rm i},ml}^{\dagger}(\omega_{\rm i},z) + {\rm H.c.}
\label{18}
\end{eqnarray}

If the nonlinear interaction is weak, we can obtain a perturbation
solution of the corresponding Schr\" odinger equation and express
the output state $ |\psi\rangle_{\rm out} $ in the form:
\begin{equation} 
 |\psi\rangle_{\rm out} = - \frac{i}{\hbar} \int_{0}^{L} dz
  \, \hat{G}_{\rm int}(z) |\psi\rangle_{\rm in},
\label{19}
\end{equation}
where $ |\psi\rangle_{\rm in} $ is the input signal and idler
state. Substitution of Eq.~(\ref{18}) into Eq.~(\ref{19}) and
consideration of the input vacuum state $ |{\rm vac}\rangle $
result in the formula
\begin{eqnarray}  
 |\psi\rangle_{\rm out} &=& t^\perp \sum_{m,l} \lambda_{ml}^\perp
  \int_{0}^{\infty} d\omega_{\rm s} \int_{0}^{\infty} d\omega_{\rm i} \,
  F_L(\omega_{\rm s},\omega_{\rm i})\nonumber \\
 & & \mbox{} \times
  \hat{a}_{{\rm s},ml}^{\dagger}(\omega_{\rm s},0)
   \hat{a}_{{\rm i},ml}^{\dagger}(\omega_{\rm i},0) |{\rm vac}\rangle,
\label{20}
\end{eqnarray}
where
\begin{eqnarray}  
 F_L(\omega_{\rm s},\omega_{\rm i}) &=& \frac{2id_{\rm eff}}{\sqrt{2\pi}^3 c^2}
  \frac{\omega_{\rm s}\omega_{\rm i}}{\sqrt{k_{\rm s} k_{\rm i}}} E_{\rm p}^\parallel
  (\omega_{\rm s}+\omega_{\rm i}) \nonumber \\
 & & \hspace{-2.5cm} \mbox{} \times  \int_{0}^{L} dz
  \exp \left( i[k_{\rm p}(\omega_{\rm s}+\omega_{\rm i})-k_{\rm s}(\omega_{\rm s})-k_{\rm i}(\omega_{\rm i})]z\right).
\label{21}
\end{eqnarray}
We note that the vacuum state $ |{\rm vac}\rangle $ is omitted in
the expression for the output state $ |\psi\rangle_{\rm out} $ in
Eq.~(\ref{20}).

Using eigenfunctions $ f_{{\rm s},q} $ and $ f_{{\rm i},q} $ and
eigenvalues $ \lambda_q^\parallel $ of the Schmidt decomposition
of the normalized function $ F_L^n $ [$ F_L = f^\parallel F_L^n $,
$ f^{\parallel 2} = \int d\omega_{\rm s} \int d\omega_{\rm i}
|F_L(\omega_{\rm s},\omega_{\rm i})|^2 $], we can rewrite
Eq.~(\ref{21}) as follows:
\begin{equation}  
 F_L(\omega_{\rm s},\omega_{\rm i}) = f^\parallel \sum_{q=0}^{\infty}
 \lambda_q^\parallel f_{{\rm s},q}(\omega_{\rm s}) f_{{\rm i},q}(\omega_{\rm i}).
\label{22}
\end{equation}
New field operators $ \hat{a}_{a,mlq} $ defined as
\begin{eqnarray}  
 \hat{a}_{a,mlq} &=& \int_{0}^{\infty} d\omega_a \, f_{a,q}^*(\omega_a)
  \hat{a}_{a,ml}(\omega_a,0) ,  \hspace{3mm} a={\rm s,i}, \nonumber \\
 & &
\label{23}
\end{eqnarray}
provide a simple formula for the output state $ |\psi\rangle_{\rm
out} $:
\begin{eqnarray}  
 |\psi\rangle_{\rm out} = t^\perp f^\parallel \sum_{m,l,q} \lambda_{ml}^\perp
  \lambda_q^\parallel \hat{a}_{{\rm s},mlq}^{\dagger}
   \hat{a}_{{\rm i},mlq}^{\dagger} |{\rm vac}\rangle .
\label{24}
\end{eqnarray}
According to Eq.~(\ref{24}) the output state $ |\psi\rangle_{\rm
out} $ is composed of photon pairs in independent paired modes
numbered by indices $ (m,l,q) $ with probability amplitudes $
t^\perp f^\parallel \lambda_{ml}^\perp \lambda_q^\parallel $.

Using the paired modes revealed both in the transverse wave-vector
plane and spectrum we rewrite the 'averaged' momentum operator $
\int_{0}^{L} dz \hat{G}_{\rm int}(z) / L $ from Eq.~(\ref{18})
into the form:
\begin{eqnarray}     
 \hat{G}_{\rm int}^{\rm av}(z) &=& - \frac{i\hbar t^\perp f^\parallel}{L}
  \sum_{m=-\infty}^{^\infty} \sum_{l,q=0}^{\infty} \lambda_{ml}^\perp
  \lambda_q^\parallel \hat{a}_{{\rm s},mlq}^{\dagger}(z)
   \hat{a}_{{\rm i},mlq}^{\dagger}(z) \nonumber \\
 & & \mbox{} + {\rm H.c.}
\label{25}
\end{eqnarray}
using the operators $ \hat{a}_{a,mlq} $ defined in Eq.~(\ref{23}).
The crucial advantage of the 'averaged' momentum operator $
\hat{G}_{\rm int}^{\rm av} $ is that it 'diagonalizes' the
nonlinear interaction leaving the separated Heisenberg equations
for each pair of modes. We note that some of the paired modes are
degenerate in certain symmetric configurations (e.g. collinear
spectrally-degenerate emission) in the sense that both the signal
and idler photons are emitted into the same spatio-spectral mode
\cite{Fedorov2014}. However, we do not consider explicitly such
modes here. Considering an $ (m,l,q) $-th mode, the Heisenberg
equations are written as follows:
\begin{eqnarray}   
 \frac{ d\hat{a}_{{\rm s},mlq}(z)}{dz} &=&
   K_{mlq}\hat{a}_{{\rm i},mlq}^\dagger(z) , \nonumber \\
 \frac{ d\hat{a}_{{\rm i},mlq}(z)}{dz} &=&
   K_{mlq}\hat{a}_{{\rm s},mlq}^\dagger(z)
\label{26}
\end{eqnarray}
using effective nonlinear coupling constants $ K_{mlq} $,
\begin{equation}  
 K_{mlq} = \frac{t^\perp f^\parallel}{L} \lambda_{ml}^\perp
 \lambda_q^\parallel .
\label{27}
\end{equation}

The solution of linear equations (\ref{26}) for an $ (m,l,q) $-th
mode and the crystal of length $ L $ takes a simple form:
\begin{eqnarray}  
 \hat{a}_{{\rm s},mlq}(L) &=& \cosh(K_{mlq}L) \hat{a}_{{\rm s},mlq}(0) \nonumber \\
 & & \mbox{} + \sinh(K_{mlq}L) \hat{a}_{{\rm i},mlq}^\dagger(0) ,  \nonumber \\
 \hat{a}_{{\rm i},mlq}(L) &=& \cosh(K_{mlq}L) \hat{a}_{{\rm i},mlq}(0) \nonumber \\
 & & \mbox{} + \sinh(K_{mlq}L) \hat{a}_{{\rm s},mlq}^\dagger(0) .
\label{28}
\end{eqnarray}

The solution for a given transverse mode $ (m,l) $ can be
conveniently expressed in the matrix form:
\begin{eqnarray}   
 \hat{\bf a}_{{\rm s},ml}(L) &=& {\bf U}_{{\rm s},ml} \hat{\bf a}_{{\rm s},ml}(0)
  + {\bf V}_{ml} \hat{\bf a}_{{\rm i},ml}^\dagger(0) , \nonumber \\
 \hat{\bf a}_{{\rm i},ml}(L) &=& {\bf U}_{{\rm i},ml} \hat{\bf a}_{{\rm i},ml}(0)
  + {\bf V}_{ml}^\dagger \hat{\bf a}_{{\rm s},ml}^\dagger(0) .
\label{29}
\end{eqnarray}
The matrices $ {\bf U}_{{\rm s},ml} $, $ {\bf U}_{{\rm i},ml} $
and $ {\bf V}_{ml} $ introduced in Eq.~(\ref{29}) are written in
their singular-valued decompositions as follows:
\begin{eqnarray}  
 {\bf U}_{a,ml} &=& {\bf F}_{a,ml} {\bf \Lambda}^U_{ml} {\bf
   F}_{a,ml}^T , \hspace{3mm} a={\rm s,i}, \nonumber \\
 {\bf V}_{ml} &=& {\bf F}_{{\rm s},ml} {\bf \Lambda}^V_{ml} {\bf
   F}_{{\rm i},ml}^\dagger .
\label{30}
\end{eqnarray}
Columns of the matrices $ {\bf F}_{{\rm s},ml} $ ($ {\bf F}_{{\rm
i},ml} $) in Eq.~(\ref{30}) are given by eigenmodes $ f_{{\rm
s},q} $ ($ f_{{\rm i},q} $) of the Schmidt decomposition written
in Eq.~(\ref{22}). Elements of the diagonal matrices $ {\bf
\Lambda}^U_{ml} $ and $ {\bf \Lambda}^V_{ml} $ are derived from
the solution given in Eq.~(\ref{28}),
\begin{eqnarray}   
 {\bf \Lambda}^U_{ml,qq} &=& U_{mlq} = \cosh(K_{mlq}L) , \nonumber \\
 {\bf \Lambda}^V_{ml,qq} &=& V_{mlq} = \sinh(K_{mlq}L) .
\label{31}
\end{eqnarray}

The solution (\ref{28}) allows to derive the mean values of
experimental physical quantities. Spectral and temporal quantities
are determined in Sec.~3 below. Spatial quantities are defined in
Appendix~A. Numbers of modes constituting the twin beam are
described in Sec.~4.

We note that the numerical results obtained in \cite{Christ2013}
show that mild broadening of the modes determined from the
perturbation solution of the Schr\"{o}dinger equation occurs for
strong pumping of the nonlinear process.

We also note that, in the considered radially symmetric
non-collinear geometry with the pump field at normal incidence,
the signal and idler fields propagate along the radial emission
angles $ \vartheta_{\rm s} $ and $ \vartheta_{\rm i} $,
respectively. The central radial emission angles $ \vartheta_{\rm
s}^0 $ and $ \vartheta_{\rm i}^0 $ corresponding to the central
frequencies $ \omega_{\rm s}^0 $ and $ \omega_{\rm i}^0 $ are
given by the conservation of energy and transverse wave vectors:
\begin{eqnarray} 
 \omega_{\rm s}^0 = \omega_{\rm p}^0 - \omega_{\rm i}^0 ,
 \hspace{5mm} k_{\rm s}^0 \sin(\vartheta_{\rm s}^0) = k_{\rm i}^0 \sin(\vartheta_{\rm i}^0) ,
\label{32}
\end{eqnarray}
$ k_a^0 = k_a(\omega_a^0) $. The central transverse wave vectors $
k_a^{\perp 0} $ are then given as $ k_a^{\perp 0} = k_a^0
\cos(\vartheta_a^0) $, $ a={\rm s,i} $. Paraxial approximation
along the radial emission angle $ \vartheta_a^0 $ provides the
following formula for wave vector $ {\bf k}_a $ ($ a={\rm s,i} $):
\begin{eqnarray}  
 & & {\bf k}_a = \Bigl( [k_a^{\perp 0} + \delta
 k_a]\cos(\varphi_a), [k_a^{\perp 0} + \delta
 k_a]\sin(\varphi_a), \nonumber \\
 & & \hspace{1cm} \left[ k_a - \frac{\delta k_a^2
 \cos(\vartheta_{\rm s}^0)^2}{2k_a} \right] \cos(\vartheta_{\rm s}^0)
 \Bigr),
\label{33}
\end{eqnarray}
where $ \delta k_a $ gives the declination of the transverse wave
vector of field $ a $. The derived formulas valid for the
close-to-collinear geometry can be applied in general also in the
non-collinear case provided that the following formal substitution
is used:
\begin{eqnarray}  
 k_a \longleftarrow k_a \cos(\vartheta_{\rm s}^0) , \,\,
 & & \delta k_a \longleftarrow \delta k_a \cos(\vartheta_{\rm s}^0)^2 .
\label{34}
\end{eqnarray}

\section{Spectral and temporal properties of twin beams}

We assume that the transverse profiles of twin beams are not
experimentally resolved and so the experimental mean values are
obtained by averaging over the transverse modes. Then the
signal-field intensity spectrum $ n_{{\rm s},\omega} $ is
expressed as follows:
\begin{eqnarray}   
 n_{{\rm s},\omega}(\omega_{\rm s}) &=& \langle \hat{a}_{\rm s}^\dagger(\omega_{\rm s},L)
  \hat{a}_{\rm s}(\omega_{\rm s},L) \rangle_\perp \nonumber \\
 &=& \sum_{ml} \sum_{q} |f_{{\rm s},q}(\omega_{\rm s})|^2 V_{mlq}^2 .
\label{35}
\end{eqnarray}
Symbol $ \langle \rangle_\perp $ denotes quantum mechanical
averaging combined with averaging in the transverse plane. The
number $ N_{\rm s} $ of generated signal photons is determined
along the formula
\begin{eqnarray}   
 N_{\rm s} &=& \int_{0}^{\infty} d\omega_{\rm s} \, n_{{\rm s},\omega}(\omega_{\rm s}) =
 \sum_{ml} \sum_{q} V_{mlq}^2 .
\label{36}
\end{eqnarray}

Averaged signal-field intensity spectral correlations are
characterized by the fourth-order correlation function $ A_{{\rm
s},\omega} $ given as:
\begin{eqnarray}   
 A_{{\rm s},\omega}(\omega_{\rm s},\omega'_{\rm s}) &=& \langle {\cal N}: \Delta[
  \hat{a}_{\rm s}^\dagger(\omega_{\rm s},L)\hat{a}_{\rm s}(\omega_{\rm s},L)]
  \nonumber \\
 & & \mbox{} \times \Delta[\hat{a}_{\rm s}^\dagger(\omega'_{\rm s},L)
  \hat{a}_{\rm s}(\omega'_{\rm s},L)]:\rangle_{\perp} \nonumber \\
 &=& \sum_{ml} |A_{{\rm s},ml,\omega}^{\rm a}(\omega_{\rm s},\omega'_{\rm s})|^2 .
\label{37}
\end{eqnarray}
The signal-field amplitude correlation function $ A_{{\rm
s},ml,\omega}^{\rm a} $ belonging to mode $ (m,l) $ is written in
the form:
\begin{eqnarray}   
 A_{{\rm s},ml,\omega}^{\rm a}(\omega_{\rm s},\omega'_{\rm s}) &=&
  \langle \hat{a}_{\rm s}^\dagger(\omega_{\rm s},L)
  \hat{a}_{\rm s}(\omega'_{\rm s},L) \rangle_{\perp,ml} \nonumber \\
 &=&
  \sum_{q} f_{{\rm s},q}^*(\omega_{\rm s}) f_{{\rm s},q}(\omega'_{\rm s})  V_{mlq}^2 .
\label{38}
\end{eqnarray}

Intensity spectral cross-correlations between the signal and idler
fields are quantified by the following fourth-order correlation
function:
\begin{eqnarray}   
 C_{\omega}(\omega_{\rm s},\omega_{\rm i}) &=& \langle {\cal N}: \Delta[
  \hat{a}_{\rm s}^\dagger(\omega_{\rm s},L) \hat{a}_{\rm s}(\omega_{\rm s},L)] \nonumber
  \\
 & & \mbox{} \times  \Delta[ \hat{a}_{\rm i}^\dagger(\omega_{\rm i},L)
  \hat{a}_{\rm i}(\omega_{\rm i},L)]: \rangle_\perp \nonumber \\
 &=& \sum_{ml} \left|
  \sum_{q} f_{{\rm s},q}(\omega_{\rm s}) f_{{\rm i},q}(\omega_{\rm i}) U_{mlq} V_{mlq}
  \right|^2 . \nonumber \\
 & &
\label{39}
\end{eqnarray}

Temporal electric-field amplitude and intensity correlations in
the twin beams outside the nonlinear crystal can be expressed,
similarly as their spectral correlations, in terms of temporal
eigenfunctions $ \tilde f_{a,q}(t_a) $ determined by the Fourier
transform:
\begin{eqnarray}   
 \tilde f_{a,q}(t_a) = \sqrt{\frac{\hbar}{2\pi}}
  \int d\omega_a \sqrt{\omega_a} f_{a,q}(\omega_a)
  \exp(-i\omega_a t_a).
\label{40}
\end{eqnarray}
The averaged signal-field photon flux $ I_{{\rm s},t} $ is then
derived in terms of functions $ \tilde f_{{\rm s},q} $,
\begin{eqnarray}   
 I_{{\rm s},t}(t_{\rm s}) &=& 2 \epsilon_0 c \langle \hat{E}_{\rm s}^{(-)}({\bf r}_{\rm s}^\perp,L,t_{\rm s})
  \hat{E}_{\rm s}^{(+)}({\bf r}_{\rm s}^\perp,L,t_{\rm s}) \rangle_\perp \nonumber \\
 &=& \sum_{ml} \sum_{q} |\tilde{f}_{{\rm s},q}(t_{\rm s})|^2 V_{mlq}^2 .
\label{41}
\end{eqnarray}
The averaged signal-field intensity temporal correlation function
$ A_{{\rm s},t} $ is expressed similarly as the spectral
correlation function $ A_{{\rm s},\omega} $ given in
Eq.~(\ref{37}),
\begin{eqnarray}   
 A_{{\rm s},t}(t_{\rm s},t'_{\rm s}) &=& (2\epsilon_0 c)^2 \langle {\cal N}: \Delta[
  \hat{E}_{\rm s}^{(-)}({\bf r}_{\rm s}^\perp,L,t_{\rm s}) \nonumber \\
 & & \hspace{-19mm} \mbox{} \times \hat{E}_{\rm s}^{(+)}({\bf r}_{\rm s}^\perp,L,t_{\rm s})]
  \Delta[\hat{E}_{\rm s}^{(-)}({\bf r}_{\rm s}^\perp,L,t'_{\rm s})
  \hat{E}_{\rm s}^{(+)}({\bf r}_{\rm s}^\perp,L,t'_{\rm s})] : \rangle_\perp \nonumber \\
 &=& \sum_{ml} \left| A_{{\rm s},ml,t}^{\rm a}(t_{\rm s},t'_{\rm s}) \right|^2 .
\label{42}
\end{eqnarray}
The signal-field amplitude temporal correlation function $ A_{{\rm
s},ml,t}^{\rm a} $ of mode $ (m,l) $ is determined along the
formula
\begin{eqnarray}   
 A_{{\rm s},ml,t}^{\rm a}(t_{\rm s},t'_{\rm s}) &=& 2\epsilon_0 c \langle
  \hat{E}_{\rm s}^{(-)}({\bf r}_{\rm s}^\perp,L,t_{\rm s})
  \hat{E}_{\rm s}^{(+)}({\bf r}_{\rm s}^\perp,L,t'_{\rm s})
  \rangle_{\perp,ml} \nonumber \\
 &=& \sum_{q} \tilde{f}_{{\rm s},q}^*(t_{\rm s}) \tilde{f}_{{\rm s},q}(t'_{\rm s})
  V_{mlq}^2.
\label{43}
\end{eqnarray}
Also the averaged intensity temporal cross-correlations between
the signal and idler fields can be quantified in the same vein as
in Eq.~(\ref{39}):
\begin{eqnarray}   
 C_{t}(t_{\rm s},t_{\rm i}) &=& (2\epsilon_0 c)^2 \langle {\cal N}: \Delta[
  \hat{E}_{\rm s}^{(-)}({\bf r}_{\rm s}^\perp,L,t_{\rm s}) \nonumber \\
  & & \hspace{-15mm} \mbox{} \times \hat{E}_{\rm s}^{(+)}({\bf r}_{\rm s}^\perp,L,t_{\rm s})]
  \Delta[ \hat{E}_{\rm i}^{(-)}({\bf r}_{\rm i}^\perp,L,t_{\rm i})
  \hat{E}_{\rm i}^{(+)}({\bf r}_{\rm i}^\perp,L,t_{\rm i})]:\rangle_\perp \nonumber \\
 &=& \sum_{ml} \left|
  \sum_{q} \tilde{f}_{{\rm s},q}(t_{\rm s}) \tilde{f}_{{\rm i},q}(t_{\rm i}) U_{mlq} V_{mlq}
  \right|^2 .
\label{44}
\end{eqnarray}

As the number of transverse modes is usually large, their
eigenvalues $ \lambda_{ml}^\perp $ form quasi-continuum. In this
case, we may introduce a suitable probability function $
\varrho_\lambda $ and make the following replacement in the above
formulas:
\begin{equation}  
 \sum_{ml} \longrightarrow \int_{0}^{1}
  d\lambda^\perp \varrho_{\lambda}(\lambda^\perp) .
\label{45}
\end{equation}
This makes the numerical computations considerably faster.

The mode $ (m,l,q) = (0,0,0) $ having the largest value of the
product $ \lambda^\perp \lambda^\parallel $ of the Schmidt
eigenvalues becomes dominant in the limit of large pump power ($
P_p \rightarrow \infty $) in the used un-depleted pump
approximation. The spectral characteristics $ n_{{\rm s},\omega}
$, $ A_{{\rm s},\omega} $ and $ C_{{\rm s},\omega} $ then attain
the simple form:
\begin{eqnarray}  
 n_{{\rm s},\omega}(\omega_{\rm s}) &=& N_{\rm s} |f_{{\rm s},0}(\omega_{\rm s})|^2, \nonumber
 \\
 A_{{\rm s},\omega}(\omega_{\rm s},\omega'_{\rm s}) &=& N_{\rm s}^2 |f_{{\rm s},0}(\omega_{\rm s})|^2
  |f_{{\rm s},0}(\omega'_{\rm s})|^2, \nonumber \\
 C_{\omega}(\omega_{\rm s},\omega_{\rm i}) &=& \lambda^{\perp 2}_{00} U_{00,0}^2
  V_{00,0}^2 |f_{{\rm s},0}(\omega_{\rm s})|^2 |f_{{\rm i},0}(\omega_{\rm i})|^2, \nonumber \\
\label{46}
\end{eqnarray}
where $ N_{\rm s} = V_{00,0}^2 $ gives the number of emitted
signal photons. According to Eqs.~(\ref{46}), the twin beam is
spectrally composed of independent single-mode signal and idler
fields in this high-intensity (classical) limit. We note that one
dominant paired mode constitutes the twin beam also in the
transverse wave-vector plane and the crystal output plane. So the
signal and idler fields are spatially and spectrally independent
but internally fully spatially and spectrally coherent.

Similar quantities as defined above in the spectral and temporal
domains are used for describing the twin beams in their transverse
wave-vector plane and crystal output plane. Modes in the crystal
output plane are determined from those of the transverse
wave-vector plane using the Fourier transform, similarly as the
temporal modes have been derived from the spectral modes. The
radial symmetry of twin beams results in harmonic azimuthal modes
in the crystal output plane. It also provides the radial modes in
the crystal output plane determined from those of the transverse
wave-vector plane using the transformation based on the Bessel
functions (for details, see Appendix~A).

\section{Dimensionality of the twin beam}

Dimensionality of a twin beam can be determined either using its
paired properties or properties of the individual signal and idler
fields. In the first case, dimensionality of entanglement is
obtained. In the second case, the numbers of independent modes
constituting the signal (or idler) field and defined in
statistical optics are reached. Entanglement dimensionality for a
general noisy twin beam is quantified via negativity
\cite{Eltschka2013}. Considering pure states of the noiseless twin
beams, the Schmidt number can be applied for quantifying
entanglement dimensionality as well \cite{Law2000,Law2004}. This
number can even be reached without making the Schmidt
decomposition \cite{Gatti2012,Horoshko2012}. The general formulas
can be recast into a simple form for quasi-monochromatic or
quasi-homogeneous fields \cite{DiLorenzoPires2009}.

Compared to weak fields, the analysis of intense twin beams is
more difficult, as the decompositions not only in the spatial and
spectral domains but also in the Hilbert spaces of individual
paired spatio-spectral modes spanned by the Fock-number states
would be needed. That is why, we apply here a simpler approach for
determining entanglement dimensionality based upon defining
creation operators for photon pairs (for details, see
\cite{PerinaJr2013}). Entanglement dimensionality $ K $ of the
twin beam is obtained in this approach as follows:
\begin{equation}   
 K = \frac{ \left(\sum_{mlq} U_{mlq}^2V_{mlq}^2 \right)^2 }{
  \sum_{mlq} U_{mlq}^4V_{mlq}^4 }.
\label{47}
\end{equation}
We note that formula (\ref{47}) reduces to the usually used
Schmidt number of spatio-spectral modes for weak twin beams.

Formula (\ref{47}) can also be applied to provide average number $
K_\omega $ of effectively populated paired spectral modes:
\begin{equation}   
 K_\omega = \sum_{ml} p_{ml}^\perp \frac{ \left(\sum_{q}
  U_{mlq}^2V_{mlq}^2 \right)^2 }{\sum_{q} U_{mlq}^4V_{mlq}^4 }.
\label{48}
\end{equation}
In Eq.~(\ref{47}), $ p_{ml}^\perp $ gives the probability of
having a photon pair in mode $ (m,l) $:
\begin{equation}  
 p_{ml}^\perp = \frac{ \sum_q V_{mlq}^2 }{ \sum_{mlq} V_{mlq}^2 }.
\label{49}
\end{equation}

Similarly as in the spectrum, average number $ K_{k\varphi} $ of
effectively populated modes in the transverse wave-vector plane is
obtained along the formula:
\begin{equation}   
 K_{k\varphi} = \sum_{q} p_{q}^\parallel \frac{ \left(\sum_{ml}
  U_{mlq}^2 V_{mlq}^2 \right)^2 }{\sum_{q} U_{mlq}^4V_{mlq}^4 }
\label{50}
\end{equation}
using probability $ p_{ml}^\parallel $ of having a photon pair in
mode $ q $;
\begin{equation}  
 p_{q}^\parallel = \frac{ \sum_{ml} V_{mlq}^2 }{ \sum_{mlq} V_{mlq}^2 }.
\label{51}
\end{equation}

The numbers $ K_\omega $ and $ K_{k\varphi} $ of paired modes in
the spectrum and transverse wave-vector plane, respectively, can
alternatively be determined as the ratio of the width $ \Delta
n_{s} $ of, say, the signal-field intensity profile and the width
of intensity cross-correlation function $ \Delta C $ in the
appropriate variable. This ratio known as the Fedorov ratio
\cite{Fedorov2005} coincides with the number $ K_\omega $ of
paired modes given in Eq.~(\ref{48}) for weak twin beams with a
Gaussian two-photon amplitude \cite{Chan2007}. It has been shown
in \cite{Mikhailova2008} that both numbers are close to each other
for general weak twin beams.

Modes in the signal and idler fields are ideally paired as well as
the signal and idler photons in individual spatio-spectral modes
for the considered noiseless twin beams. That is why,
dimensionality of the twin beam can also be determined from the
number of modes and their populations counted either in the signal
or idler field. Applying the coherence theory \cite{Perina1985},
an effective number of independent modes (degrees of freedom) in
the signal (or idler) field can be obtained from its photon-number
statistics. The resulting number $ K^n $ of modes constituting,
e.g., the signal field is given by the formula valid for a
multimode thermal field \cite{Perina2005}:
\begin{eqnarray}  
 K^n &=& \frac{ \left( \sum_{mlq} \langle \hat{n}_{s,mlq}
  \rangle \right)^2 }{ \sum_{mlq} \left( \langle {\cal N} :
  \hat{n}_{s,mlq}^2: \rangle - \langle \hat{n}_{s,mlq} \rangle^2 \right) }
  \nonumber \\
 &=& \frac{ \left( \sum_{mlq} V_{mlq}^2 \right)^2 }{ \sum_{mlq}
  V_{mlq}^4 } ;
\label{52}
\end{eqnarray}
$ \hat{n}_{s,mlq} \equiv \hat{a}_{s,mlq}^\dagger(L)
\hat{a}_{s,mlq}(L) $.

Also, the formula for averaged number $ K^n_{\omega} $ of spectral
modes can be written, in analogy with the derivation of
Eq.~(\ref{48}) from Eq.~(\ref{47}):
\begin{eqnarray}  
 K^n_{\omega} &=& \sum_{ml} p_{ml}^\perp K^n_{\omega,ml},
\label{53} \\
 K^n_{\omega,ml} &=& \frac{ \left( \sum_{q} \langle \hat{n}_{s,mlq}
  \rangle \right)^2 }{ \sum_{q} \left( \langle {\cal N} :
  \hat{n}_{s,mlq}^2: \rangle - \langle \hat{n}_{s,mlq} \rangle^2 \right) }
  \nonumber \\
 &=& \frac{ \left( \sum_{q} V_{mlq}^2 \right)^2 }{ \sum_{q}
  V_{mlq}^4 } . \nonumber
\end{eqnarray}
The averaged number $ K^n_{k\varphi} $ of modes in the transverse
wave-vector plane can be determined by a formula analogous to that
written in Eq.~({\ref{53}) [compare Eqs.~(\ref{48}) and
(\ref{50}}].

The ratio $ K^\Delta_{\rm s} $ of the width $ \Delta n_{\rm s} $
of a signal-field intensity profile and the width $ \Delta A_{\rm
s}^{\rm a} $ of the appropriate signal-field amplitude
auto-correlation function,
\begin{equation} 
 K^\Delta_{\rm s} = \frac{\Delta n_{\rm s}}{\Delta A_{\rm s}^{\rm a}} ,
\label{54}
\end{equation}
defined in any variable represents also a good quantifier of the
number of independent modes of a twin beam in this variable. We
compare different quantifiers of dimensionality of the twin beam
under real experimental conditions below.

\section{Spectral and temporal properties of intense twin beams}

In the numerical analysis, we consider a BBO crystal 8-mm long cut
for non-collinear type I process (eoo) for the
spectrally-degenerate interaction pumped by the pulse at
wavelength $ \lambda_{\rm p} = 349 $~nm with spectral width $
\Delta\lambda_{\rm p} = 0.1 $~nm, transverse profile with radius $
w_{\rm p} = 1 $~mm and repetition rate $ f = 400 $~s$ {}^{-1} $.
This pulse is provided by the third harmonics of the Nd:YLF laser
at wavelength 1.047~$\mu $m. Assuming the pump field at normal
incidence, the signal and idler fields at the central wavelengths
$ \lambda_{\rm s}^0 = \lambda_{\rm i}^0 = 698 $~nm ($
\vartheta_{\rm BBO} = 36.3 $~deg) propagate outside the crystal
under the radial emission angles $ \vartheta_{\rm s}^0 =
\vartheta_{\rm i}^0 = 8.45 $~deg. As this configuration is
symmetric for the signal and idler fields, we further discuss only
the properties of signal field. We assume that the conditions are
such that the spectral and spatial properties of the twin beam
factorize.

\begin{figure}         
 \resizebox{0.47\hsize}{!}{\includegraphics{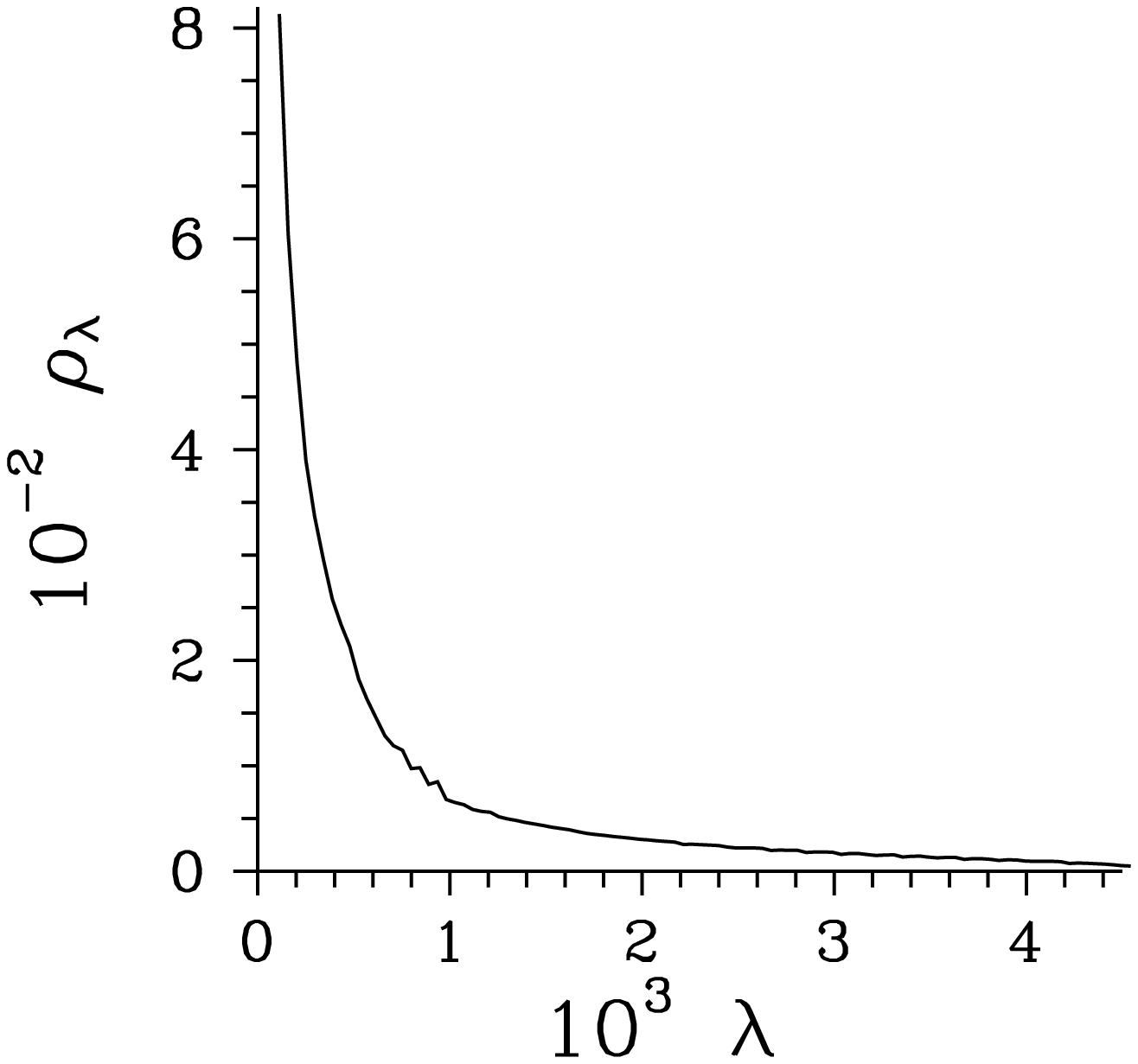}}
  \hspace{1mm}
 \resizebox{0.47\hsize}{!}{\includegraphics{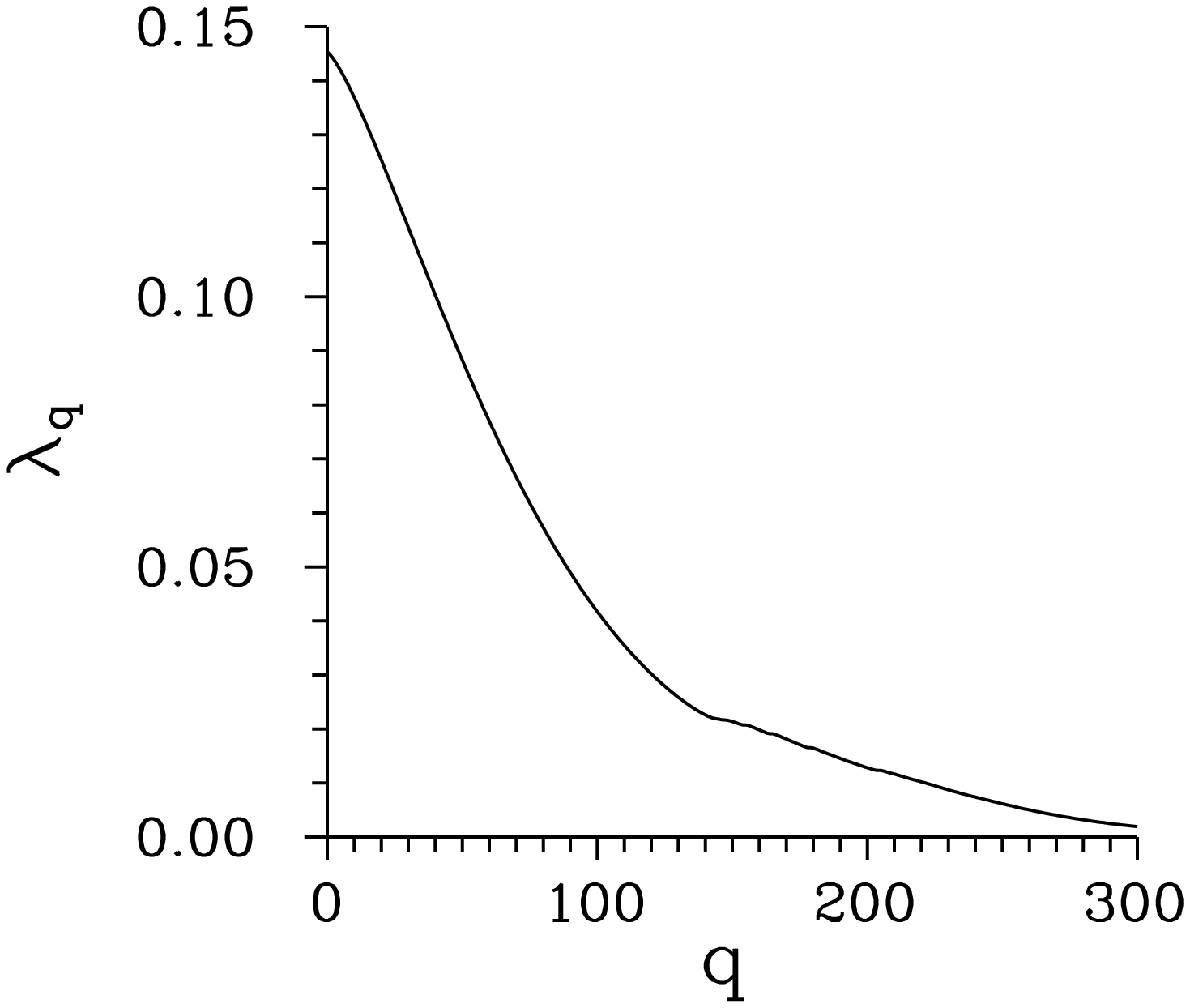}}

 (a) \hspace{.5\hsize} (b)

 \caption{(a) Probability function $ \varrho_{\lambda} $ of
  eigenvalues $ \lambda^\perp_{ml} $ in the transverse wave-vector plane and (b) spectral
  eigenvalues $ \lambda_q $; $ w_{\rm p} = 1 \times 10^{-3} $~m, $ \Delta\lambda_{\rm
  p} = 1 \times 10^{-10} $~m.}
 \label{fig1}
\end{figure}

The generated twin beam is composed of roughly 80 thousand
transverse modes at low intensity. It contains 34 modes in radial
direction and 2350 modes in azimuthal direction (for more details,
see \cite{PerinaJr2015}). As the number of transverse modes is
large, we can introduce quasi-continuum of the Schmidt eigenvalues
with its probability function $ \varrho_{\lambda} $ defined in
Eq.~(\ref{45}). The probability function $ \varrho_{\lambda} $ is
plotted in Fig.~\ref{fig1}(a). It reflects the fact that the
smaller the eigenvalue the larger the number of such eigenvalues.
There occur around 80 independent spectral modes in the
low-intensity regime, as shown in Fig.~\ref{fig1}(b).

\begin{figure}         
 \resizebox{.85\hsize}{!}{\includegraphics{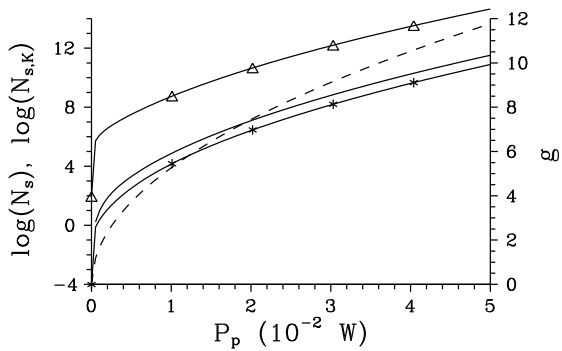}}
  \caption{Number $ N_{\rm s} $ of emitted signal photons (solid curve with $ \triangle $),
   number $ N_{{\rm s},K^n} $ of emitted signal photons per mode defined by
   photon-number statistics (solid curve with $ \ast $),
   number $ N_{{\rm s},K} $ of emitted signal photons per mode given by
   Eq.~(\ref{54}) (plain curve),
   and gain $ g $ (dashed curve) as functions of
   pump power $ P_{\rm p} $; $ \log $ denotes the decimal logarithm;
   $ w_{\rm p} = 1 \times 10^{-3} $~m, $ \Delta\lambda_{\rm p}
   = 1 \times 10^{-10} $~m.}
\label{fig2}
\end{figure}

The number $ N_{\rm s} $ of emitted signal photons increases
roughly exponentially with the increasing pump power $ P_{\rm p} $
for more intense fields \cite{Allevi2014}, as shown in
Fig.~\ref{fig2}.
The curves in Fig.~\ref{fig2} giving the number of emitted signal
photons per one mode show that the probabilities of spontaneous
and stimulated emissions of a signal photon (together with its
idler twin) become comparable for pump powers $ P_{\rm p} $ around
0.5~mW. Exponential increase of the number $ N_{\rm s} $ of
emitted signal photons occurs already for pump powers $ P_{\rm p}
$ one order of magnitude lower. It is useful to define gain $ g $
of the nonlinear interaction and to use it instead of the pump
power $ P_{\rm p} $ when comparing the theoretical quantities with
their experimental counterparts. The gain $ g $ arises from a
simplified model of the nonlinear interaction that assumes only
one effective mode and the initial vacuum state of the signal and
idler fields. Formulas (\ref{27}), (\ref{28}) and (\ref{35}) of
Sec.~II provide in this case the expression ($ \lambda^\perp_{00}
= \lambda^\parallel_{0} = 1 $)
\begin{equation}   
 N_{\rm s} = \sinh ( t^\perp f^\parallel \sqrt{P_{\rm p}} )^2 .
\label{55}
\end{equation}
It suggests the following approximative formula for fitting the
experimental dependence of the number $ N_{\rm s} $ of signal
photons:
\begin{equation}  
 N_{\rm s} = N_{{\rm s},0} \sinh(g)^2 ,
\label{56}
\end{equation}
where $ g = g_0\sqrt{P_{\rm p}} $ and $ N_{{\rm s},0} $ and $ g_0
$ are suitable constants. The values of gain $ g $ assigned to
pump powers $ P_{\rm p} $ are plotted in Fig.~\ref{fig2}. They
show the advantage of this parametrization: Stimulated emission of
photon pairs begins to dominate over spontaneous emission for the
values of gain $ g $ around one and the transition from quantum to
classical regimes (mesoscopic regime) occurs for the values of $ g
$ around 10.

\begin{figure}         
 \resizebox{.8\hsize}{!}{\includegraphics{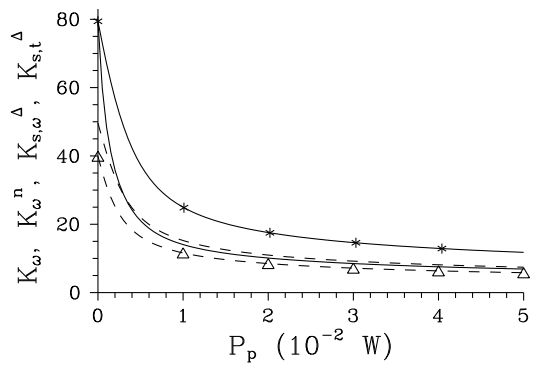}}
 \caption{Spectral entanglement dimensionality $ K_\omega $ (plain solid curve),
  number $ K_{\omega}^n $ of modes determined from photon-number statistics
  (solid curve with $ \ast $) and number $ K_{{\rm s},\omega}^\Delta $ [$ K_{{\rm s},t}^\Delta $]
  of spectral [temporal] modes given by Eq.~(\ref{54}) (dashed curve [dashed curve with $ \triangle $])
  as functions of pump power $ P_{\rm p} $; $ w_{\rm p} = 1 \times 10^{-3} $~m,
  $ \Delta\lambda_{\rm p} = 1 \times 10^{-10} $~m.}
\label{fig3}
\end{figure}

Spectral entanglement dimensionality $ K_\omega $ determined by
formula (\ref{48}) decreases with the increasing values of pump
power $ P_{\rm p} $ \cite{Allevi2014a} (see Fig.~\ref{fig3}).
The number $ K_{{\rm s},\omega}^\Delta $ of spectral signal-field
modes as well as the number $ K_{{\rm s},t}^\Delta $ of temporal
signal-field modes given in Eq.~(\ref{54}) by the ratios of
appropriate widths and plotted in Fig.~\ref{fig3} are lower than
the spectral entanglement dimensionality $ K_{\omega} $. The
comparison of curves in Fig.~\ref{fig3} shows that the
experimentally available values of $ K_{{\rm s},\omega}^\Delta $
and $ K_{{\rm s},t}^\Delta $ can successfully be used for
quantifying dimensionality of the twin beam, together with the
theoretical entanglement dimensionality $ K_\omega $. The number
of modes constituting the twin beam can also be derived from the
photon-number statistics in the signal (or idler) field
\cite{Perina1991,PerinaJr2012a,PerinaJr2013a}. In this case, the
number $ K_{\omega}^n $ of modes is given by formula (\ref{53}).
It provides systematically greater numbers of modes, as the curves
in Fig.~\ref{fig3} show. The values of entanglement dimensionality
$ K_\omega $ and number $ K_{\omega}^n $ of modes nearly coincide
in the low-intensity regime. This immediately follows from the
comparison of Eqs.~(\ref{48}) and (\ref{53}) in the limit $
U_{mlq} \approx 1 $. Also, the numbers $ K_{{\rm s},\omega}^\Delta
$, $ K_{{\rm s},t}^\Delta $ and $ K_{\omega}^n $ of modes equal to
the entanglement dimensionality $ K_\omega $ in the high-intensity
limit ($ P_{\rm p} \rightarrow \infty $). This occurs because the
strongest mode completely dominates over the other modes in this
limit.

\begin{figure}         
 \resizebox{0.47\hsize}{!}{\includegraphics{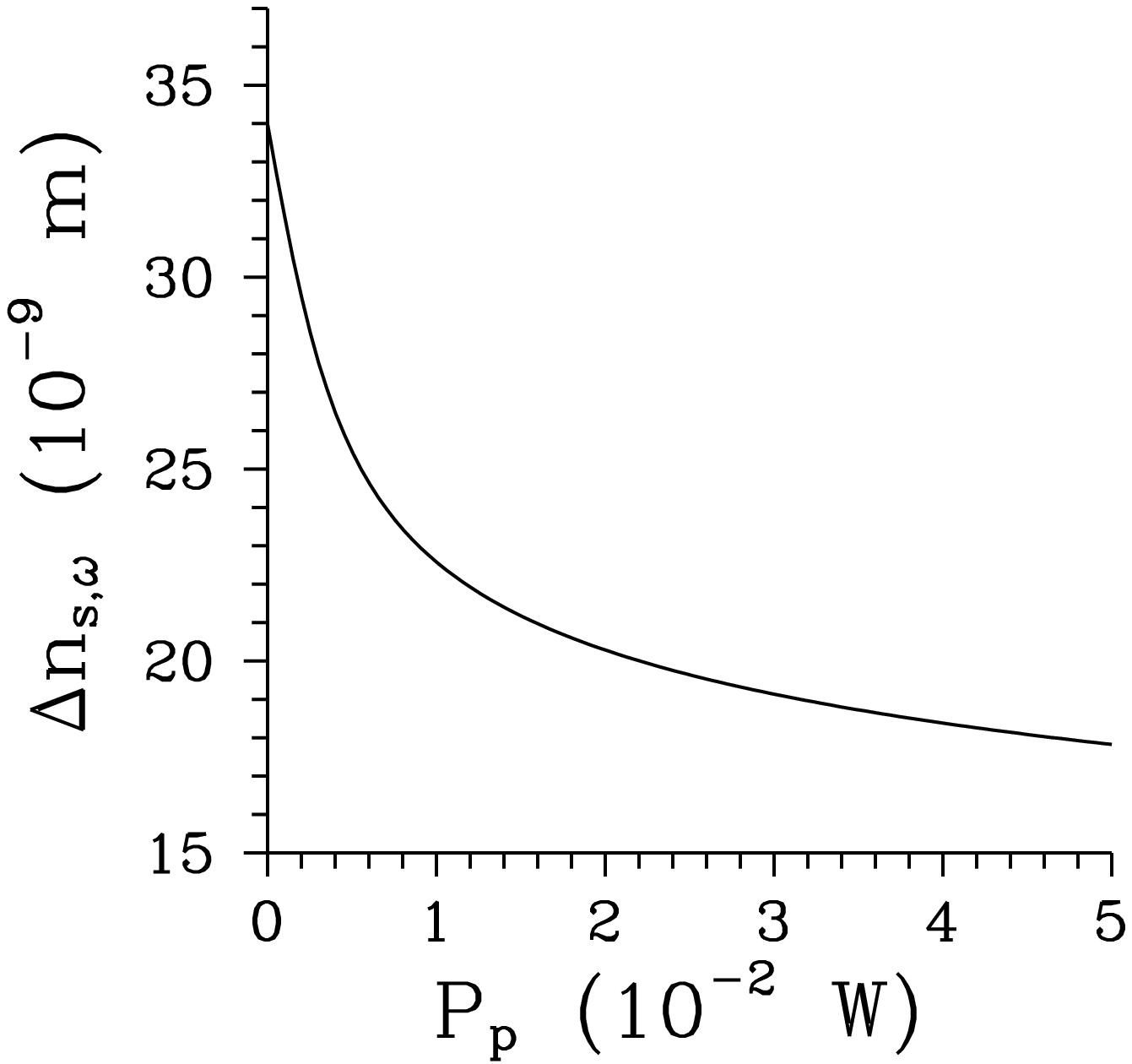}}
  \hspace{1mm}
 \resizebox{0.47\hsize}{!}{\includegraphics{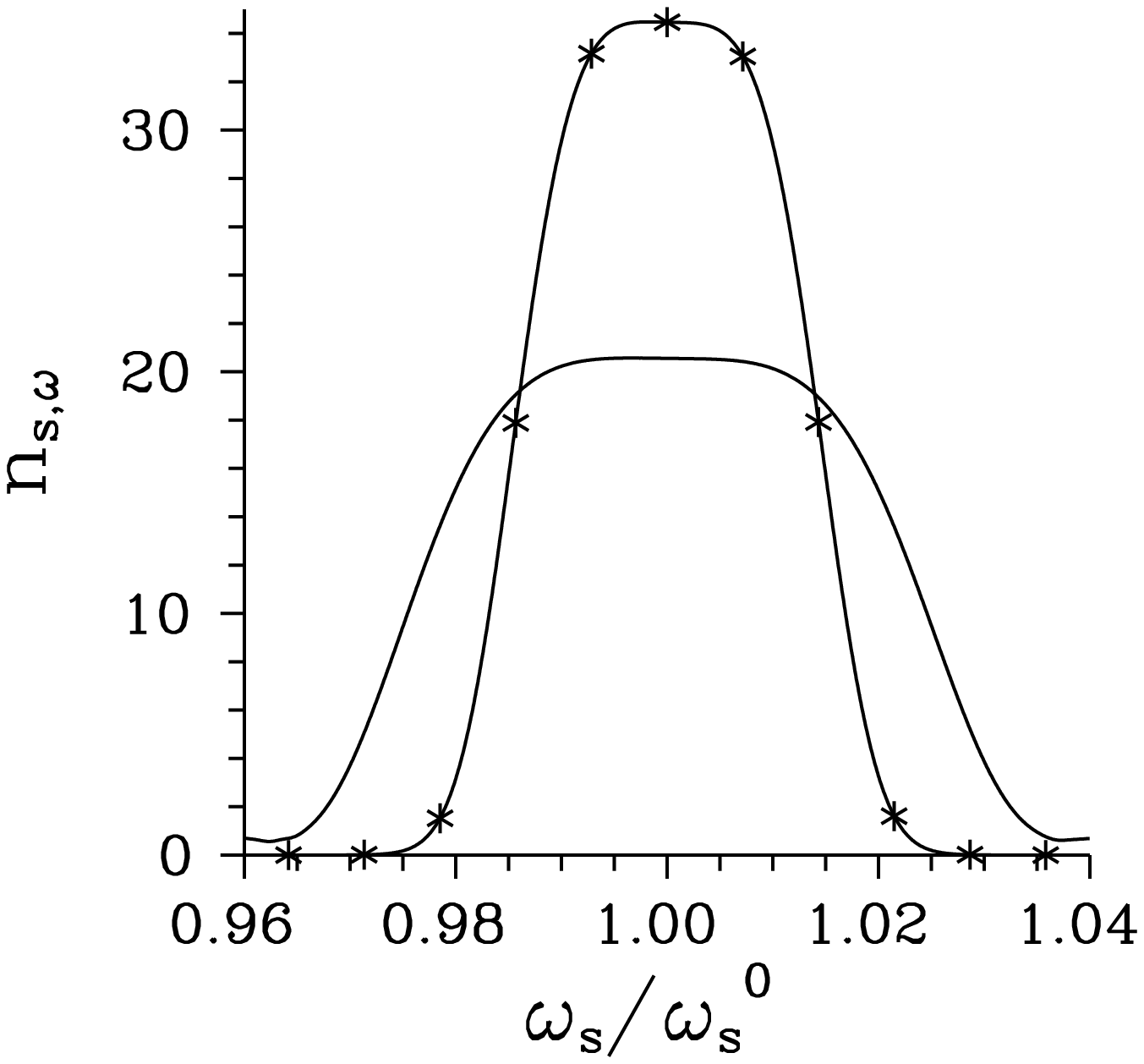}}

 (a) \hspace{.5\hsize} (b)

 \caption{(a) Width $ \Delta n_{{\rm s},\omega} $ of signal-field intensity
  spectrum as a function of pump power $ P_{\rm p} $ and (b) spectrum $ n_{{\rm s},\omega} $
  for $ P_{\rm p} = 1 \times 10^{-7} $~W (plain curve) and $ P_{\rm p} = 2 \times 10^{-2} $~W
  (solid curve with $ \ast $); $ w_{\rm p} = 1 \times 10^{-3} $~m,
  $ \Delta\lambda_{\rm p} = 1 \times 10^{-10} $~m. Spectrum $ n_{{\rm s},\omega} $ is normalized
  according to $ \int d\omega_{\rm s} n_{{\rm s},\omega}(\omega_{\rm s})/\omega_{\rm s}^0 = 1 $.}
 \label{fig4}
\end{figure}

\begin{figure}         
 \resizebox{0.47\hsize}{!}{\includegraphics{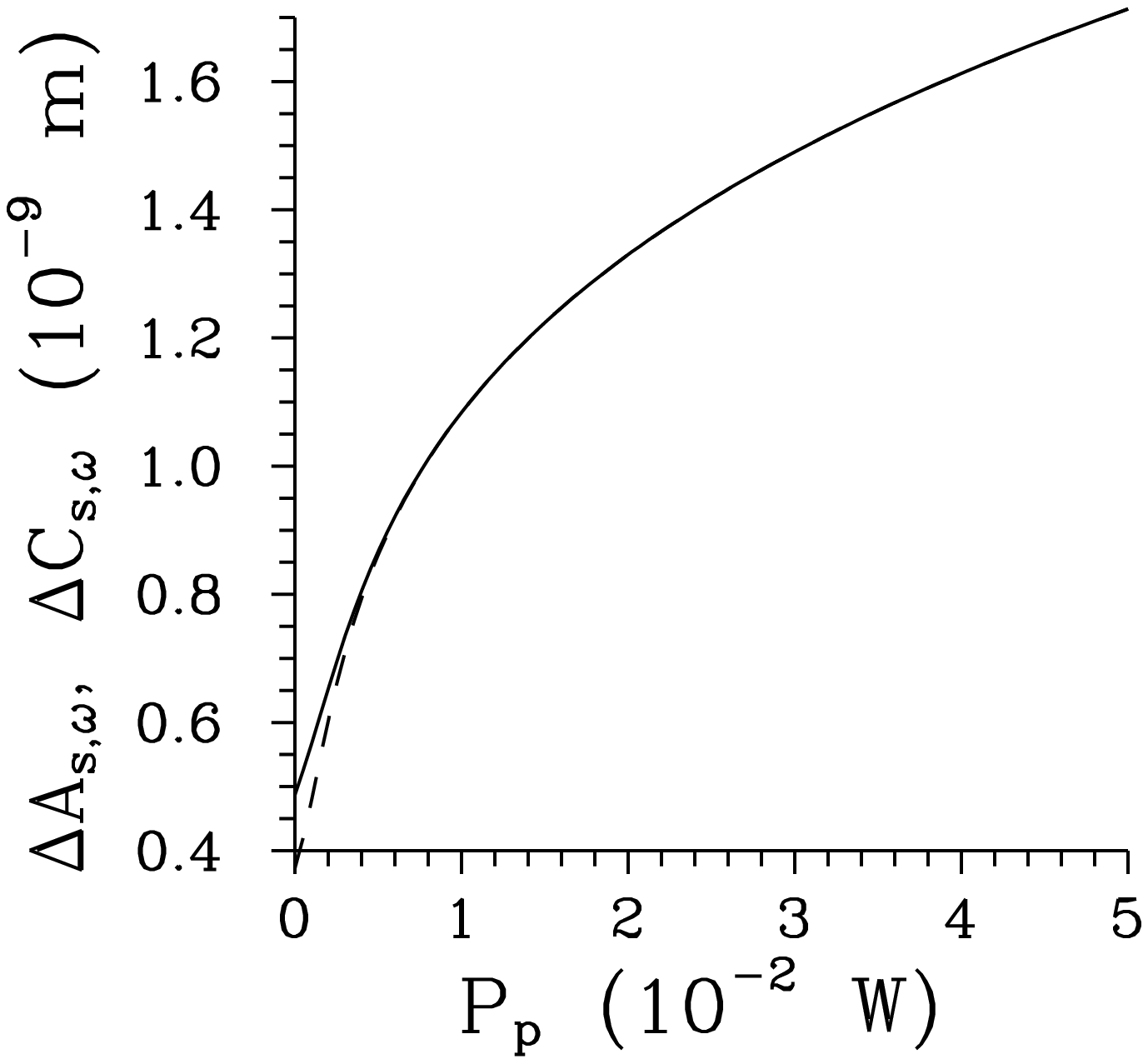}}
  \hspace{1mm}
 \resizebox{0.47\hsize}{!}{\includegraphics{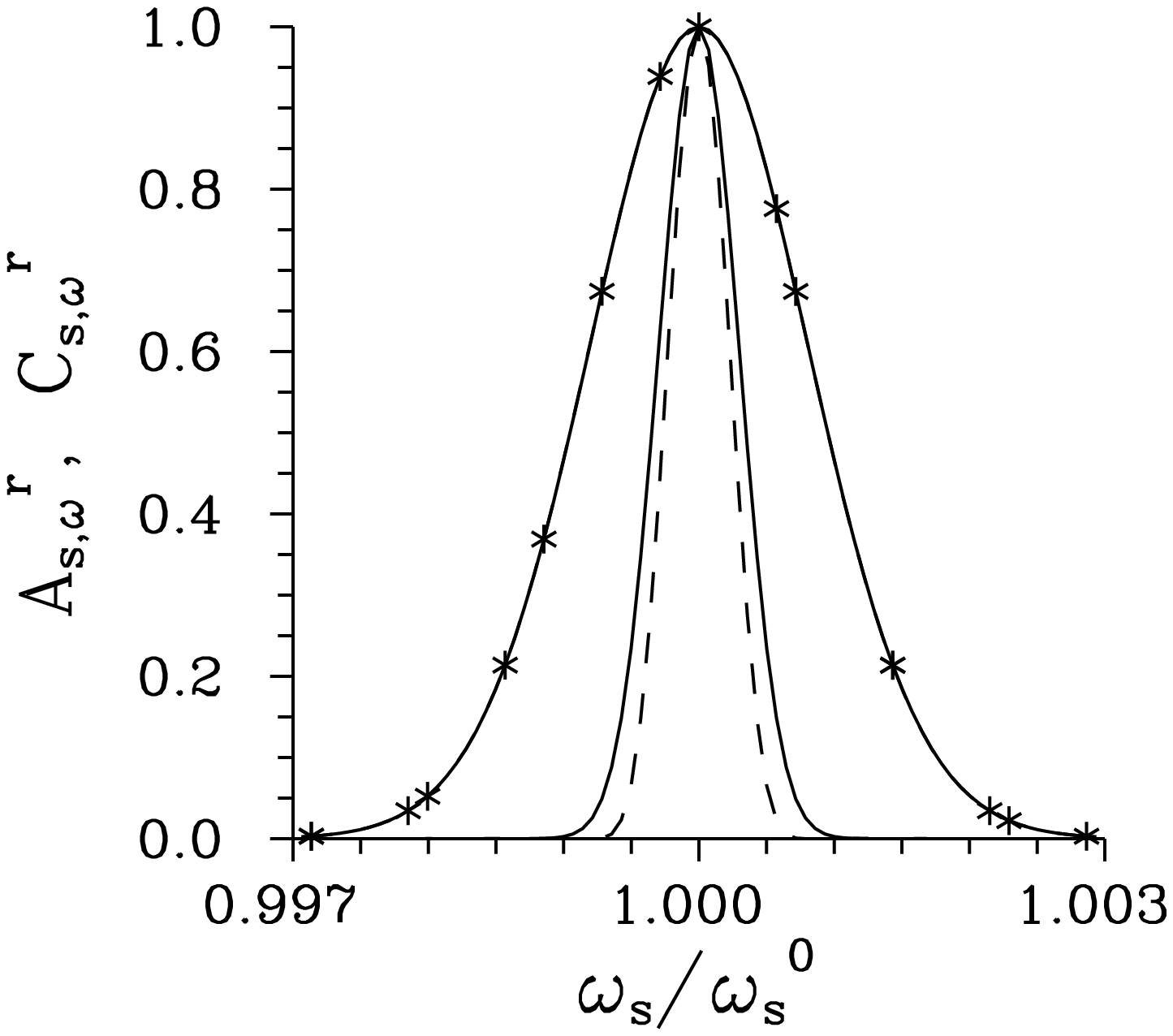}}

 (a) \hspace{.5\hsize} (b)

 \caption{(a) Widths $ \Delta A_{{\rm s},\omega} $ of signal-field
  intensity auto-correlation function (solid curve) and $ \Delta C_{{\rm s},\omega} $ of intensity
  cross-correlation function (dashed curve) as functions of pump
  power $ P_{\rm p} $. In (b), intensity auto-correlation
  function $ A_{{\rm s},\omega}^{\rm r}(\omega_{\rm s}) \equiv A_{{\rm s},\omega}(\omega_{\rm s},\omega_{\rm s}^0)/
  A_{{\rm s},\omega}(\omega_{\rm s}^0,\omega_{\rm s}^0) $ and cross-correlation
  function $ C_{{\rm s},\omega}^{\rm r}(\omega_{\rm s}) \equiv C_{{\rm s},\omega}(\omega_{\rm s},\omega_{\rm i}^0)/
  C_{{\rm s},\omega}(\omega_{\rm s}^0,\omega_{\rm i}^0) $ are plotted for
  $ P_{\rm p} = 1 \times 10^{-7} $~W (plain curves) and $ P_{\rm p} = 2 \times 10^{-2} $~W
  (nearly coinciding curves with $ \ast $); $ w_{\rm p} = 1 \times 10^{-3} $~m,
  $ \Delta\lambda_{\rm p} = 1 \times 10^{-10} $~m.}
 \label{fig5}
\end{figure}

Decrease of the number $ K_{{\rm s},\omega}^\Delta $ of
signal-field modes with the increasing pump power $ P_{\rm p} $
originates in the behavior of spectral widths $ \Delta n_{{\rm
s},\omega} $ and $ \Delta A_{{\rm s},\omega}^{\rm a} $. Whereas
the spectral width $ \Delta n_{{\rm s},\omega} $ of the
signal-field intensity profile decreases with the increasing pump
power $ P_{\rm p} $ [see Fig.~\ref{fig4}(a)], the width $ \Delta
A_{{\rm s},\omega}^{\rm a} $ of signal-field amplitude
auto-correlation function increases [for the width $ \Delta
A_{{\rm s},\omega} $ of intensity auto-correlation function, see
Fig.~\ref{fig5}(a)].
This occurs because the spectral modes with greater eigenvalues $
\lambda_q $ become more and more important with the increasing
pump power $ P_{\rm p} $. Hand in hand, the role of modes with
small eigenvalues $ \lambda_q $ is suppressed. As the modes with
large eigenvalues $ \lambda_q $ are localized more in the middle
of the spectrum (for more details, see, e.g., \cite{Christ2011}),
narrowing of the signal-field intensity spectrum is naturally
observed. This is accompanied by reshaping of the spectrum $
n_{{\rm s},\omega} $ that looses small oscillating tails present
in the low-intensity regime [see Fig.~\ref{fig4}(b)]. As the
generation of photons by stimulated emission increases with the
increasing pump power $ P_{\rm p} $, coherence in the twin beam
increases. This leads to broadening of the widths $ \Delta A_{{\rm
s},\omega} $ and $ \Delta C_{{\rm s},\omega} $ of intensity auto-
and cross-correlation functions \cite{Allevi2014a}. As the curves
in Fig.~\ref{fig5}(b) drawn for two different pump powers $ P_{\rm
p} $ show, the intensity auto-correlation function $ A_{{\rm
s},\omega} $ is wider than its cross-correlation counterpart $
C_{{\rm s},\omega} $ in the low-intensity regime (for the
explanation, see \cite{PerinaJr2015}). When stimulated emission
begins to dominate over spontaneous emission [see
Fig.~\ref{fig5}(a)], the auto-correlation and cross-correlation
functions approach each other. In the high-intensity limit $
P_{\rm p} \rightarrow \infty $, the signal and idler fields are
single-mode and so they are spectrally coherent.

\begin{figure}         
 \resizebox{0.47\hsize}{!}{\includegraphics{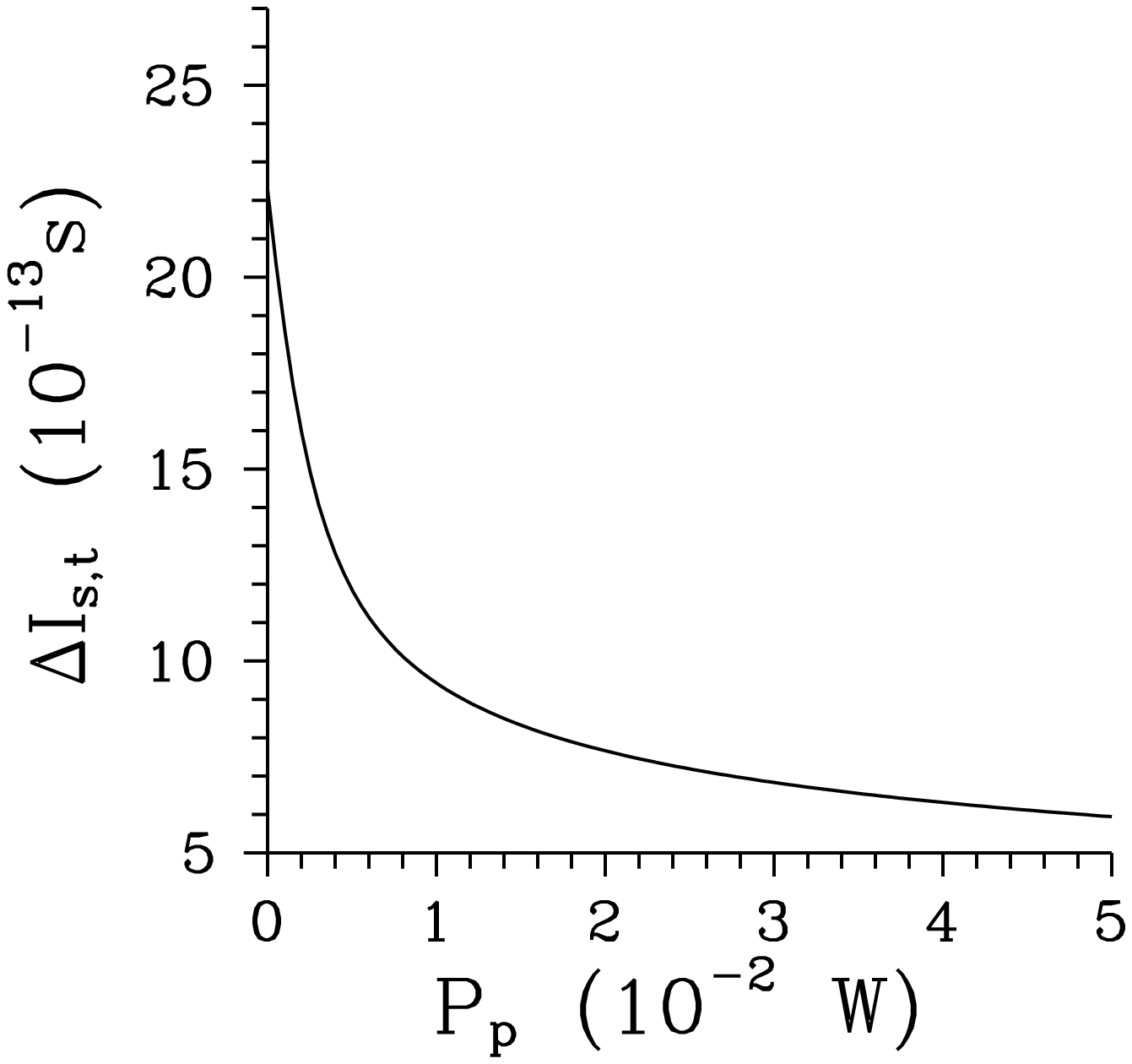}}
  \hspace{1mm}
 \resizebox{0.47\hsize}{!}{\includegraphics{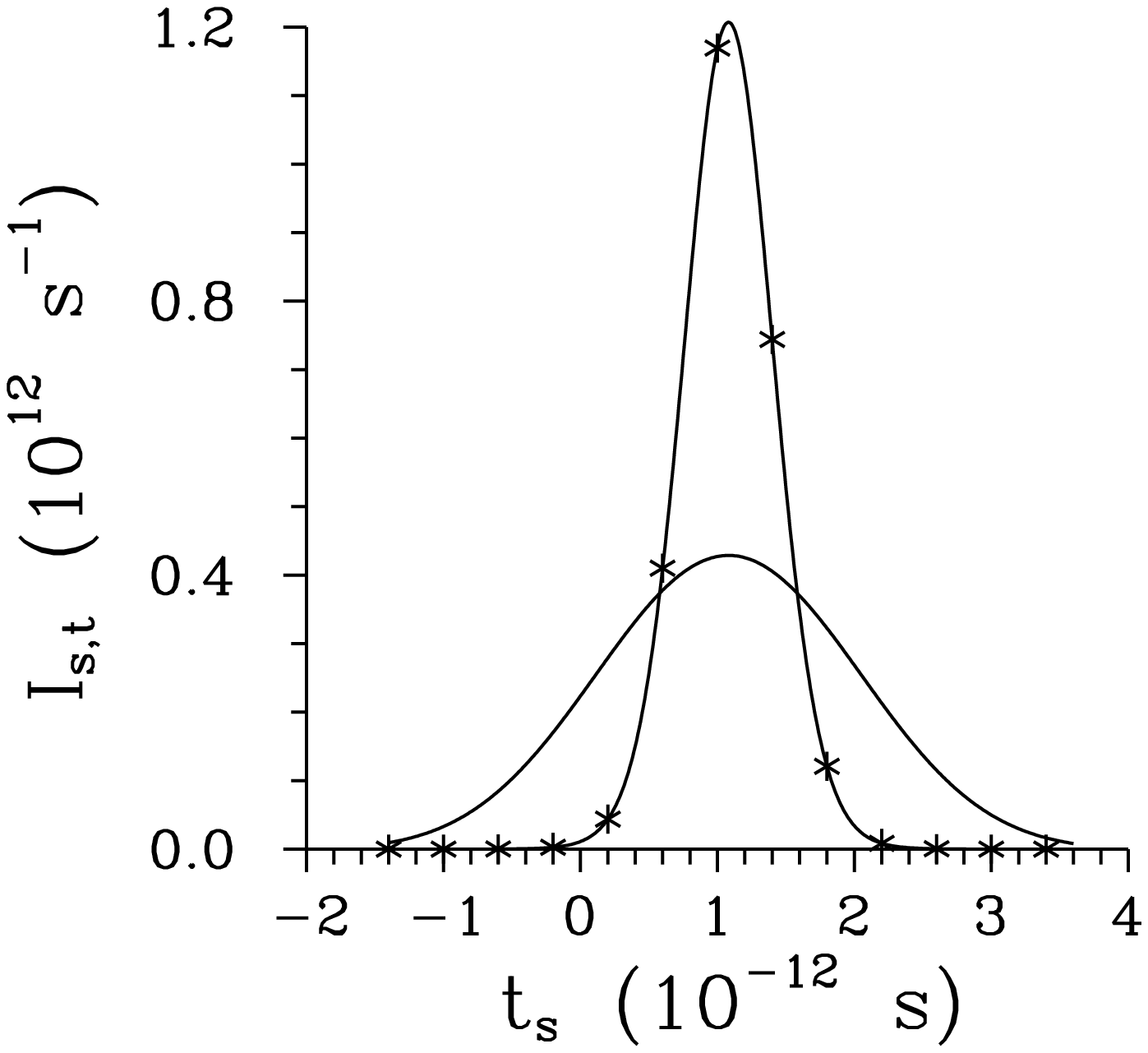}}

 (a) \hspace{.5\hsize} (b)

 \caption{(a) Width $ \Delta I_{{\rm s},t} $ (FWHM) of signal-field
  photon flux as a function of pump power $ P_{\rm p} $ and (b)
  photon flux $ I_{{\rm s},t} $
  for $ P_{\rm p} = 1 \times 10^{-7} $~W (plain curve) and $ P_{\rm p} = 2 \times 10^{-2} $~W
  (solid curve with $ \ast $); $ w_{\rm p} = 1 \times 10^{-3} $~m,
  $ \Delta\lambda_{\rm p} = 1 \times 10^{-10} $~m. The curves in (b) are normalized
  such that $ \int dt_{\rm s} I_{{\rm s},t}(t_{\rm s}) = 1 $.}
 \label{fig6}
\end{figure}

\begin{figure}         
 \resizebox{0.47\hsize}{!}{\includegraphics{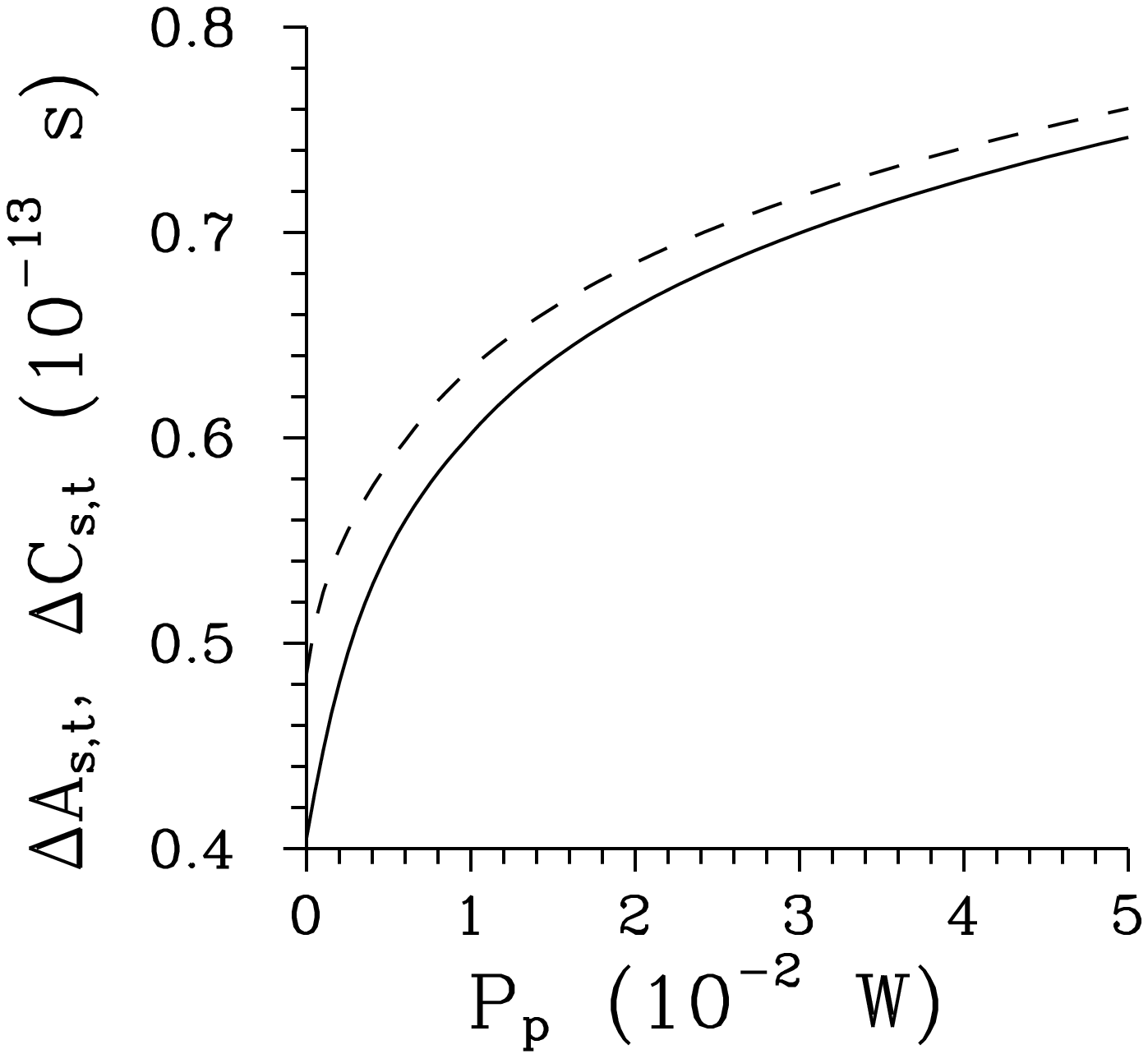}}
  \hspace{1mm}
 \resizebox{0.47\hsize}{!}{\includegraphics{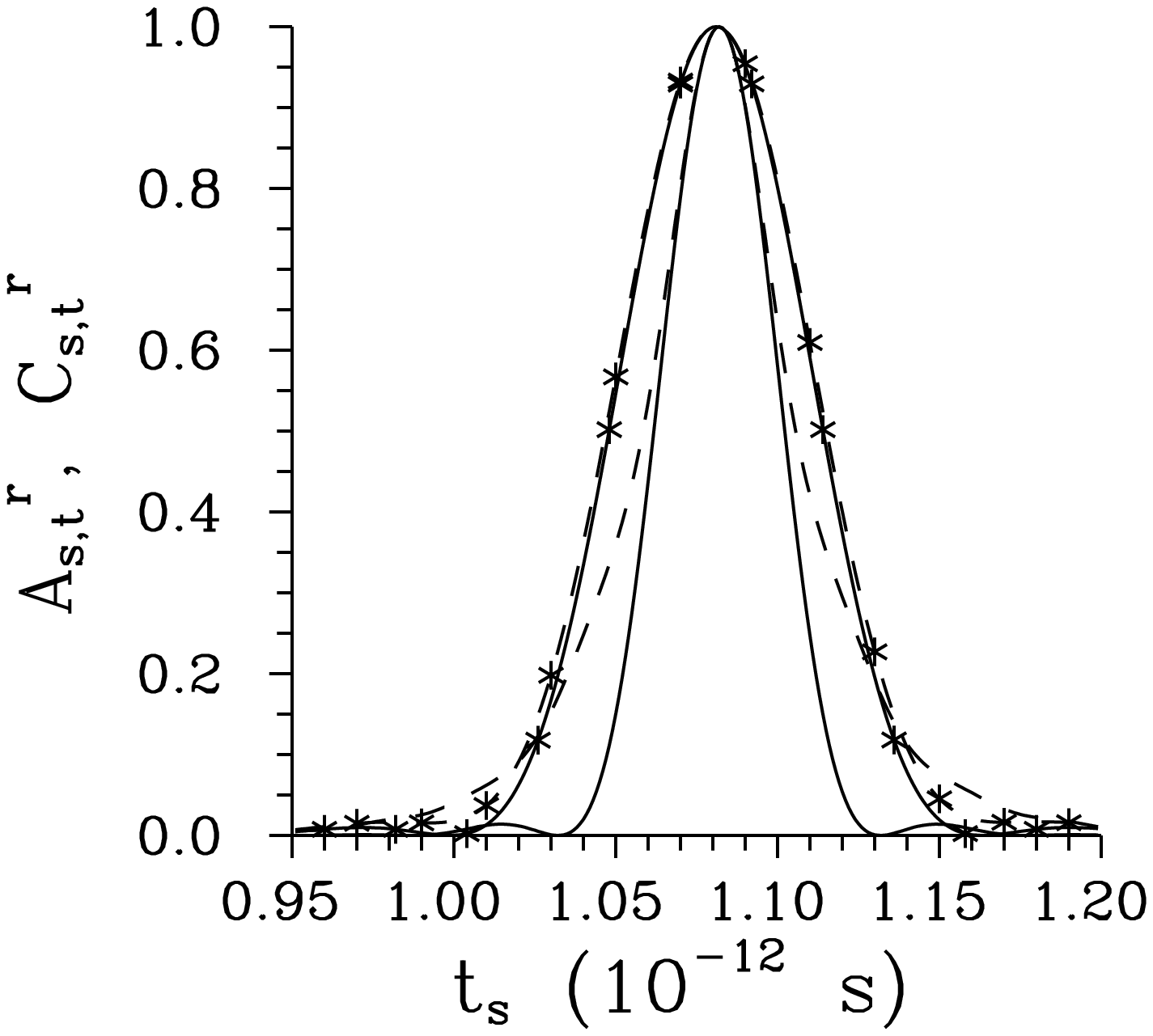}}

 (a) \hspace{.5\hsize} (b)

 \caption{(a) Widths $ \Delta A_{{\rm s},t} $ (FWHM) of signal-field
  intensity auto-correlation function (solid curve) and $ \Delta C_{{\rm s},t} $ of intensity
  cross-correlation function (dashed curve) depending on pump
  power $ P_{\rm p} $. In (b), intensity auto-correlation
  function $ A_{{\rm s},t}^{\rm r}(t_{\rm s}) \equiv A_{{\rm s},t}(t_{\rm s},t_{\rm s}^{\rm max})
  / A_{{\rm s},t}(t_{\rm s}^{\rm max},t_{\rm s}^{\rm max}) $
  and cross-correlation function $ C_{{\rm s},t}^{\rm r}(t_{\rm s}) \equiv C_{{\rm s},t}(t_{\rm s},t_{\rm i}^{\rm max})/
  C_{{\rm s},t}(t_{\rm s}^{\rm max},t_{\rm i}^{\rm max}) $ are plotted for
  $ P_{\rm p} = 1 \times 10^{-7} $~W (plain curves) and $ P_{\rm p} = 2 \times 10^{-2} $~W
  (nearly coinciding curves with $ \ast $); $ t_{\rm s}^{\rm max} $ and $ t_{\rm i}^{\rm max} $
  give the times with maximal photon fluxes in the signal and idler
  fields, respectively; $ w_{\rm p} = 1 \times 10^{-3} $~m,
  $ \Delta\lambda_{\rm p} = 1 \times 10^{-10} $~m.}
 \label{fig7}
\end{figure}

The signal field is emitted in the form of a short pulse. It is
composed of the temporal modes $ \tilde{f}_{{\rm s},q} $ given in
Eq.~(\ref{40}). These modes behave similarly as their spectral
counterparts \cite{Brecht2014}. Thus, a $ q $-th mode has $ q $
maxima and $q-1$ zeroes in its intensity temporal profile. Also,
the greater the number $ q $, the wider the mode. The field
transition to the high-intensity regime looks as follows. The
signal pulse is in general longer than the pump pulse in the
low-intensity regime \cite{PerinaJr1999a}. However, as shown in
Fig.~\ref{fig6}(a) the signal pulse shortens with the increasing
pump power $ P_{\rm p} $ due to the nonlinear interaction
described in momentum operator $ \hat{G}_{\rm int} $ written in
Eq.~(\ref{1}).
The signal pulse is also delayed with respect to the pump pulse
[see Fig.~\ref{fig6}(b)] as a consequence of different group
velocities of two pulses inside the crystal \cite{Brambilla2010}.
Coherence in the signal field as well as coherence between the
signal and idler fields increase with the increasing pump power $
P_{\rm p} $ due to stimulated emission, as documented in
Fig.~\ref{fig7}(a).
The intensity auto-correlation function $ A_{{\rm s},t} $ is
narrower than the intensity cross-correlation function $ C_{{\rm
s},t} $ in the time domain and low-intensity regime [see
Fig.~\ref{fig7}(b)]. This is opposed to the behavior of spectral
correlation functions. It originates in properties of the Fourier
transform. The cross-correlation and auto-correlation functions
are close to each other for greater values of pump power $ P_{\rm
p} $, as shown in Fig.~\ref{fig7}(b). In the high-intensity limit
$ P_{\rm p} \rightarrow \infty $, the twin beam is found in a
separable temporally coherent state composed of the signal- and
idler-field temporal modes $ \tilde{f}_{{\rm s},0} $ and $
\tilde{f}_{{\rm i},0} $ written in Eq.~(\ref{40}).

\section{Properties of intense twin beams in the transverse wave-vector and crystal output
planes}

We analyze properties of the twin beam in the wave-vector
transverse plane (far field) and the crystal output plane (near
field) assuming spectral (or temporal) averaging.

\begin{figure}         
 \resizebox{.8\hsize}{!}{\includegraphics{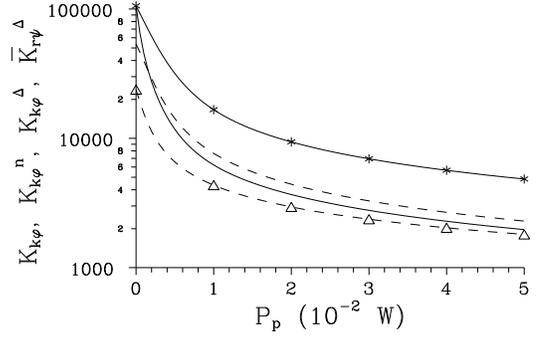}}
 \caption{Entanglement dimensionality $ K_{k\varphi} $ (plain solid curve),
  number $ K_{k\varphi}^n $ of signal-field modes determined from photon-number statistics
  (solid curve with $ \ast $) and number $ K_{{\rm s},k\varphi}^\Delta $ [$ \tilde{K}_{{\rm s},r\psi}^\Delta $]
  of signal-field modes in the transverse wave-vector [crystal output] plane
  given by Eq.~(\ref{54}) (plain dashed curve [dashed curve with $ \triangle $])
  as they depend on pump power $ P_{\rm p} $. Number $ \tilde{K}_{{\rm s},r\psi}^\Delta $ equals
  $ \tilde{K}_{{\rm s},r}^{\Delta 2} $ given in Eq.~(\ref{54}) in
  which the width $ \Delta \tilde{A}_{{\rm s},r}^{\rm a} = 2 \int_{r'_{\rm s}}^\infty dr_{\rm s} \, (r_{\rm s}-r'_{\rm s})
  A_{{\rm s},r}^{\rm a}(r_{\rm s},r'_{\rm s}) / \int_{r'_{\rm s}}^\infty dr_{\rm s} \, A_{s,r}^{\rm a}(r_{\rm s},r'_{\rm s})
  $; $ w_{\rm p} = 1 \times 10^{-3} $~m,
  $ \Delta\lambda_{\rm p} = 1 \times 10^{-10} $~m.}
\label{fig8}
\end{figure}

Entanglement dimensionality $ K_{k\varphi} $ in the transverse
plane gives around 80 thousand modes in the low-intensity regime.
It decreases with the increasing pump power $ P_{\rm p} $ (see
Fig.~\ref{fig8}) \cite{Allevi2014,Allevi2014a}. This behavior is
similar to that found in the spectral and temporal domains. It can
be explained in the same way. Entanglement dimensionality $
K_{k\varphi} $ in the transverse plane and number $ K_{k\varphi}^n
$ of signal-field transverse modes provided by the photon-number
statistics are close to each other, as shown in Fig.~\ref{fig8}.
These numbers can alternatively be experimentally estimated using
the product $ K_k^\Delta K_\varphi^\Delta $ of ratios of intensity
widths and widths of amplitude auto-correlation functions both in
radial and azimuthal transverse wave-vector directions applying
formula~(\ref{54}). Factorization of the number of modes into its
radial and azimuthal contributions is valid in our geometry in
which the photons are emitted into a narrow ring in the
wave-vector transverse plane. In the low-intensity limit, there is
around 34 [2350] modes in radial [azimuthal] wave-vector
direction. Around 10 [1000] modes are found in radial [azimuthal]
wave-vector direction for the pump power $ P_{\rm p} = 50 $~mW. On
the other hand, the signal and idler photons form a disc centered
around $ x=y=0 $~m in the crystal output plane. As the correlated
areas in the crystal output plane are radially symmetric and
practically do not change with intensity (see below), we can
estimate the number of transverse modes also by the squared ratio
$ K_r^{\Delta 2} $ determined from the appropriate widths in
radial direction. As the curves in Fig.~\ref{fig8} confirm, all
these quantities give reasonable numbers of modes of the analyzed
twin beam close to the entanglement dimensionality $ K_{k\varphi}
$.

\begin{figure}         
 \resizebox{0.47\hsize}{!}{\includegraphics{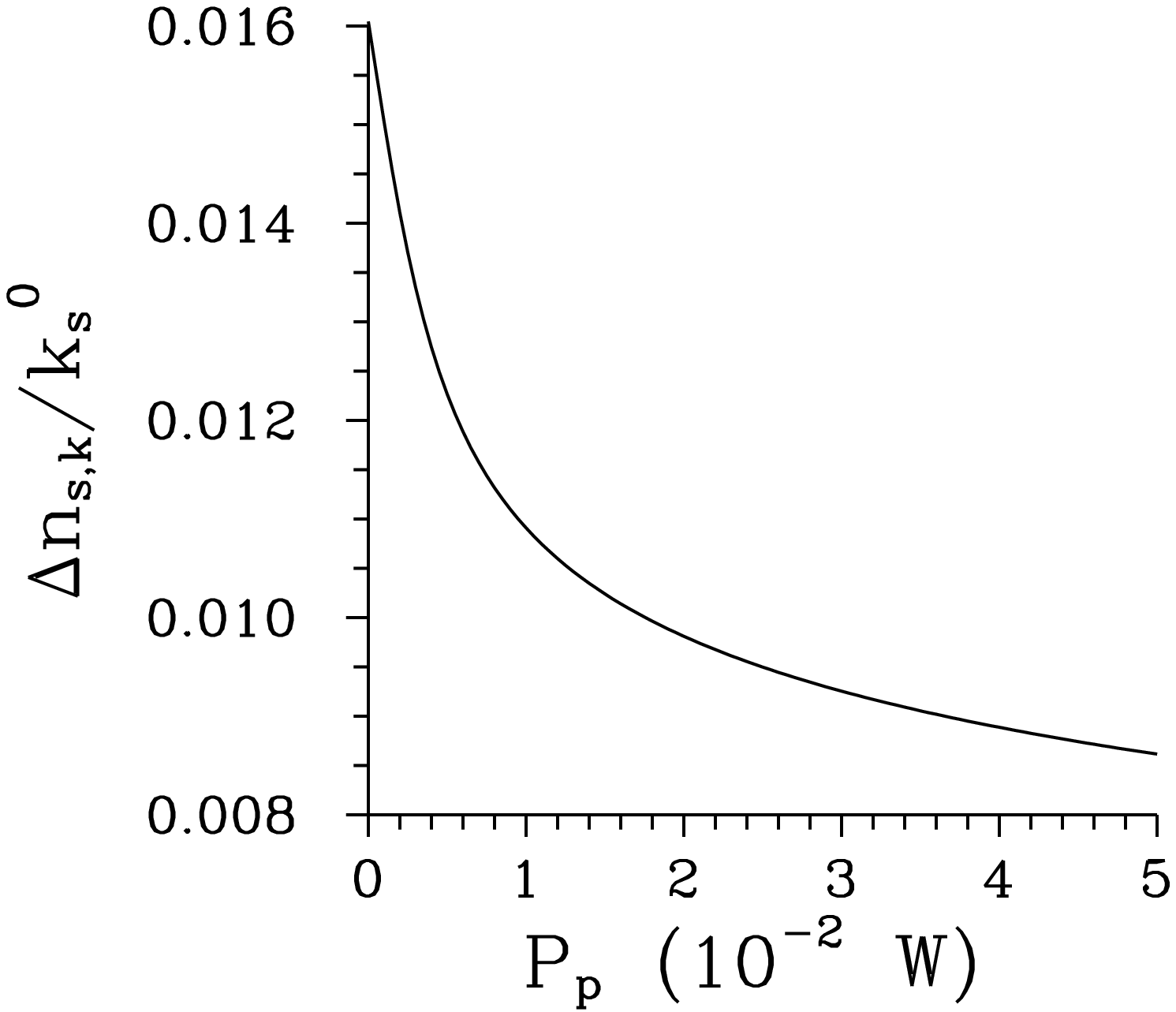}}
  \hspace{1mm}
 \resizebox{0.47\hsize}{!}{\includegraphics{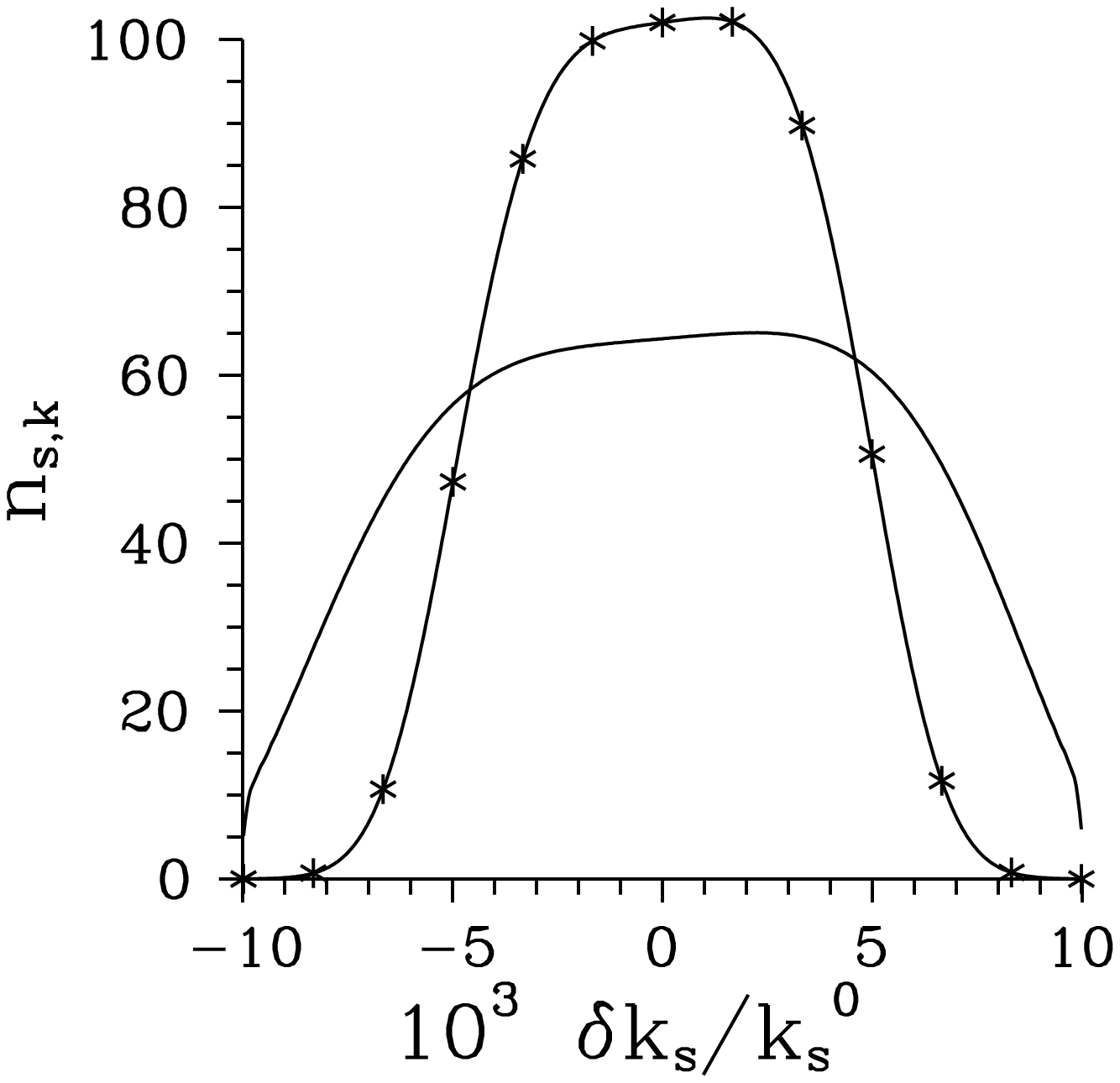}}

 (a) \hspace{.5\hsize} (b)

 \caption{(a) Width $ \Delta n_{{\rm s},k} $ of radial signal-field
  intensity profile as a function of pump power $ P_{\rm p} $ and (b)
  radial intensity profile $ n_{{\rm s},k} $ for $ P_{\rm p} = 1 \times 10^{-7} $~W (plain curve) and $ P_{\rm p} = 2 \times 10^{-2} $~W
  (solid curve with $ \ast $); $ w_{\rm p} = 1 \times 10^{-3} $~m,
  $ \Delta\lambda_{\rm p} = 1 \times 10^{-10} $~m. Intensity profile $ n_{{\rm s},k} $ is normalized
  such that $ \int dk_{\rm s} n_{{\rm s},k}(k_{\rm s})/k_{\rm s}^0 = 1 $.}
 \label{fig9}
\end{figure}

\begin{figure}         
 \resizebox{0.47\hsize}{!}{\includegraphics{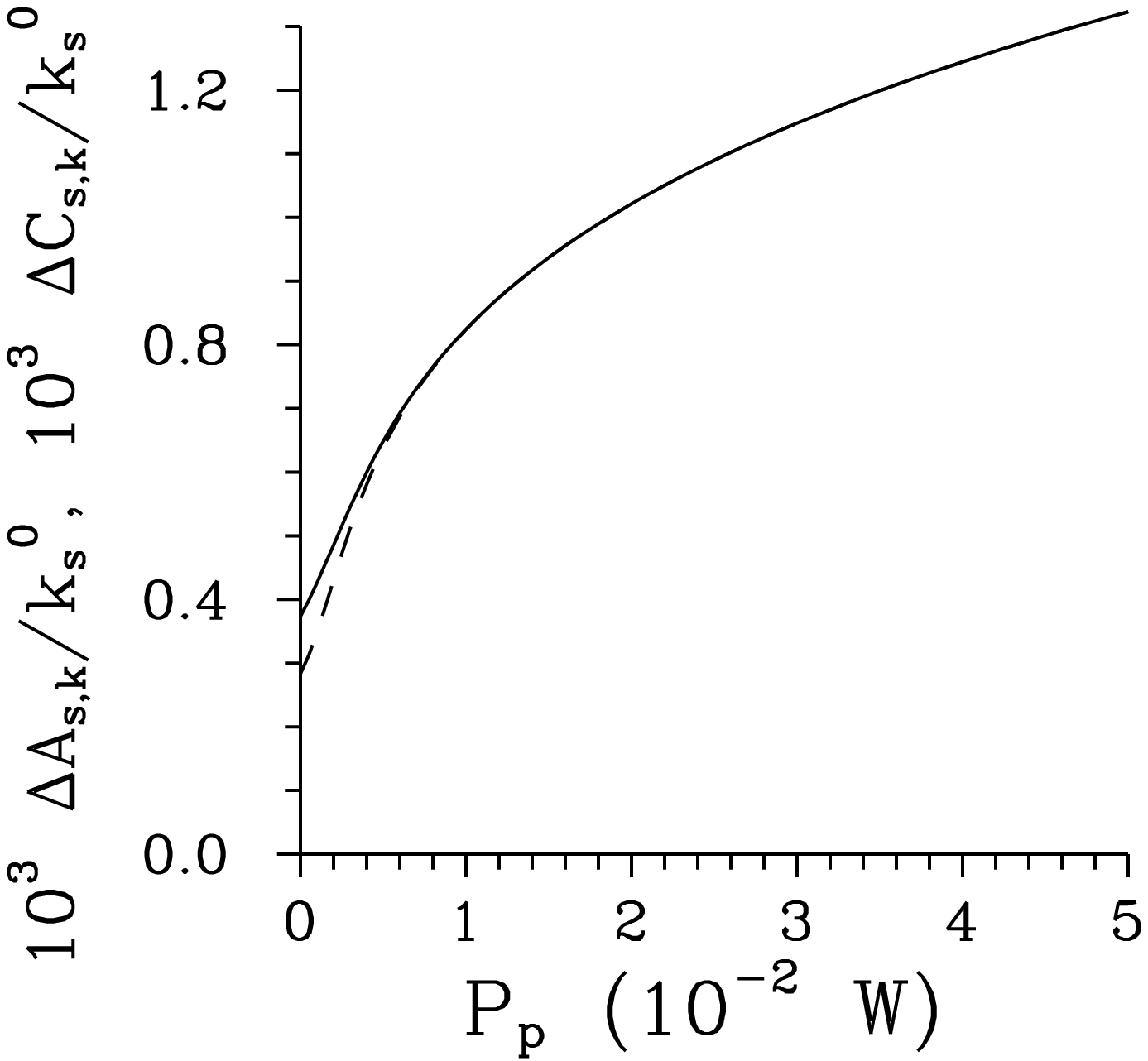}}
  \hspace{1mm}
 \resizebox{0.47\hsize}{!}{\includegraphics{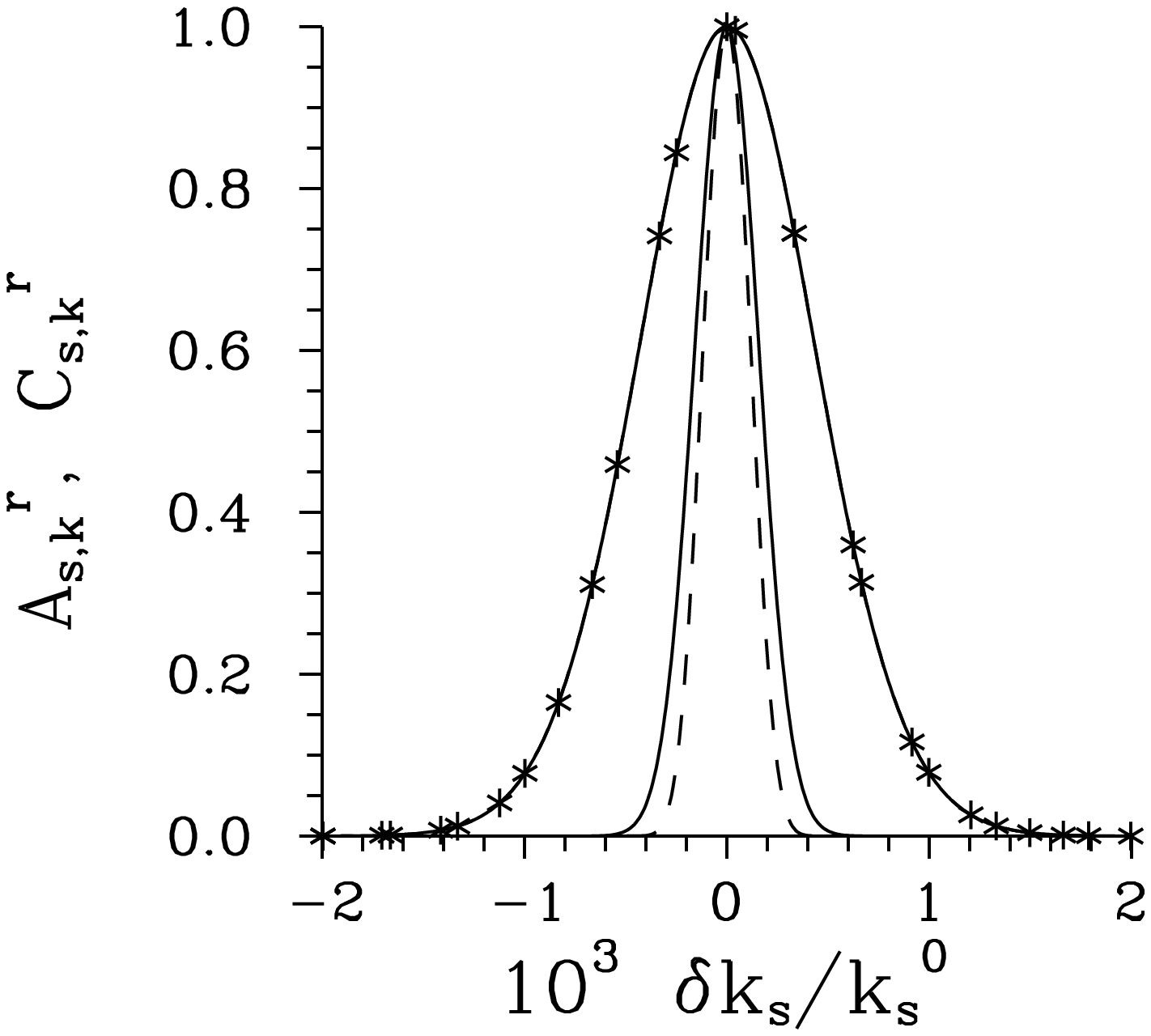}}

 (a) \hspace{.5\hsize} (b)

 \caption{(a) Widths $ \Delta A_{{\rm s},k} $ of signal-field
  intensity auto-correlation function (solid curve) and $ \Delta C_{{\rm s},k} $ of intensity
  cross-correlation function (dashed curve) in radial wave-vector direction as they depend on pump
  power $ P_{\rm p} $. In (b), signal-field intensity auto-correlation
  function $ A_{{\rm s},k}^{\rm r}(\delta k_{\rm s}) \equiv A_{{\rm s},k}(k_{\rm s}^0+\delta k_{\rm s},k_{\rm s}^0)/
  A_{{\rm s},k}(k_{\rm s}^0,k_{\rm s}^0) $ and cross-correlation
  function $ C_{{\rm s},k}^{\rm r}(\delta k_{\rm s}) \equiv C_{{\rm s},k}(k_{\rm s}^0+\delta k_{\rm s},k_{\rm i}^0)/
  C_{{\rm s},k}(k_{\rm s}^0,k_{\rm i}^0) $ are plotted for
  $ P_{\rm p} = 1 \times 10^{-7} $~W (plain curves) and $ P_{\rm p} = 2 \times 10^{-2} $~W
  (nearly coinciding curves with $ \ast $); $ w_{\rm p} = 1 \times 10^{-3} $~m,
  $ \Delta\lambda_{\rm p} = 1 \times 10^{-10} $~m.}
 \label{fig10}
\end{figure}

In the wave-vector transverse plane, decrease of entanglement
dimensionality $ K_{k\varphi} $ with the increasing pump power $
P_{\rm p} $ is explained by decrease of the width $ \Delta n_{{\rm
s},k} $ of intensity profile in radial wave-vector direction (see
Fig.~\ref{fig9}) accompanied by increase of widths $ \Delta
A_{{\rm s},k}^{\rm a} $ and $ \Delta A_{{\rm s},\varphi}^{\rm a} $
of amplitude auto-correlation functions in radial and azimuthal
wave-vector directions, respectively
\cite{Brida2009b,Allevi2014,Allevi2014a}. The ring in the
transverse wave-vector plane formed by the signal photons
\cite{Brambilla2004} thus becomes narrower with the increasing
pump power $ P_{\rm p} $, as confirmed by the radial signal-field
intensity profiles $ n_{{\rm s},k} $ plotted in
Fig.~\ref{fig9}(b). The behavior of intensity profile $ n_{{\rm
s},k} $ and intensity auto- ($ A_{{\rm s},k} $) and
cross-correlation ($ C_{{\rm s},k} $) functions in radial
wave-vector direction (see Fig.~\ref{fig10}) resembles that found
in the frequency domain. Also here the auto-correlation functions
$ A_{{\rm s},k} $ are broader than their cross-correlation
counterparts $ C_{{\rm s},k} $ for low intensities, but they
approach each other for more intense twin beams (see
Fig.~\ref{fig10}). This behavior follows from qualitative
similarity of mode profiles in both variables. We remind that an $
l $-th mode in radial wave-vector direction has $ l $ maxima and $
l-1 $ zeroes in its intensity profile. The behavior of auto- ($
A_{{\rm s},\varphi} $) and cross-correlation ($ C_{{\rm
s},\varphi} $) functions in the azimuthal wave-vector direction is
similar to that found in the radial wave-vector direction (see
Fig.~\ref{11}).
\begin{figure}         
 \resizebox{0.47\hsize}{!}{\includegraphics{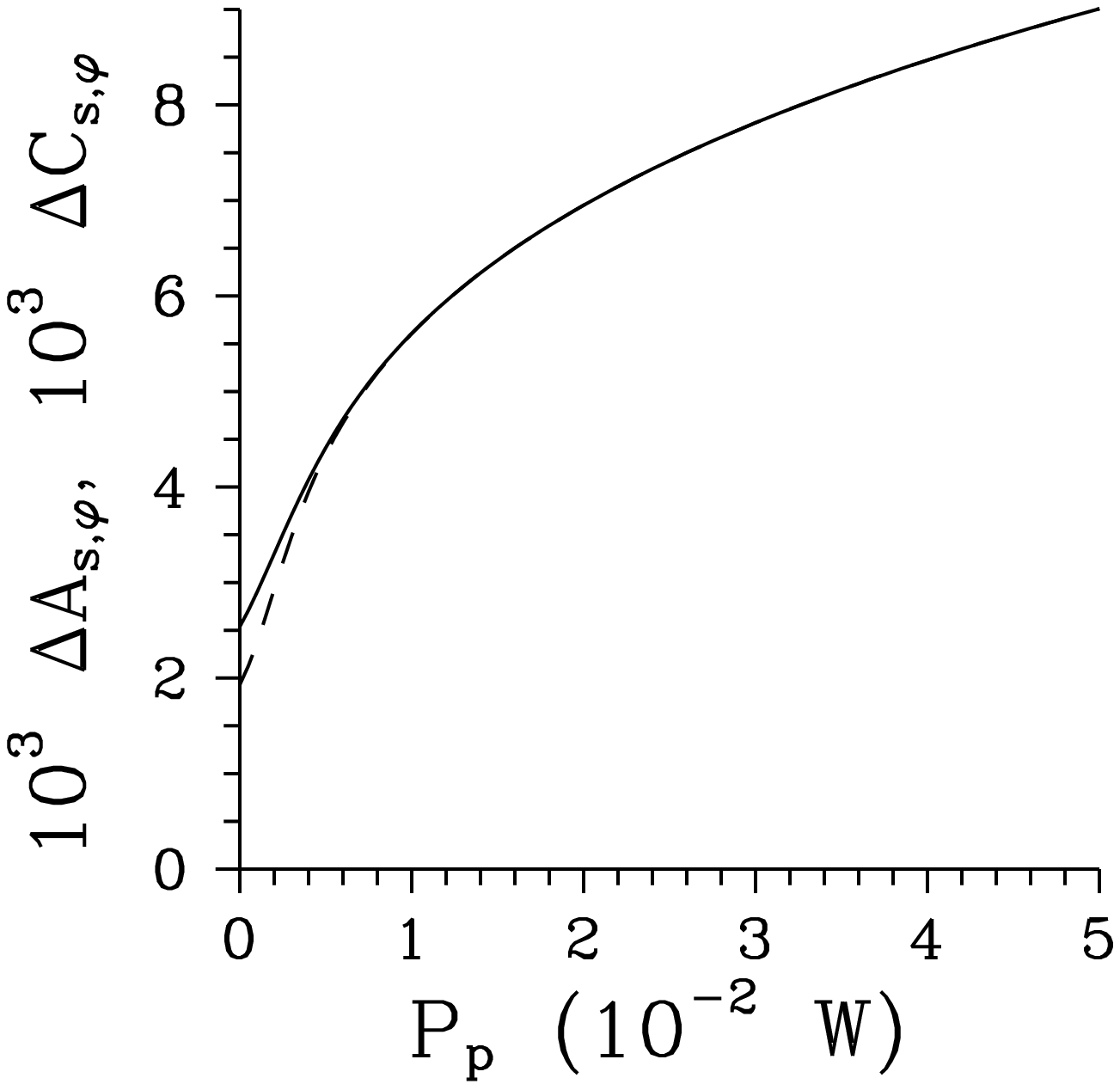}}
  \hspace{1mm}
 \resizebox{0.47\hsize}{!}{\includegraphics{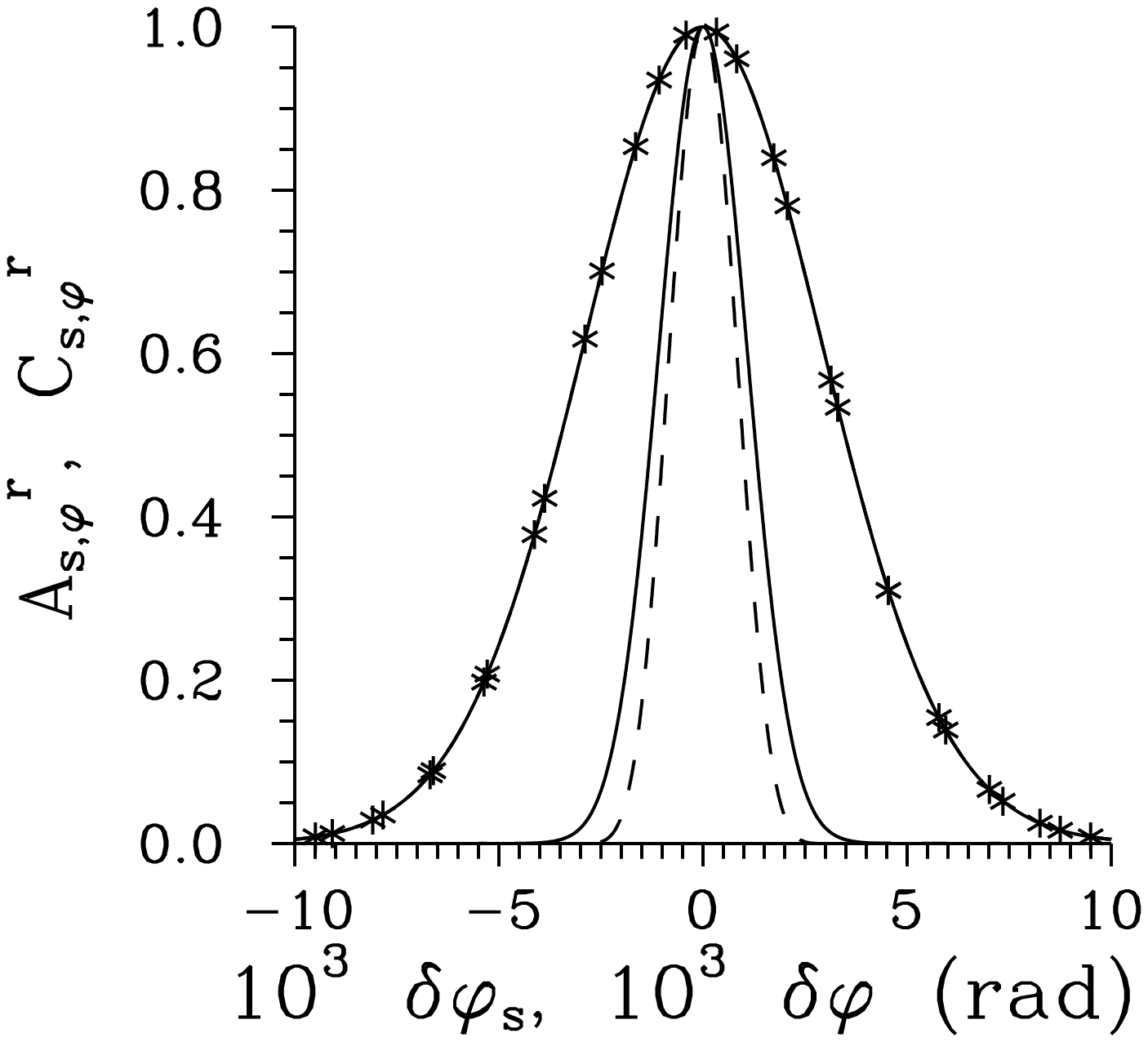}}

 (a) \hspace{.5\hsize} (b)

 \caption{(a) Widths $ \Delta A_{{\rm s},\varphi} $ of signal-field
  intensity auto-correlation function (solid curve) and $ \Delta C_{{\rm s},\varphi} $ of intensity
  cross-correlation function (dashed curve) in azimuthal wave-vector direction in dependence on pump
  power $ P_{\rm p} $. In (b), signal-field intensity auto-correlation
  function $ A_{{\rm s},\varphi}^{\rm r}(\delta\varphi_{\rm s}) \equiv
  A_{{\rm s},\varphi}(\varphi_{\rm s}^0+\delta\varphi_{\rm s},\varphi_{\rm s}^0) / A_{{\rm s},\varphi}(\varphi_{\rm s}^0,\varphi_{\rm s}^0) $
  valid for an arbitrary $ \varphi_{\rm s}^0 $ and cross-correlation function $ C_{{\rm s},\varphi}^{\rm r}(\delta\varphi)
  \equiv C_{{\rm s},\varphi}(\varphi_{\rm s}^0+\delta\varphi,\varphi_{\rm i}^0) /
  C_{{\rm s},\varphi}(\varphi_{\rm s}^0,\varphi_{\rm i}^0) $ given for an arbitrary $ \varphi_{\rm i}^0 =\varphi_{\rm s}^0+\pi $
  are shown for $ P_{\rm p} = 1 \times 10^{-7} $~W (plain curves) and $ P_{\rm p} = 2 \times 10^{-2} $~W
  (nearly coinciding curves with $ \ast $); $ w_{\rm p} = 1 \times 10^{-3} $~m,
  $ \Delta\lambda_{\rm p} = 1 \times 10^{-10} $~m.}
 \label{fig11}
\end{figure}

\begin{figure}         
 \resizebox{0.47\hsize}{!}{\includegraphics{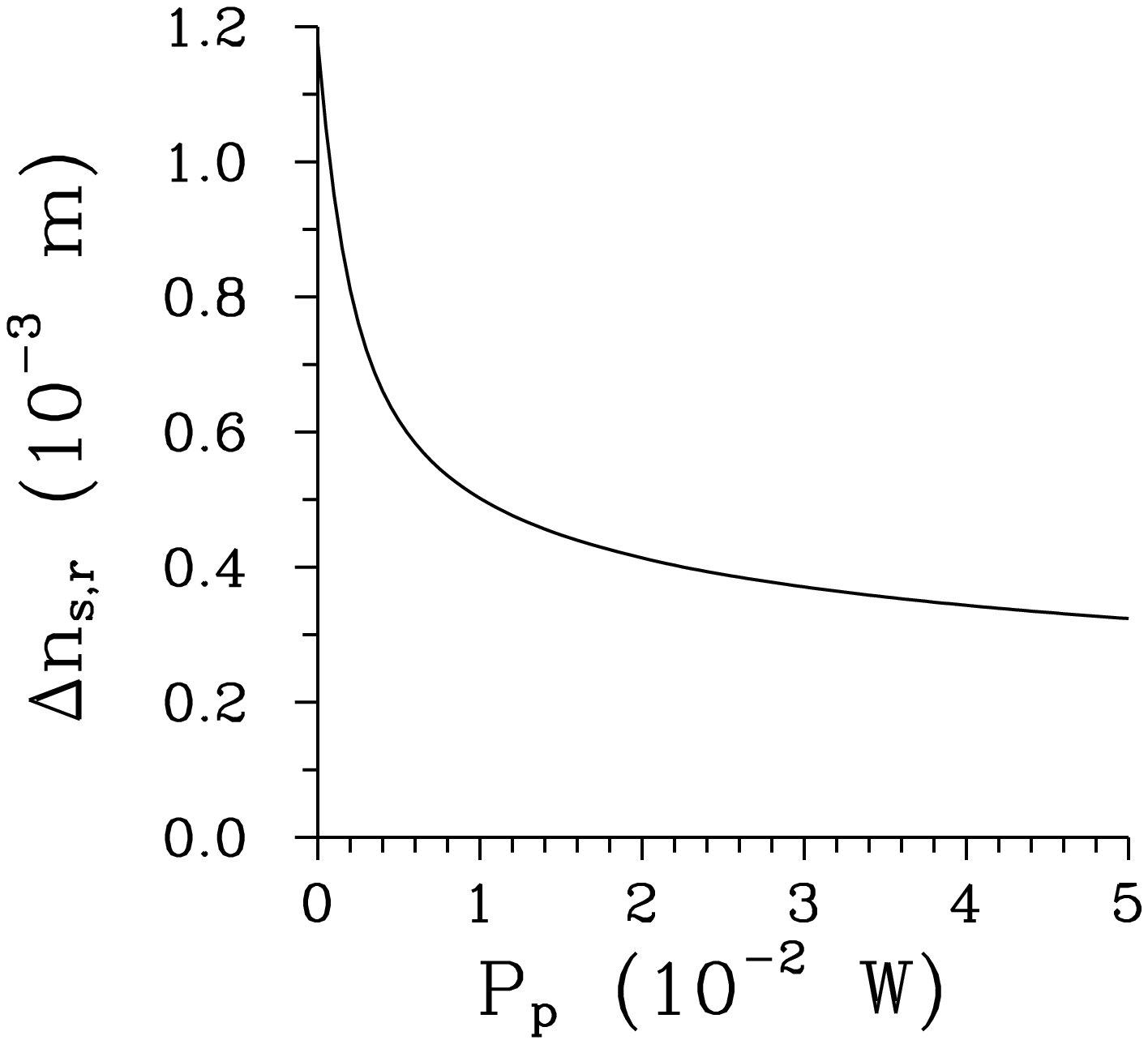}}
  \hspace{1mm}
 \resizebox{0.47\hsize}{!}{\includegraphics{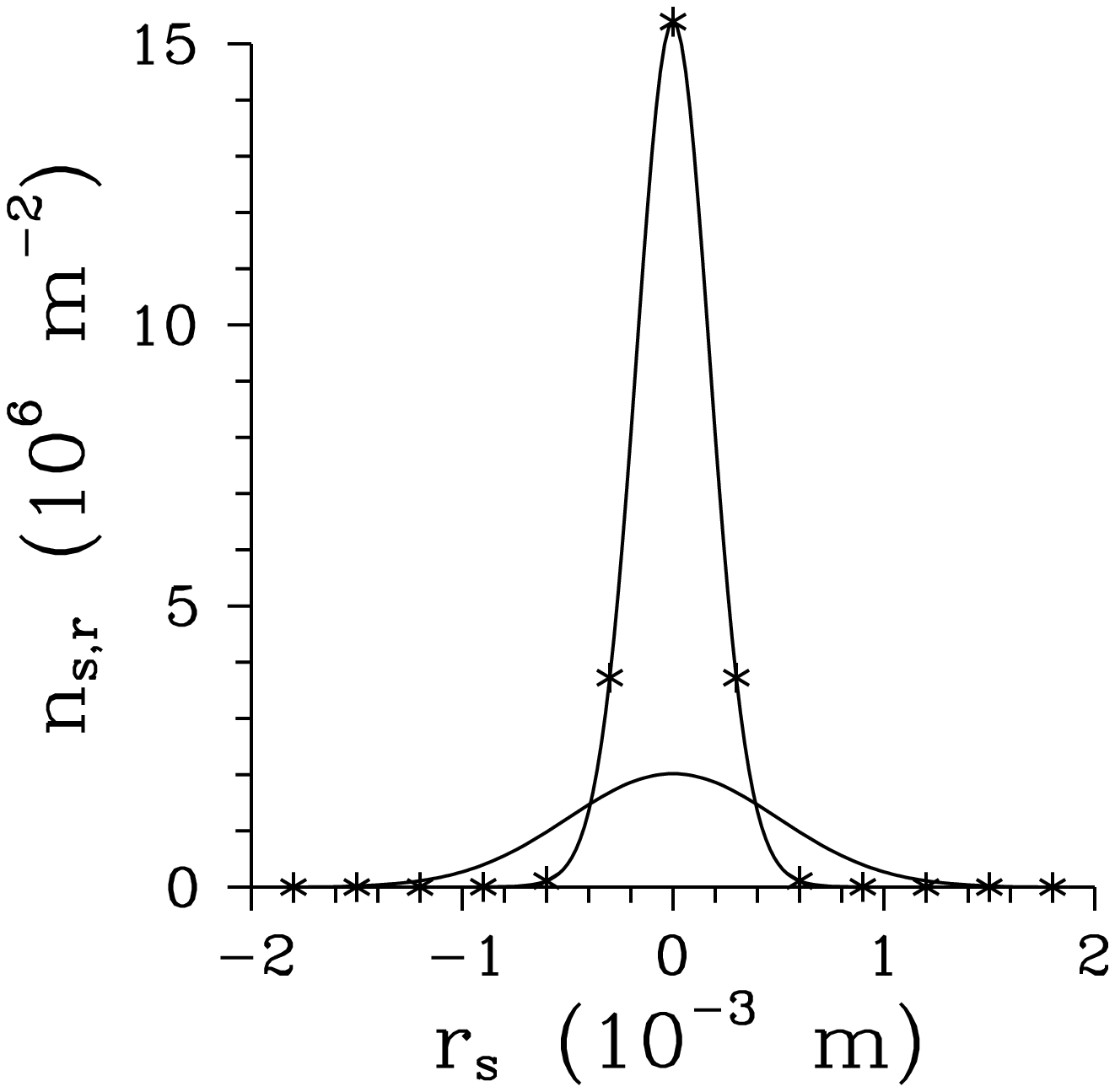}}

 (a) \hspace{.5\hsize} (b)

 \caption{(a) Width $ \Delta n_{{\rm s},r} $ of radial signal-field
  intensity profile in the crystal output plane of as a function
  of pump power $ P_{\rm p} $ and (b)
  radial intensity profile $ n_{{\rm s},r} $ for $ P_{\rm p} = 1 \times 10^{-7} $~W (plain curve) and $ P_{\rm p} = 2 \times 10^{-2} $~W
  (solid curve with $ \ast $); $ w_{\rm p} = 1 \times 10^{-3} $~m,
  $ \Delta\lambda_{\rm p} = 1 \times 10^{-10} $~m. Intensity profile $ n_{s,r} $ is normalized
  such that $ \int_0^\infty dr_{\rm s} r_{\rm s} n_{s,r}(r_{\rm s}) = 1/2 $.}
 \label{fig12}
\end{figure}

\begin{figure}         
 \resizebox{0.47\hsize}{!}{\includegraphics{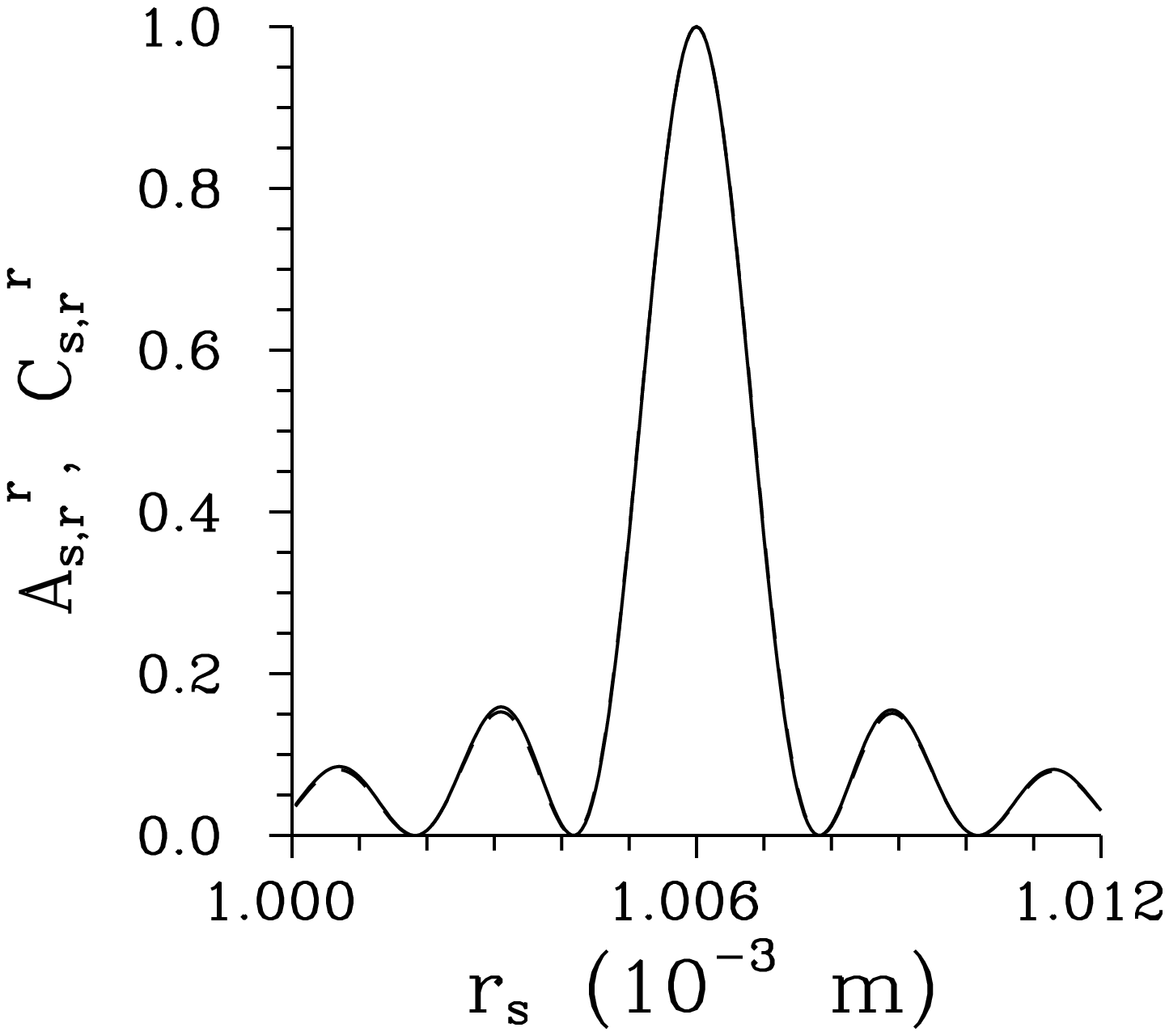}}
  \hspace{1mm}
 \resizebox{0.47\hsize}{!}{\includegraphics{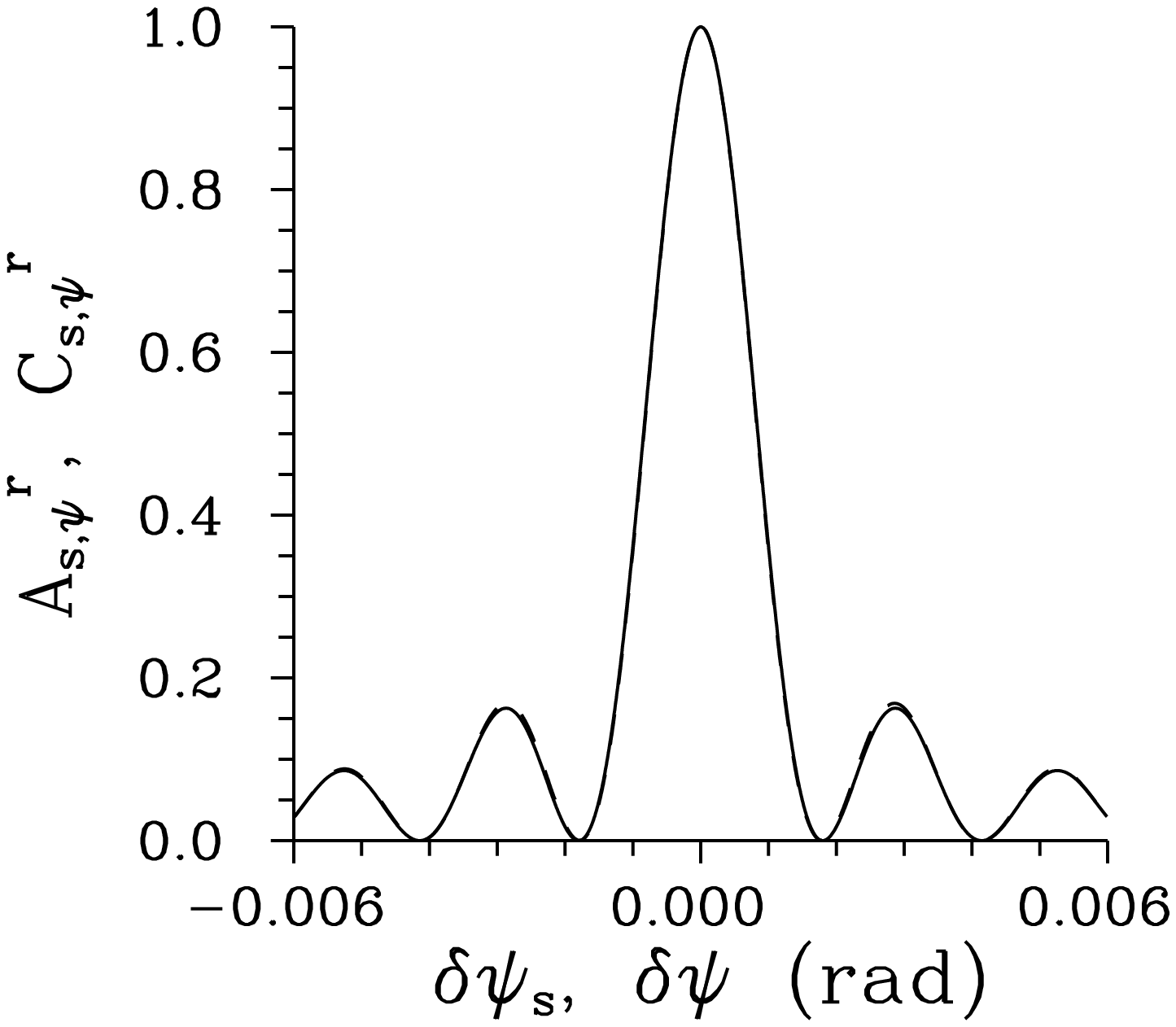}}

 (a) \hspace{.5\hsize} (b)

 \caption{(a) [(b)] Signal-field intensity auto-correlation
  function $ A_{{\rm s},r}^{\rm r} $ [$ A_{{\rm s},\psi}^r $] (solid curve) and cross-correlation
  function $ C_{{\rm s},r}^{\rm r} $ [$ C_{{\rm s},\psi}^r $] (dashed curve) are shown for
  $ P_{\rm p} = 1 \times 10^{-7} $~W (both curves nearly coincide);
  $ A_{{\rm s},r}^{\rm r}(r_{\rm s}) \equiv A_{{\rm s},r}(r_{\rm s},r_{\rm s}^0)/A_{{\rm s},r}(r_{\rm s}^0,r_{\rm s}^0) $,
  $ C_{{\rm s},r}^{\rm r}(r_{\rm s}) \equiv C_{{\rm s},r}(r_{\rm s},r_{\rm i}^0)/C_{{\rm s},r}(r_{\rm s}^0,r_{\rm i}^0) $,
  $ A_{{\rm s},\psi}^{\rm r}(\delta\psi_{\rm s}) \equiv A_{{\rm s},\psi}(\psi_{\rm s}^0+\delta\psi_{\rm s},\psi_{\rm s}^0;r_{\rm s}^0)/
  A_{{\rm s},\psi}(\psi_{\rm s}^0,\psi_{\rm s}^0;r_{\rm s}^0) $,
  $ C_{{\rm s},\psi}^{\rm r}(\delta\psi) \equiv C_{{\rm s},\psi}(\psi_{\rm s}^0+\delta\psi,\psi_{\rm i}^0;r_{\rm s}^0,r_{\rm i}^0)/
  C_{{\rm s},\psi}(\psi_{\rm s}^0,\psi_{\rm i}^0;r_{\rm s}^0,r_{\rm i}^0) $;
  $ r_{\rm s}^0 = r_{\rm i}^0 = 1.006~\times 10^{-3} $~m; $ \psi_{\rm s}^0 =
  \psi_{\rm i}^0 $; $ w_{\rm p} = 1 \times 10^{-3} $~m,
  $ \Delta\lambda_{\rm p} = 1 \times 10^{-10} $~m.}
\label{fig13}
\end{figure}

Contrary to the transverse wave-vector plane, decrease of
entanglement dimensionality $ K_{r\psi} $ with the increasing pump
power $ P_{\rm p} $ manifests itself solely by decrease of the
width $ \Delta n_{s,r} $ of radial signal-field intensity profile
in the crystal output plane (see Fig.~\ref{fig12}).
Whereas the width $ \Delta n_{{\rm s},r} $ of radial signal-field
intensity profile coincides with the width of pump beam for
low-intensity twin beams \cite{Caspani2010}, it is narrower for
more intense twin beams. This is explained by more intense
amplification of the modes localized close to the pump-beam center
relative to those occurring at the tails of the beam. Widths of
auto- ($ \Delta A_{{\rm s},r} $ and $ \Delta A_{{\rm s},\psi} $)
and cross-correlation ($ \Delta C_{{\rm s},r} $ and $ \Delta
C_{{\rm s},\psi} $) functions as well as their shapes are nearly
identical in the crystal output plane, as shown in
Fig.~\ref{fig13}.
Moreover, they do not practically depend on the pump power $
P_{\rm p} $.

Whereas both auto- and cross-correlation functions in the
frequency, time and transverse wave-vector domains have compact
shapes, long tails and oscillations are characteristic for the
correlation functions in the crystal output plane (see
Fig.~\ref{fig13}) \cite{Caspani2010}. This stems from a different
mode structure found in this case and discussed below. The
analysis has shown that the correlation functions $ A_{{\rm
s},r\psi} $ and $ C_{{\rm s},r\psi} $ are rotationally symmetric
and more-less independent on the position inside the emission
disc. This originates in the used experimental configuration in
which $ \Delta C_{{\rm s},k}/k_{\rm s}^{\perp 0} \approx 0.01 $.
This value is so low that it does not allow to develop variations
with the varying position in the crystal output plane. We note
that photon pairs emitted at the crystal end contribute to the
center of correlation functions, whereas photon pairs generated at
the beginning of the crystal are observed at the tails of the
correlation functions. Thus, the width $ \Delta A_{{\rm s},r}^{\rm
a} $ of radial signal-field amplitude auto-correlation function is
sufficient for the characterization of coherence properties ($
\Delta A_{{\rm s},r}^{\rm a} = 2.297 \times 10^{-6}$~m). We note
that the width $ \Delta A_{{\rm s},\psi}^{\rm a} $ in the
azimuthal direction depends on the distance $ r_{\rm s} $ from the
disc center. It attains its maximal value ($ \Delta A_{{\rm
s},\psi} = 2\pi $) for $ r_{\rm s} = 0 $~m and then monotonously
decreases with the increasing distance $ r_{\rm s} $, in accord
with the polar geometry. However, the presence of oscillations in
the correlation functions shown in Fig.~\ref{fig13} disqualifies
the width $ \Delta A_{{\rm s},r}^{\rm a} $ (FWHM) as a suitable
quantifier of the extension of field's correlations. Width $
\Delta \tilde{A}_{{\rm s},r}^{\rm a} $ determined from the first
moments of position and defined in the caption to Fig.~\ref{fig8}
has been found suitable in this case. It has also been used in the
determination of the number $ \tilde{K}_{r\psi} $ of signal-field
modes in the crystal output plane plotted in Fig.~\ref{fig8} ($
\tilde{K}_{{\rm s},r}^{\rm a} \approx 3.17 K_{{\rm s},r}^{\rm a}
$).

\begin{figure}         
 \resizebox{0.47\hsize}{!}{\includegraphics{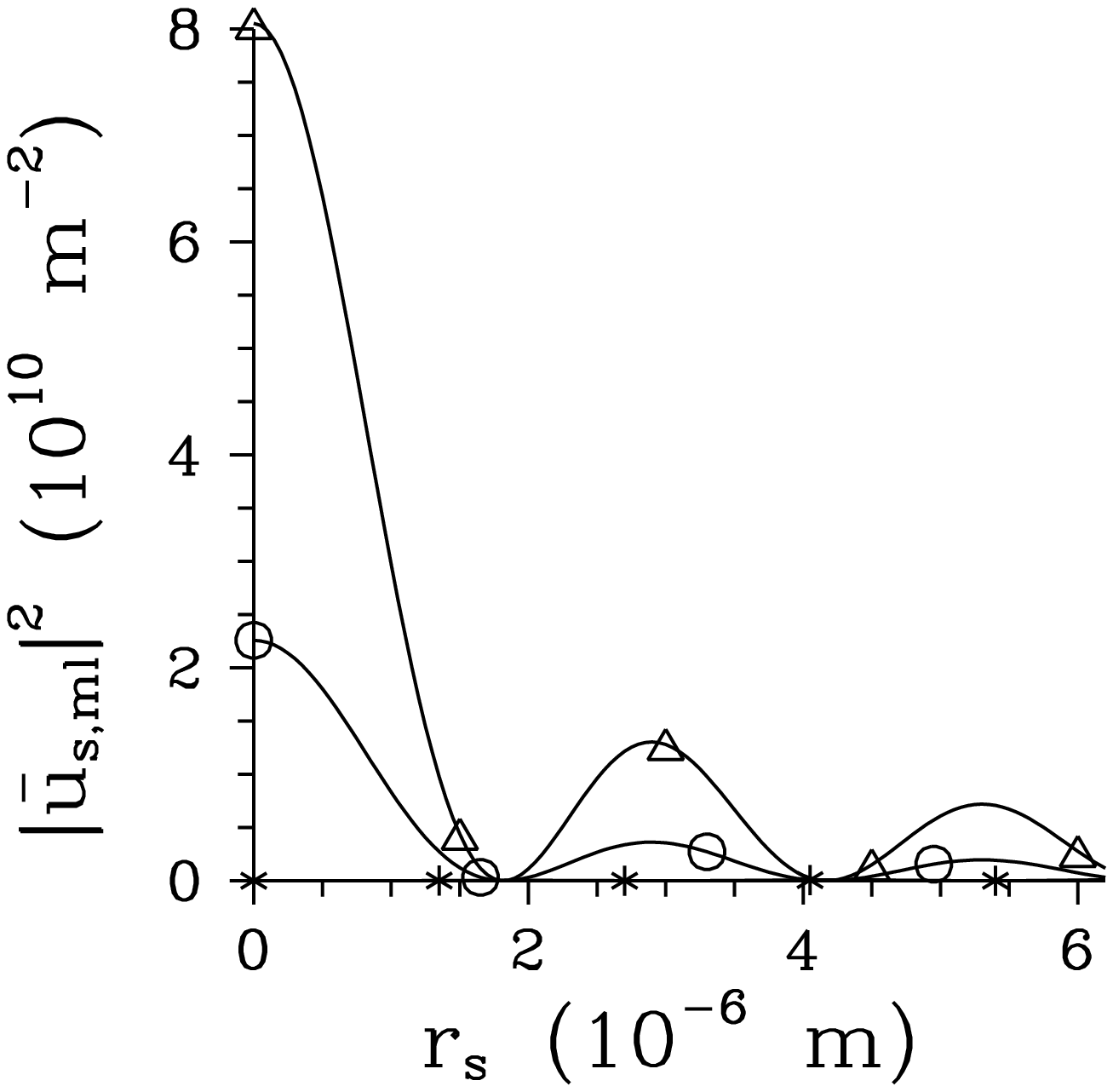}}
  \hspace{1mm}
 \resizebox{0.47\hsize}{!}{\includegraphics{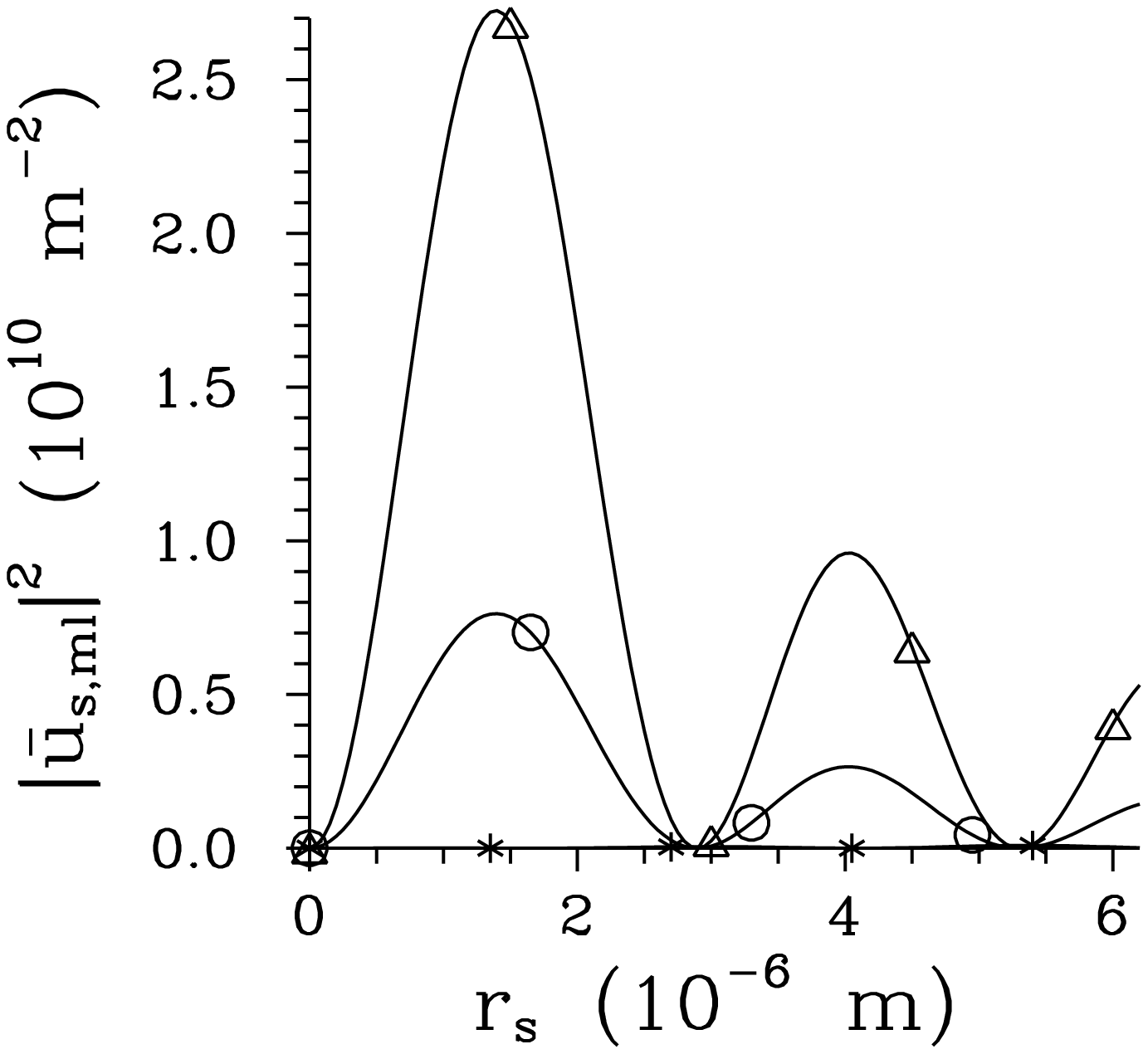}}

 (a) \hspace{.5\hsize} (b)

 \caption{Intensity profiles $ |\tilde{u}_{{\rm s},ml}|^2 $ of signal-field radial
  modes in the crystal output plane for $ l=0 $ (curve with $ \triangle $),
  $ 1 $ ($ \ast $) and $ 2 $ ($ \circ
  $) assuming (a) $ m=0 $ and (b) $ m=1 $. Modes $ \tilde{u}_{s,ml} $
  are normalized such that $ \int_0^\infty dr_{\rm s} r_{\rm s} |\tilde{u}_{{\rm s},ml}(r_{\rm s})|^2 = 1
  $; $ w_{\rm p} = 1 \times 10^{-3} $~m, $ \Delta\lambda_{\rm p} = 1 \times 10^{-10} $~m.}
\label{fig14}
\end{figure}

The oscillatory behavior of correlation functions and their
independence on pump power $ P_{\rm p} $ originate in the form of
radial modes $ \tilde{u}_{{\rm s},ml}(r_{\rm s}) $ and $
\tilde{u}_{{\rm i},ml}(r_{\rm i}) $ given by the transformation
from the wave-vector transverse plane based on the Bessel
functions [see Eq.~(\ref{A9}) in Appendix~A]. There exist two
types of modes. Modes obtained for the azimuthal number $ m=0 $
have their maximum at $ r=0 $~m [see Fig.~\ref{fig14}(a)]. They
are indispensable for describing the central part of emission
disc. On the other hand, modes with $ m \neq 0 $ have zero
intensity for $ r=0 $~m and attain their maximal values for $
r_{s,\rm max} > 0 $ [for $ m=1 $, see Fig.~\ref{fig14}(b)]. The
larger the azimuthal number $ m $, the greater the value of $
r_{s,\rm max} $. Fixing the azimuthal number $ m $, all radial
modes with different radial numbers $ l $ have zeroes in their
intensity profiles at the same positions. This property leads to
practical independence of the correlation functions on pump power
$ P_{\rm p} $. As the graphs in Fig.~\ref{fig14} indicate, the
modes $ \tilde{u}_{{\rm s},ml} $ with odd radial numbers $ l $
have small intensities compared to those with even radial numbers
$ l $. So, the modes with odd numbers $ l $ have to be very
delocalized in radial direction $ r $ and their influence to the
properties of twin beams is practically negligible. This behavior
has its origin in the shapes of modes $ u_{{\rm s},ml} $ in the
radial wave-vector direction that are close to odd functions in $
\delta k^\perp $.

\section{Conclusions}

We have analyzed the properties of general spatio-spectral twin
beams in the paraxial and parametric approximations. Considering
their spatial and spectral degrees of freedom in their common
evolution during the nonlinear interaction, we have investigated
the properties of twin beams as they depend on the pump intensity.
We have determined auto- and cross-correlation functions of a twin
beam in the spectral and temporal domains as well as the
transverse wave-vector and crystal output planes in terms of the
appropriate paired Schmidt modes. We have mutually compared their
behavior. Whereas the spectral and temporal coherence and the
coherence in the transverse wave-vector plane increase with the
increasing pump intensity, the coherence in the crystal output
plane is practically independent on the pump intensity. Whereas
the spectral and transverse wave-vector auto-correlation functions
are broader than their cross-correlation counterparts for lower
pump intensities, the opposed is true for the temporal correlation
functions. However, auto- and cross-correlation functions approach
each other for higher pump intensities.

Entanglement dimensionality of a twin beam as a function of the
pump intensity has been determined and compared with the numbers
of modes derived from solely the signal field using either its
photon-number statistics or widths of appropriate auto-correlation
functions. The numbers of signal-field modes have been confirmed
as good quantifiers of entanglement dimensionality of the twin
beam.

Practical independence of auto- and cross-correlation functions on
the pump intensity in the crystal output plane has been explained
by the special structure of paired modes in this plane
qualitatively different from the common one occurring, e.g., in
the spectral or temporal domains. Moreover, only every second
paired mode contributes significantly to the structure of a twin
beam for non-collinear geometries.

We believe that this comprehensive analysis of intense twin beams
will stimulate further experimental investigations of intense twin
beams. Moreover, as all spatio-spectral modes of a twin beam are
taken into account, the model allows for its extension to pump
intensities at which pump depletion is observed
\cite{Allevi2014,Allevi2014a}.

\acknowledgments

The author thanks M. Bondani, J. Pe\v{r}ina, O. Haderka and A.
Allevi for stimulating discussions. He gratefully acknowledges the
support by project LO1305 of the Ministry of Education, Youth and
Sports of the Czech Republic and project P205/15-08971S of the
Grant Agency of the Czech Republic.

\appendix

\section{Properties of twin beams in the transverse wave-vector and crystal output planes}

In Appendix~A, we define quantities in the transverse wave-vector
space determined by spectral averaging as well as quantities in
the crystal output plane obtained after temporal averaging. This
averaging provides transverse intensity profiles as well as
intensity auto- and cross-correlation functions.

The signal-field intensity profile $ n_{{\rm s},k\varphi} $ in the
transverse wave-vector plane is obtained as follows:
\begin{eqnarray}   
 n_{{\rm s},k\varphi}(k_{\rm s}^\perp,\varphi_{\rm s}) &=& \langle \hat{a}_{\rm s}^\dagger(k_{\rm s}^\perp,\varphi_{\rm s},\omega_{\rm s},L)
  \hat{a}_{\rm s}(k_{\rm s}^\perp,\varphi_{\rm s},\omega_{\rm s},L) \rangle_\parallel \nonumber \\
 &=& \sum_{q} \sum_{ml} |t_{{\rm s},ml}(k_{\rm s}^\perp,\varphi_{\rm s})|^2 V_{mlq}^2 .
\label{A1}
\end{eqnarray}
In Eq.~(\ref{A1}), symbol $ \langle \rangle_\parallel $ means
spectral averaging. The radial signal-field intensity profile $
n_{{\rm s},k} $ is then gives by a cut from the intensity profile
$ n_{{\rm s},k\varphi} $:
\begin{equation}  
 n_{{\rm s},k}(k_{\rm s}^\perp) = n_{{\rm s},k\varphi}(k_{\rm s}^\perp,\varphi_{\rm s}^0=0) .
\label{A2}
\end{equation}

Averaged signal-field intensity correlations in the transverse
wave-vector plane are described by the following fourth-order
auto-correlation function $ A_{{\rm s},k\varphi} $:
\begin{eqnarray}   
 A_{{\rm s},k\varphi}(k_{\rm s}^\perp,\varphi_{\rm s},k_{\rm s}^{'\perp},\varphi'_{\rm s}) &=& \langle {\cal N}: \Delta[
  \hat{a}_{\rm s}^\dagger(k_{\rm s}^\perp,\varphi_{\rm s},\omega_{\rm s},L)
  \nonumber \\
 & & \hspace{-15mm} \mbox{} \times \hat{a}_{\rm s}(k_{\rm s}^\perp,\varphi_{\rm s},\omega_{\rm s},L)]
  \Delta[\hat{a}_{\rm s}^\dagger(k_{\rm s}^{'\perp},\varphi'_{\rm s},\omega'_{\rm s},L) \nonumber \\
 & & \hspace{-15mm} \mbox{} \times
  \hat{a}_{\rm s}(k_{\rm s}^{'\perp},\varphi'_{\rm s},\omega'_{\rm s},L)]: \rangle_{\parallel} \nonumber \\
 &=& \sum_{q} |A_{{\rm s},q,k\varphi}^{\rm a}(k_{\rm s}^\perp,\varphi_{\rm s},k_{\rm s}^{'\perp},\varphi'_{\rm s})|^2 .
  \nonumber \\
 & &
\label{A3}
\end{eqnarray}
The signal-field amplitude auto-correlation function $ A_{{\rm
s},q,k\varphi}^{\rm a} $ of mode $ q $ is determined as follows:
\begin{eqnarray}   
 A_{{\rm s},q,k\varphi}^{\rm a}(k_{\rm s}^\perp,\varphi_{\rm s},k_{\rm s}^{'\perp},\varphi'_{\rm s})
 & & \nonumber \\
 & & \hspace{-25mm} =
  \langle \hat{a}_{\rm s}^\dagger(k_{\rm s}^\perp,\varphi_{\rm s},\omega_s,L)
  \hat{a}_{\rm s}(k_{\rm s}^{'\perp},\varphi'_{\rm s},\omega'_s,L) \rangle_{\parallel,q} \nonumber \\
 & & \hspace{-25mm} =
  \sum_{ml} t_{{\rm s},ml}^*(k_{\rm s}^\perp,\varphi_{\rm s}) t_{{\rm s},ml}(k_{\rm s}^{'\perp},\varphi'_{\rm s})  V_{mlq}^2 .
\label{A4}
\end{eqnarray}
Radial ($ A_{{\rm s},k} $) and azimuthal ($ A_{{\rm s},\varphi} $)
signal-field intensity correlation functions are derived from
Eq.~(\ref{A4}) along the relations:
\begin{eqnarray}   
 A_{{\rm s},k}(k_{\rm s}^\perp,k_{\rm s}^{'\perp}) &=&
  A_{{\rm s},k\varphi}(k_{\rm s}^\perp,\varphi_{\rm s}^0=0,k'_{\rm s}{}^{\perp},\varphi'_{\rm s}{}^{0}=0),
  \nonumber \\
 A_{{\rm s},\varphi}(\varphi_{\rm s},\varphi'_{\rm s}) &=&
  A_{{\rm s},k\varphi}(k_{\rm s}^{\perp 0},\varphi_{\rm s},k_{\rm s}^{\perp 0},\varphi'_{\rm s}).
\label{A5}
\end{eqnarray}

Similarly, intensity cross-correlations in the wave-vector signal
and idler transverse planes are quantified by the fourth-order
correlation function $ C_{k\varphi} $:
\begin{eqnarray}   
 C_{k\varphi}(k_{\rm s}^\perp,\varphi_{\rm s},k_{\rm i}^\perp,\varphi_{\rm i}) \hspace{-22mm} & & \nonumber \\
 &=& \langle {\cal N}: \Delta[
  \hat{a}_{\rm s}^\dagger(k_{\rm s}^\perp,\varphi_{\rm s},\omega_{\rm s},L) \hat{a}_{\rm s}(k_{\rm s}^\perp,\varphi_{\rm s},\omega_{\rm s},L)] \nonumber
  \\
 & & \mbox{} \times  \Delta[ \hat{a}_{\rm i}^\dagger(k_{\rm i}^\perp,\varphi_{\rm i},\omega_{\rm i},L)
  \hat{a}_{\rm i}(k_{\rm i}^\perp,\varphi_{\rm i},\omega_{\rm i},L)] : \rangle_\parallel \nonumber \\
 &=& \sum_{q} \left|
  \sum_{ml} t_{{\rm s},ml}(k_{\rm s}^\perp,\varphi_{\rm s}) t_{{\rm i},ml}(k_{\rm i}^\perp,\varphi_{\rm i}) U_{mlq} V_{mlq}
  \right|^2 . \nonumber \\
 & &
\label{A6}
\end{eqnarray}
Radial ($ C_{{\rm s},k} $) and azimuthal ($ C_{{\rm s},\varphi} $)
intensity cross-correlation functions are easily determined from
the cross-correlation function $ C_{k\varphi} $:
\begin{eqnarray}   
 C_{{\rm s},k}(k_{\rm s}^\perp,k_{\rm i}^\perp) &=&
  C_{{\rm s},k\varphi}(k_{\rm s}^\perp,\varphi_{\rm s}^0=0,k_{\rm i}^\perp,\varphi_{\rm i}^0=\pi),
  \nonumber \\
 C_{{\rm s},\varphi}(\varphi_{\rm s},\varphi_{\rm i}) &=&
  C_{{\rm s},k\varphi}(k_{\rm s}^{\perp 0},\varphi_{\rm s},k_{\rm i}^{\perp 0},\varphi_{\rm i}).
\label{A7}
\end{eqnarray}

On the other hand, properties of the twin beams at the crystal
output plane (near field) are described by the two-dimensional
Fourier transform of eigenmodes $ t $ written in Eq.~(\ref{16})
and defined in the transverse wave-vector plane. This transform
applied to the radially symmetric geometry leaves us with
eigenmodes $ \tilde{t}_a $ [$ x_a = r_a\cos(\psi_a) $, $ y_a =
r_a\sin(\psi_a) $],
\begin{eqnarray}  
 \tilde{t}_{{\rm s},ml}(r_{\rm s},\psi_{\rm s}) &=& \frac{\tilde{u}_{{\rm s},ml}(r_{\rm s})\exp(im\psi_{\rm s})}{
  \sqrt{2\pi}} ,
  \nonumber \\
 \tilde{t}_{{\rm i},ml}(r_{\rm i},\psi_{\rm i}) &=&  \frac{\tilde{u}_{{\rm i},ml}(r_{\rm i})\exp(-im\psi_{\rm i})}{
  \sqrt{2\pi}} ,
\label{A8}
\end{eqnarray}
where
\begin{eqnarray}  
 \tilde{u}_{a,ml}(r_a) &=& i^m \int_{0}^{\infty}
  dk_a^\perp \sqrt{k_a^\perp} u_{a,ml}(k_a^\perp) {\rm
  J}_m(k_a^\perp r_a), \nonumber \\
 & & \hspace{10mm} a=s,i
\label{A9}
\end{eqnarray}
and $ {\rm J}_m $ stands for the Bessel function of $ m$-th order.

Using eigenmodes $ \tilde{t}_{{\rm s},ml} $ defined in
Eq.~(\ref{A8}), the averaged signal-field photon flux $ I_{{\rm
s},r\varphi} $ in the crystal output plane is expressed as:
\begin{eqnarray}   
 I_{{\rm s},r\psi}(r_{\rm s},\psi_{\rm s})
 &=& 2 \epsilon_0 c \langle \hat{E}_{\rm s}^{(-)}(r_{\rm s},\psi_{\rm s},L,t_{\rm s})
  \hat{E}_{\rm s}^{(+)}(r_{\rm s},\psi_{\rm s},L,t'_{\rm s}) \rangle_\parallel \nonumber \\
 &=& \sum_{q} \sum_{ml} |\tilde{t}_{{\rm s},ml}(r_{\rm s},\psi_{\rm s})|^2 V_{mlq}^2 .
\label{A10}
\end{eqnarray}
The corresponding radial signal-field intensity profile $ I_{{\rm
s},r} $ is determined as:
\begin{equation}  
 I_{{\rm s},r}(r_{\rm s}) = I_{{\rm s},r\psi}(r_{\rm s},\psi_{\rm s}^0=0) .
\label{A11}
\end{equation}

The averaged signal-field intensity auto-correlation function $
A_{s,r\psi} $ in the crystal output plane is obtained by the
formula analogous to that written in Eq.~(\ref{A3}),
\begin{eqnarray}   
 A_{{\rm s},r\psi}(r_{\rm s},\psi_{\rm s},r'_{\rm s},\psi'_{\rm s}) &=& (2\epsilon_0 c)^2 \langle {\cal N}:\Delta[
  \hat{E}_{\rm s}^{(-)}(r_{\rm s},\psi_{\rm s},L,t_{\rm s})  \nonumber \\
 & & \hspace{-15mm} \mbox{} \times \hat{E}_{\rm s}^{(+)}(r_{\rm s},\psi_{\rm s},L,t_{\rm s})]
  \Delta[\hat{E}_{\rm s}^{(-)}(r'_{\rm s},\psi'_{\rm s},L,t'_{\rm s}) \nonumber \\
 & & \hspace{-15mm} \mbox{} \times
  \hat{E}_{\rm s}^{(+)}(r'_{\rm s},\psi'_{\rm s},L,t'_{\rm s})] : \rangle_\parallel \nonumber \\
 &=& \sum_{q} \left| A_{{\rm s},q,r\psi}^{\rm a}(r_{\rm s},\psi_{\rm s},r'_{\rm s},\psi'_{\rm s}) \right|^2 .
\label{A12}
\end{eqnarray}
In Eq.~(\ref{A12}), the signal-field amplitude auto-correlation
function $ A_{{\rm s},q,r\psi}^{\rm a} $ characterizing mode $ q $
is determined as:
\begin{eqnarray}   
 A_{{\rm s},q,r\psi}^{\rm a}(r_{\rm s},\psi_{\rm s},r'_{\rm s},\psi'_{\rm s}) \hspace{-20mm} & & \nonumber \\
  &=& 2\epsilon_0 c \langle
  \hat{E}_{\rm s}^{(-)}(r_{\rm s},\psi_{\rm s},L,t_{\rm s})\hat{E}_{\rm s}^{(+)}(r'_{\rm s},\psi'_{\rm s},L,t'_{\rm s})
  \rangle_{\parallel,q} \nonumber \\
 &=& \sum_{ml} \tilde{t}_{{\rm s},ml}^*(r_{\rm s},\psi_{\rm s}) \tilde{t}_{{\rm s},ml}(r'_{\rm s},\psi'_{\rm s})
  V_{mlq}^2.
\label{A13}
\end{eqnarray}
The radial ($ A_{{\rm s},r} $) and azimuthal ($ A_{{\rm s},\psi}
$) signal-field intensity auto-correlation functions are easily
derived as follows:
\begin{eqnarray}   
 A_{{\rm s},r}(r_{\rm s},r'_{\rm s}) &=&
  A_{{\rm s},r\psi}(r_{\rm s},\psi_{\rm s}^0=0,r'_{\rm s},\psi_{\rm s}^0=0) ,
  \nonumber \\
 A_{{\rm s},\psi}(\psi_{\rm s},\psi'_{\rm s}) &=&
  A_{{\rm s},r\psi}(r_{\rm s}^0,\psi_{\rm s},r_{\rm s}^0,\psi'_{\rm s}).
\label{A14}
\end{eqnarray}

Finally, averaged intensity cross-correlations between the signal
and idler fields are described by the following fourth-order
cross-correlation function:
\begin{eqnarray}   
 C_{r\psi}(r_{\rm s},\psi_{\rm s},r_{\rm i},\psi_{\rm i}) \hspace{-23mm} & & \nonumber \\
  &=& (2\epsilon_0 c)^2 \langle {\cal N}: \Delta[
  \hat{E}_{\rm s}^{(-)}(r_{\rm s},\psi_{\rm s},L,t_{\rm s}) \hat{E}_{\rm s}^{(+)}(r_{\rm s},\psi_{\rm s},L,t_{\rm s})]
  \nonumber \\
  & & \mbox{} \times
  \Delta[ \hat{E}_{\rm i}^{(-)}(r_{\rm i},\psi_{\rm i},L,t_{\rm i})\hat{E}_{\rm i}^{(+)}(r_{\rm i},\psi_{\rm i},L,t_{\rm i})]
  : \rangle_\parallel \nonumber \\
 &=& \sum_{q} \left|
  \sum_{ml} \tilde{t}_{{\rm s},q}(r_{\rm s},\psi_{\rm s}) \tilde{t}_{{\rm i},q}(r_{\rm i},\psi_{\rm i}) U_{mlq} V_{mlq}
  \right|^2 .
\label{A15}
\end{eqnarray}
The corresponding radial ($ C_{{\rm s},r} $) and azimuthal ($
C_{{\rm s},\psi} $) intensity cross-correlation functions are
defined as:
\begin{eqnarray}   
 C_{{\rm s},r}(r_{\rm s},r_{\rm i}) &=&
  C_{{\rm s},r\psi}(r_{\rm s},\psi_{\rm s}^0=0,r_{\rm i},\psi_{\rm i}^0=0) ,
  \nonumber \\
 C_{{\rm s},\psi}(\psi_{\rm s},\psi_{\rm i}) &=&
  C_{{\rm s},r\psi}(r^0,\psi_{\rm s},r^0,\psi_{\rm i}).
\label{A16}
\end{eqnarray}

\bibliography{perina}

\begin{thebibliography}{69}
\expandafter\ifx\csname natexlab\endcsname\relax\def\natexlab#1{#1}\fi
\expandafter\ifx\csname bibnamefont\endcsname\relax
  \def\bibnamefont#1{#1}\fi
\expandafter\ifx\csname bibfnamefont\endcsname\relax
  \def\bibfnamefont#1{#1}\fi
\expandafter\ifx\csname citenamefont\endcsname\relax
  \def\citenamefont#1{#1}\fi
\expandafter\ifx\csname url\endcsname\relax
  \def\url#1{\texttt{#1}}\fi
\expandafter\ifx\csname urlprefix\endcsname\relax\def\urlprefix{URL }\fi
\providecommand{\bibinfo}[2]{#2}
\providecommand{\eprint}[2][]{\url{#2}}

\bibitem[{\citenamefont{Boyd}(2003)}]{Boyd2003}
\bibinfo{author}{\bibfnamefont{R.~W.} \bibnamefont{Boyd}},
  \emph{\bibinfo{title}{Nonlinear Optics, 2nd edition}}
  (\bibinfo{publisher}{Academic Press, New York}, \bibinfo{year}{2003}).

\bibitem[{\citenamefont{Mandel and Wolf}(1995)}]{Mandel1995}
\bibinfo{author}{\bibfnamefont{L.}~\bibnamefont{Mandel}} \bibnamefont{and}
  \bibinfo{author}{\bibfnamefont{E.}~\bibnamefont{Wolf}},
  \emph{\bibinfo{title}{Optical Coherence and Quantum Optics}}
  (\bibinfo{publisher}{Cambridge Univ. Press, Cambridge},
  \bibinfo{year}{1995}).

\bibitem[{\citenamefont{Bouwmeester et~al.}(2000)\citenamefont{Bouwmeester,
  Ekert, and Zeilinger}}]{Bouwmeester2000}
\bibinfo{author}{\bibfnamefont{D.}~\bibnamefont{Bouwmeester}},
  \bibinfo{author}{\bibfnamefont{A.}~\bibnamefont{Ekert}}, \bibnamefont{and}
  \bibinfo{author}{\bibfnamefont{A.}~\bibnamefont{Zeilinger}},
  \emph{\bibinfo{title}{The Physics of Quantum Information}}
  (\bibinfo{publisher}{Springer, Berlin}, \bibinfo{year}{2000}).

\bibitem[{\citenamefont{Carrasco et~al.}(2004)\citenamefont{Carrasco, Torres,
  Torner, Sergienko, Saleh, and Teich}}]{Carrasco2004}
\bibinfo{author}{\bibfnamefont{S.}~\bibnamefont{Carrasco}},
  \bibinfo{author}{\bibfnamefont{J.~P.} \bibnamefont{Torres}},
  \bibinfo{author}{\bibfnamefont{L.}~\bibnamefont{Torner}},
  \bibinfo{author}{\bibfnamefont{A.~V.} \bibnamefont{Sergienko}},
  \bibinfo{author}{\bibfnamefont{B.~E.~A.} \bibnamefont{Saleh}},
  \bibnamefont{and} \bibinfo{author}{\bibfnamefont{M.~C.} \bibnamefont{Teich}},
  \bibinfo{journal}{Opt. Lett.} \textbf{\bibinfo{volume}{29}},
  \bibinfo{pages}{2429} (\bibinfo{year}{2004}).

\bibitem[{\citenamefont{Kolobov and Sokolov}(1989{\natexlab{a}})}]{Kolobov1989}
\bibinfo{author}{\bibfnamefont{M.~I.} \bibnamefont{Kolobov}} \bibnamefont{and}
  \bibinfo{author}{\bibfnamefont{I.~V.} \bibnamefont{Sokolov}},
  \bibinfo{journal}{Zh. Eksp. Teor. Fiz.} \textbf{\bibinfo{volume}{96}},
  \bibinfo{pages}{1945} (\bibinfo{year}{1989}{\natexlab{a}}).

\bibitem[{\citenamefont{Kolobov and
  Sokolov}(1989{\natexlab{b}})}]{Kolobov1989a}
\bibinfo{author}{\bibfnamefont{M.~I.} \bibnamefont{Kolobov}} \bibnamefont{and}
  \bibinfo{author}{\bibfnamefont{I.~V.} \bibnamefont{Sokolov}},
  \bibinfo{journal}{Phys. Lett. A} \textbf{\bibinfo{volume}{140}},
  \bibinfo{pages}{101} (\bibinfo{year}{1989}{\natexlab{b}}).

\bibitem[{\citenamefont{Jedrkiewicz et~al.}(2004)\citenamefont{Jedrkiewicz,
  Jiang, Brambilla, Gatti, Bache, Lugiato, and {Di~Trapani}}}]{Jedrkiewicz2004}
\bibinfo{author}{\bibfnamefont{O.}~\bibnamefont{Jedrkiewicz}},
  \bibinfo{author}{\bibfnamefont{Y.~K.} \bibnamefont{Jiang}},
  \bibinfo{author}{\bibfnamefont{E.}~\bibnamefont{Brambilla}},
  \bibinfo{author}{\bibfnamefont{A.}~\bibnamefont{Gatti}},
  \bibinfo{author}{\bibfnamefont{M.}~\bibnamefont{Bache}},
  \bibinfo{author}{\bibfnamefont{L.~A.} \bibnamefont{Lugiato}},
  \bibnamefont{and}
  \bibinfo{author}{\bibfnamefont{P.}~\bibnamefont{{Di~Trapani}}},
  \bibinfo{journal}{Phys. Rev. Lett.} \textbf{\bibinfo{volume}{93}},
  \bibinfo{pages}{243601} (\bibinfo{year}{2004}).

\bibitem[{\citenamefont{Bondani et~al.}(2007)\citenamefont{Bondani, Allevi,
  Zambra, Paris, and Andreoni}}]{Bondani2007}
\bibinfo{author}{\bibfnamefont{M.}~\bibnamefont{Bondani}},
  \bibinfo{author}{\bibfnamefont{A.}~\bibnamefont{Allevi}},
  \bibinfo{author}{\bibfnamefont{G.}~\bibnamefont{Zambra}},
  \bibinfo{author}{\bibfnamefont{M.~G.~A.} \bibnamefont{Paris}},
  \bibnamefont{and} \bibinfo{author}{\bibfnamefont{A.}~\bibnamefont{Andreoni}},
  \bibinfo{journal}{Phys. Rev. A} \textbf{\bibinfo{volume}{76}},
  \bibinfo{pages}{013833} (\bibinfo{year}{2007}).

\bibitem[{\citenamefont{Blanchet et~al.}(2008)\citenamefont{Blanchet, Devaux,
  Furfaro, and Lantz}}]{Blanchet2008}
\bibinfo{author}{\bibfnamefont{J.-L.} \bibnamefont{Blanchet}},
  \bibinfo{author}{\bibfnamefont{F.}~\bibnamefont{Devaux}},
  \bibinfo{author}{\bibfnamefont{L.}~\bibnamefont{Furfaro}}, \bibnamefont{and}
  \bibinfo{author}{\bibfnamefont{E.}~\bibnamefont{Lantz}},
  \bibinfo{journal}{Phys. Rev. Lett.} \textbf{\bibinfo{volume}{101}},
  \bibinfo{pages}{233604} (\bibinfo{year}{2008}).

\bibitem[{\citenamefont{Brida et~al.}(2009{\natexlab{a}})\citenamefont{Brida,
  Caspani, Gatti, Genovese, Meda, and Berchera}}]{Brida2009a}
\bibinfo{author}{\bibfnamefont{G.}~\bibnamefont{Brida}},
  \bibinfo{author}{\bibfnamefont{L.}~\bibnamefont{Caspani}},
  \bibinfo{author}{\bibfnamefont{A.}~\bibnamefont{Gatti}},
  \bibinfo{author}{\bibfnamefont{M.}~\bibnamefont{Genovese}},
  \bibinfo{author}{\bibfnamefont{A.}~\bibnamefont{Meda}}, \bibnamefont{and}
  \bibinfo{author}{\bibfnamefont{I.~R.} \bibnamefont{Berchera}},
  \bibinfo{journal}{Phys. Rev. Lett.} \textbf{\bibinfo{volume}{102}},
  \bibinfo{pages}{213602} (\bibinfo{year}{2009}{\natexlab{a}}).

\bibitem[{\citenamefont{Brida et~al.}(2010)\citenamefont{Brida, Degiovanni,
  Genovese, Rastello, and Berchera}}]{Brida2010}
\bibinfo{author}{\bibfnamefont{G.}~\bibnamefont{Brida}},
  \bibinfo{author}{\bibfnamefont{I.~P.} \bibnamefont{Degiovanni}},
  \bibinfo{author}{\bibfnamefont{M.}~\bibnamefont{Genovese}},
  \bibinfo{author}{\bibfnamefont{M.~L.} \bibnamefont{Rastello}},
  \bibnamefont{and} \bibinfo{author}{\bibfnamefont{I.~R.}
  \bibnamefont{Berchera}}, \bibinfo{journal}{Opt. Express}
  \textbf{\bibinfo{volume}{18}}, \bibinfo{pages}{20572} (\bibinfo{year}{2010}).

\bibitem[{\citenamefont{Boyer et~al.}(2008)\citenamefont{Boyer, Marino, Pooser,
  and Lett}}]{Boyer2008}
\bibinfo{author}{\bibfnamefont{V.}~\bibnamefont{Boyer}},
  \bibinfo{author}{\bibfnamefont{A.~M.} \bibnamefont{Marino}},
  \bibinfo{author}{\bibfnamefont{R.~C.} \bibnamefont{Pooser}},
  \bibnamefont{and} \bibinfo{author}{\bibfnamefont{P.~D.} \bibnamefont{Lett}},
  \bibinfo{journal}{Science} \textbf{\bibinfo{volume}{321}},
  \bibinfo{pages}{544} (\bibinfo{year}{2008}).

\bibitem[{\citenamefont{Rubin et~al.}(1994)\citenamefont{Rubin, Klyshko, Shih,
  and Sergienko}}]{Rubin1994}
\bibinfo{author}{\bibfnamefont{M.~H.} \bibnamefont{Rubin}},
  \bibinfo{author}{\bibfnamefont{D.~N.} \bibnamefont{Klyshko}},
  \bibinfo{author}{\bibfnamefont{Y.~H.} \bibnamefont{Shih}}, \bibnamefont{and}
  \bibinfo{author}{\bibfnamefont{A.~V.} \bibnamefont{Sergienko}},
  \bibinfo{journal}{Phys. Rev. A} \textbf{\bibinfo{volume}{50}},
  \bibinfo{pages}{5122} (\bibinfo{year}{1994}).

\bibitem[{\citenamefont{{Pe\v{r}ina~Jr.}
  et~al.}(1999)\citenamefont{{Pe\v{r}ina~Jr.}, Sergienko, Jost, Saleh, and
  Teich}}]{PerinaJr1999a}
\bibinfo{author}{\bibfnamefont{J.}~\bibnamefont{{Pe\v{r}ina~Jr.}}},
  \bibinfo{author}{\bibfnamefont{A.~V.} \bibnamefont{Sergienko}},
  \bibinfo{author}{\bibfnamefont{B.~M.} \bibnamefont{Jost}},
  \bibinfo{author}{\bibfnamefont{B.~E.~A.} \bibnamefont{Saleh}},
  \bibnamefont{and} \bibinfo{author}{\bibfnamefont{M.~C.} \bibnamefont{Teich}},
  \bibinfo{journal}{Phys. Rev. A} \textbf{\bibinfo{volume}{59}},
  \bibinfo{pages}{2359} (\bibinfo{year}{1999}).

\bibitem[{\citenamefont{{Pe\v{r}ina~Jr.}}(2014)}]{PerinaJr2014}
\bibinfo{author}{\bibfnamefont{J.}~\bibnamefont{{Pe\v{r}ina~Jr.}}}, in
  \emph{\bibinfo{booktitle}{Progress in Optics, Vol. 59}}, edited by
  \bibinfo{editor}{\bibfnamefont{E.}~\bibnamefont{Wolf}}
  (\bibinfo{publisher}{Elsevier, Amsterdam}, \bibinfo{year}{2014}), pp.
  \bibinfo{pages}{89---158}.

\bibitem[{\citenamefont{Jedrkiewicz et~al.}(2012)\citenamefont{Jedrkiewicz,
  Gatti, Brambilla, and {Di~Trapani}}}]{Jedrkiewicz2012}
\bibinfo{author}{\bibfnamefont{O.}~\bibnamefont{Jedrkiewicz}},
  \bibinfo{author}{\bibfnamefont{A.}~\bibnamefont{Gatti}},
  \bibinfo{author}{\bibfnamefont{E.}~\bibnamefont{Brambilla}},
  \bibnamefont{and}
  \bibinfo{author}{\bibfnamefont{P.}~\bibnamefont{{Di~Trapani}}},
  \bibinfo{journal}{Phys. Rev. Lett.} \textbf{\bibinfo{volume}{109}},
  \bibinfo{pages}{243901} (\bibinfo{year}{2012}).

\bibitem[{\citenamefont{Machulka et~al.}(2014)\citenamefont{Machulka, Haderka,
  {Pe\v{r}ina Jr}, Lamperti, Allevi, and Bondani}}]{Machulka2014}
\bibinfo{author}{\bibfnamefont{R.}~\bibnamefont{Machulka}},
  \bibinfo{author}{\bibfnamefont{O.}~\bibnamefont{Haderka}},
  \bibinfo{author}{\bibfnamefont{J.}~\bibnamefont{{Pe\v{r}ina Jr}}},
  \bibinfo{author}{\bibfnamefont{M.}~\bibnamefont{Lamperti}},
  \bibinfo{author}{\bibfnamefont{A.}~\bibnamefont{Allevi}}, \bibnamefont{and}
  \bibinfo{author}{\bibfnamefont{M.}~\bibnamefont{Bondani}},
  \bibinfo{journal}{Opt. Express} \textbf{\bibinfo{volume}{22}},
  \bibinfo{pages}{13374} (\bibinfo{year}{2014}).

\bibitem[{\citenamefont{Gatti et~al.}(2003)\citenamefont{Gatti, Zambrini,
  {San~Miguel}, and Lugiato}}]{Gatti2003}
\bibinfo{author}{\bibfnamefont{A.}~\bibnamefont{Gatti}},
  \bibinfo{author}{\bibfnamefont{R.}~\bibnamefont{Zambrini}},
  \bibinfo{author}{\bibfnamefont{M.}~\bibnamefont{{San~Miguel}}},
  \bibnamefont{and} \bibinfo{author}{\bibfnamefont{L.~A.}
  \bibnamefont{Lugiato}}, \bibinfo{journal}{Phys. Rev. A}
  \textbf{\bibinfo{volume}{68}}, \bibinfo{eid}{053807} (\bibinfo{year}{2003}).

\bibitem[{\citenamefont{Brambilla et~al.}(2010)\citenamefont{Brambilla,
  Caspani, Lugiato, and Gatti}}]{Brambilla2010}
\bibinfo{author}{\bibfnamefont{E.}~\bibnamefont{Brambilla}},
  \bibinfo{author}{\bibfnamefont{L.}~\bibnamefont{Caspani}},
  \bibinfo{author}{\bibfnamefont{L.~A.} \bibnamefont{Lugiato}},
  \bibnamefont{and} \bibinfo{author}{\bibfnamefont{A.}~\bibnamefont{Gatti}},
  \bibinfo{journal}{Phys. Rev. A} \textbf{\bibinfo{volume}{82}},
  \bibinfo{pages}{013835} (\bibinfo{year}{2010}).

\bibitem[{\citenamefont{Caspani et~al.}(2010)\citenamefont{Caspani, Brambilla,
  and Gatti}}]{Caspani2010}
\bibinfo{author}{\bibfnamefont{L.}~\bibnamefont{Caspani}},
  \bibinfo{author}{\bibfnamefont{E.}~\bibnamefont{Brambilla}},
  \bibnamefont{and} \bibinfo{author}{\bibfnamefont{A.}~\bibnamefont{Gatti}},
  \bibinfo{journal}{Phys. Rev. A} \textbf{\bibinfo{volume}{81}},
  \bibinfo{pages}{033808} (\bibinfo{year}{2010}).

\bibitem[{\citenamefont{Dayan et~al.}(2007)\citenamefont{Dayan, Bromberg, Afek,
  and Silberberg}}]{Dayan2007}
\bibinfo{author}{\bibfnamefont{B.}~\bibnamefont{Dayan}},
  \bibinfo{author}{\bibfnamefont{Y.}~\bibnamefont{Bromberg}},
  \bibinfo{author}{\bibfnamefont{I.}~\bibnamefont{Afek}}, \bibnamefont{and}
  \bibinfo{author}{\bibfnamefont{Y.}~\bibnamefont{Silberberg}},
  \bibinfo{journal}{Phys. Rev. A} \textbf{\bibinfo{volume}{75}},
  \bibinfo{pages}{043804} (\bibinfo{year}{2007}).

\bibitem[{\citenamefont{Brambilla et~al.}(2004)\citenamefont{Brambilla, Gatti,
  Bache, and Lugiato}}]{Brambilla2004}
\bibinfo{author}{\bibfnamefont{E.}~\bibnamefont{Brambilla}},
  \bibinfo{author}{\bibfnamefont{A.}~\bibnamefont{Gatti}},
  \bibinfo{author}{\bibfnamefont{M.}~\bibnamefont{Bache}}, \bibnamefont{and}
  \bibinfo{author}{\bibfnamefont{L.~A.} \bibnamefont{Lugiato}},
  \bibinfo{journal}{Phys. Rev. A} \textbf{\bibinfo{volume}{69}},
  \bibinfo{pages}{023802} (\bibinfo{year}{2004}).

\bibitem[{\citenamefont{Law et~al.}(2000)\citenamefont{Law, Walmsley, and
  Eberly}}]{Law2000}
\bibinfo{author}{\bibfnamefont{C.~K.} \bibnamefont{Law}},
  \bibinfo{author}{\bibfnamefont{I.~A.} \bibnamefont{Walmsley}},
  \bibnamefont{and} \bibinfo{author}{\bibfnamefont{J.~H.}
  \bibnamefont{Eberly}}, \bibinfo{journal}{Phys. Rev. Lett.}
  \textbf{\bibinfo{volume}{84}}, \bibinfo{pages}{5304} (\bibinfo{year}{2000}).

\bibitem[{\citenamefont{Law and Eberly}(2004)}]{Law2004}
\bibinfo{author}{\bibfnamefont{C.~K.} \bibnamefont{Law}} \bibnamefont{and}
  \bibinfo{author}{\bibfnamefont{J.~H.} \bibnamefont{Eberly}},
  \bibinfo{journal}{Phys. Rev. Lett.} \textbf{\bibinfo{volume}{92}},
  \bibinfo{eid}{127903} (\bibinfo{year}{2004}).

\bibitem[{\citenamefont{Christ et~al.}(2011)\citenamefont{Christ, Laiho,
  Eckstein, Cassemiro, and Silberhorn}}]{Christ2011}
\bibinfo{author}{\bibfnamefont{A.}~\bibnamefont{Christ}},
  \bibinfo{author}{\bibfnamefont{K.}~\bibnamefont{Laiho}},
  \bibinfo{author}{\bibfnamefont{A.}~\bibnamefont{Eckstein}},
  \bibinfo{author}{\bibfnamefont{K.~N.} \bibnamefont{Cassemiro}},
  \bibnamefont{and}
  \bibinfo{author}{\bibfnamefont{C.}~\bibnamefont{Silberhorn}},
  \bibinfo{journal}{New J. Phys.} \textbf{\bibinfo{volume}{13}},
  \bibinfo{eid}{033027} (\bibinfo{year}{2011}).

\bibitem[{\citenamefont{Avella et~al.}(2014)\citenamefont{Avella, Gramegna,
  Shurupov, Brida, Chekhova, and Genovese}}]{Avella2014}
\bibinfo{author}{\bibfnamefont{A.}~\bibnamefont{Avella}},
  \bibinfo{author}{\bibfnamefont{M.}~\bibnamefont{Gramegna}},
  \bibinfo{author}{\bibfnamefont{A.}~\bibnamefont{Shurupov}},
  \bibinfo{author}{\bibfnamefont{G.}~\bibnamefont{Brida}},
  \bibinfo{author}{\bibfnamefont{M.}~\bibnamefont{Chekhova}}, \bibnamefont{and}
  \bibinfo{author}{\bibfnamefont{M.}~\bibnamefont{Genovese}},
  \bibinfo{journal}{Phys. Rev. A} \textbf{\bibinfo{volume}{89}},
  \bibinfo{pages}{023808} (\bibinfo{year}{2014}).

\bibitem[{\citenamefont{Shapiro and Shakeel}(1997)}]{Shapiro1997}
\bibinfo{author}{\bibfnamefont{J.~H.} \bibnamefont{Shapiro}} \bibnamefont{and}
  \bibinfo{author}{\bibfnamefont{A.}~\bibnamefont{Shakeel}},
  \bibinfo{journal}{J. Opt. Soc. Am. B} \textbf{\bibinfo{volume}{14}},
  \bibinfo{pages}{232} (\bibinfo{year}{1997}).

\bibitem[{\citenamefont{Bennink and Boyd}(2002)}]{Bennink2002}
\bibinfo{author}{\bibfnamefont{R.~S.} \bibnamefont{Bennink}} \bibnamefont{and}
  \bibinfo{author}{\bibfnamefont{R.~W.} \bibnamefont{Boyd}},
  \bibinfo{journal}{Phys. Rev. A} \textbf{\bibinfo{volume}{66}},
  \bibinfo{pages}{053815} (\bibinfo{year}{2002}).

\bibitem[{\citenamefont{Bobrov et~al.}(2013)\citenamefont{Bobrov, Straupe,
  Kovlakov, and Kulik}}]{Bobrov2013}
\bibinfo{author}{\bibfnamefont{I.~B.} \bibnamefont{Bobrov}},
  \bibinfo{author}{\bibfnamefont{S.~S.} \bibnamefont{Straupe}},
  \bibinfo{author}{\bibfnamefont{E.~V.} \bibnamefont{Kovlakov}},
  \bibnamefont{and} \bibinfo{author}{\bibfnamefont{S.~P.} \bibnamefont{Kulik}},
  \bibinfo{journal}{N. J. Phys.} \textbf{\bibinfo{volume}{15}},
  \bibinfo{pages}{073016} (\bibinfo{year}{2013}).

\bibitem[{\citenamefont{Brecht et~al.}(2014)\citenamefont{Brecht, Eckstein,
  Ricken, Quiring, Suche, Sansoni, and Silberhorn}}]{Brecht2014}
\bibinfo{author}{\bibfnamefont{B.}~\bibnamefont{Brecht}},
  \bibinfo{author}{\bibfnamefont{A.}~\bibnamefont{Eckstein}},
  \bibinfo{author}{\bibfnamefont{R.}~\bibnamefont{Ricken}},
  \bibinfo{author}{\bibfnamefont{V.}~\bibnamefont{Quiring}},
  \bibinfo{author}{\bibfnamefont{H.}~\bibnamefont{Suche}},
  \bibinfo{author}{\bibfnamefont{L.}~\bibnamefont{Sansoni}}, \bibnamefont{and}
  \bibinfo{author}{\bibfnamefont{C.}~\bibnamefont{Silberhorn}},
  \bibinfo{journal}{Phys. Rev. A} \textbf{\bibinfo{volume}{90}},
  \bibinfo{pages}{030302(R)} (\bibinfo{year}{2014}).

\bibitem[{\citenamefont{Annamalai et~al.}(2011)\citenamefont{Annamalai,
  Stelmakh, Vasilyev, and Kumar}}]{Annamalai2011}
\bibinfo{author}{\bibfnamefont{M.}~\bibnamefont{Annamalai}},
  \bibinfo{author}{\bibfnamefont{N.}~\bibnamefont{Stelmakh}},
  \bibinfo{author}{\bibfnamefont{M.}~\bibnamefont{Vasilyev}}, \bibnamefont{and}
  \bibinfo{author}{\bibfnamefont{P.}~\bibnamefont{Kumar}},
  \bibinfo{journal}{Opt. Express} \textbf{\bibinfo{volume}{19}},
  \bibinfo{pages}{26710} (\bibinfo{year}{2011}).

\bibitem[{\citenamefont{Wasilewski et~al.}(2006)\citenamefont{Wasilewski,
  Lvovsky, Banaszek, and Radzewicz}}]{Wasilewski2006}
\bibinfo{author}{\bibfnamefont{W.}~\bibnamefont{Wasilewski}},
  \bibinfo{author}{\bibfnamefont{A.~I.} \bibnamefont{Lvovsky}},
  \bibinfo{author}{\bibfnamefont{K.}~\bibnamefont{Banaszek}}, \bibnamefont{and}
  \bibinfo{author}{\bibfnamefont{C.}~\bibnamefont{Radzewicz}},
  \bibinfo{journal}{Phys. Rev. A} \textbf{\bibinfo{volume}{73}},
  \bibinfo{pages}{063819} (\bibinfo{year}{2006}).

\bibitem[{\citenamefont{Lvovsky et~al.}(2007)\citenamefont{Lvovsky, Wasilevski,
  and Banaszek}}]{Lvovsky2007}
\bibinfo{author}{\bibfnamefont{A.~I.} \bibnamefont{Lvovsky}},
  \bibinfo{author}{\bibfnamefont{W.}~\bibnamefont{Wasilevski}},
  \bibnamefont{and} \bibinfo{author}{\bibfnamefont{K.}~\bibnamefont{Banaszek}},
  \bibinfo{journal}{J. Mod. Opt.} \textbf{\bibinfo{volume}{54}},
  \bibinfo{pages}{721} (\bibinfo{year}{2007}).

\bibitem[{\citenamefont{Christ et~al.}(2013)\citenamefont{Christ, Brecht,
  Mauerer, and Silberhorn}}]{Christ2013}
\bibinfo{author}{\bibfnamefont{A.}~\bibnamefont{Christ}},
  \bibinfo{author}{\bibfnamefont{B.}~\bibnamefont{Brecht}},
  \bibinfo{author}{\bibfnamefont{W.}~\bibnamefont{Mauerer}}, \bibnamefont{and}
  \bibinfo{author}{\bibfnamefont{C.}~\bibnamefont{Silberhorn}},
  \bibinfo{journal}{New J. Phys.} \textbf{\bibinfo{volume}{15}},
  \bibinfo{pages}{053038} (\bibinfo{year}{2013}).

\bibitem[{\citenamefont{Sharapova et~al.}(2015)\citenamefont{Sharapova, {P\'
  erez}, Tikhonova, and Chekhova}}]{Sharapovova2015}
\bibinfo{author}{\bibfnamefont{P.}~\bibnamefont{Sharapova}},
  \bibinfo{author}{\bibfnamefont{A.~M.} \bibnamefont{{P\' erez}}},
  \bibinfo{author}{\bibfnamefont{O.~V.} \bibnamefont{Tikhonova}},
  \bibnamefont{and} \bibinfo{author}{\bibfnamefont{M.~V.}
  \bibnamefont{Chekhova}}, \bibinfo{journal}{Phys. Rev. A}
  \textbf{\bibinfo{volume}{91}}, \bibinfo{pages}{043816}
  (\bibinfo{year}{2015}).

\bibitem[{\citenamefont{{Pe\v{r}ina~Jr.}}(2013)}]{PerinaJr2013}
\bibinfo{author}{\bibfnamefont{J.}~\bibnamefont{{Pe\v{r}ina~Jr.}}},
  \bibinfo{journal}{Phys. Rev. A} \textbf{\bibinfo{volume}{87}},
  \bibinfo{eid}{013833} (\bibinfo{year}{2013}).

\bibitem[{\citenamefont{{Stobi\'nska} et~al.}(2012)\citenamefont{{Stobi\'nska},
  {T\"{o}ppel}, Sekatski, and Chekhova}}]{Stobinska2012}
\bibinfo{author}{\bibfnamefont{M.}~\bibnamefont{{Stobi\'nska}}},
  \bibinfo{author}{\bibfnamefont{F.}~\bibnamefont{{T\"{o}ppel}}},
  \bibinfo{author}{\bibfnamefont{P.}~\bibnamefont{Sekatski}}, \bibnamefont{and}
  \bibinfo{author}{\bibfnamefont{M.~V.} \bibnamefont{Chekhova}},
  \bibinfo{journal}{Phys. Rev. A} \textbf{\bibinfo{volume}{86}},
  \bibinfo{eid}{022323} (\bibinfo{year}{2012}).

\bibitem[{\citenamefont{Chekhova et~al.}(2015)\citenamefont{Chekhova, G.Leuchs,
  and Zukowski}}]{Chekhova2015}
\bibinfo{author}{\bibfnamefont{M.~V.} \bibnamefont{Chekhova}},
  \bibinfo{author}{\bibnamefont{G.Leuchs}}, \bibnamefont{and}
  \bibinfo{author}{\bibfnamefont{M.}~\bibnamefont{Zukowski}},
  \bibinfo{journal}{Opt. Comm.} \textbf{\bibinfo{volume}{337}},
  \bibinfo{pages}{27} (\bibinfo{year}{2015}).

\bibitem[{\citenamefont{Hum and Fejer}(2007)}]{Hum2007}
\bibinfo{author}{\bibfnamefont{D.~S.} \bibnamefont{Hum}} \bibnamefont{and}
  \bibinfo{author}{\bibfnamefont{M.~M.} \bibnamefont{Fejer}},
  \bibinfo{journal}{Comptes Rendus Physique} \textbf{\bibinfo{volume}{8}},
  \bibinfo{pages}{180} (\bibinfo{year}{2007}).

\bibitem[{\citenamefont{Svozil{\'i}k and
  {Pe\v{r}ina~Jr.}}(2009)}]{Svozilik2009}
\bibinfo{author}{\bibfnamefont{J.}~\bibnamefont{Svozil{\'i}k}}
  \bibnamefont{and}
  \bibinfo{author}{\bibfnamefont{J.}~\bibnamefont{{Pe\v{r}ina~Jr.}}},
  \bibinfo{journal}{Phys. Rev. A} \textbf{\bibinfo{volume}{80}},
  \bibinfo{eid}{023819} (\bibinfo{year}{2009}).

\bibitem[{\citenamefont{Kolchin et~al.}(2006)\citenamefont{Kolchin, Du,
  Belthangady, Yin, and Harris}}]{Kolchin2006}
\bibinfo{author}{\bibfnamefont{P.}~\bibnamefont{Kolchin}},
  \bibinfo{author}{\bibfnamefont{S.}~\bibnamefont{Du}},
  \bibinfo{author}{\bibfnamefont{C.}~\bibnamefont{Belthangady}},
  \bibinfo{author}{\bibfnamefont{G.~Y.} \bibnamefont{Yin}}, \bibnamefont{and}
  \bibinfo{author}{\bibfnamefont{S.~E.} \bibnamefont{Harris}},
  \bibinfo{journal}{Phys. Rev. Lett.} \textbf{\bibinfo{volume}{97}},
  \bibinfo{pages}{113602} (\bibinfo{year}{2006}).

\bibitem[{\citenamefont{Glorieux et~al.}(2010)\citenamefont{Glorieux, Dubessy,
  Guibal, Guidoni, Likforman, Coudreau, and Arimondo}}]{Glorieux2010}
\bibinfo{author}{\bibfnamefont{Q.}~\bibnamefont{Glorieux}},
  \bibinfo{author}{\bibfnamefont{R.}~\bibnamefont{Dubessy}},
  \bibinfo{author}{\bibfnamefont{S.}~\bibnamefont{Guibal}},
  \bibinfo{author}{\bibfnamefont{L.}~\bibnamefont{Guidoni}},
  \bibinfo{author}{\bibfnamefont{J.-P.} \bibnamefont{Likforman}},
  \bibinfo{author}{\bibfnamefont{T.}~\bibnamefont{Coudreau}}, \bibnamefont{and}
  \bibinfo{author}{\bibfnamefont{E.}~\bibnamefont{Arimondo}},
  \bibinfo{journal}{Phys. Rev. A} \textbf{\bibinfo{volume}{82}},
  \bibinfo{pages}{033819} (\bibinfo{year}{2010}).

\bibitem[{\citenamefont{Corzo et~al.}(2012)\citenamefont{Corzo, Marino, Jones,
  and Lett}}]{Corzo2012}
\bibinfo{author}{\bibfnamefont{N.~V.} \bibnamefont{Corzo}},
  \bibinfo{author}{\bibfnamefont{A.~M.} \bibnamefont{Marino}},
  \bibinfo{author}{\bibfnamefont{K.~M.} \bibnamefont{Jones}}, \bibnamefont{and}
  \bibinfo{author}{\bibfnamefont{P.~D.} \bibnamefont{Lett}},
  \bibinfo{journal}{Phys. Rev. Lett.} \textbf{\bibinfo{volume}{109}},
  \bibinfo{pages}{043602} (\bibinfo{year}{2012}).

\bibitem[{\citenamefont{Allevi and Bondani}(2014)}]{Allevi2014}
\bibinfo{author}{\bibfnamefont{A.}~\bibnamefont{Allevi}} \bibnamefont{and}
  \bibinfo{author}{\bibfnamefont{M.}~\bibnamefont{Bondani}},
  \bibinfo{journal}{J. Opt. Soc. Am. B} \textbf{\bibinfo{volume}{31}},
  \bibinfo{pages}{B14} (\bibinfo{year}{2014}).

\bibitem[{\citenamefont{Allevi et~al.}(2014)\citenamefont{Allevi, Jedrkiewicz,
  Brambilla, Gatti, {Pe\v{r}ina~Jr.}, Haderka, and Bondani}}]{Allevi2014a}
\bibinfo{author}{\bibfnamefont{A.}~\bibnamefont{Allevi}},
  \bibinfo{author}{\bibfnamefont{O.}~\bibnamefont{Jedrkiewicz}},
  \bibinfo{author}{\bibfnamefont{E.}~\bibnamefont{Brambilla}},
  \bibinfo{author}{\bibfnamefont{A.}~\bibnamefont{Gatti}},
  \bibinfo{author}{\bibfnamefont{J.}~\bibnamefont{{Pe\v{r}ina~Jr.}}},
  \bibinfo{author}{\bibfnamefont{O.}~\bibnamefont{Haderka}}, \bibnamefont{and}
  \bibinfo{author}{\bibfnamefont{M.}~\bibnamefont{Bondani}},
  \bibinfo{journal}{Phys. Rev. A} \textbf{\bibinfo{volume}{90}},
  \bibinfo{pages}{063812} (\bibinfo{year}{2014}).

\bibitem[{\citenamefont{Allevi et~al.}(2015{\natexlab{a}})\citenamefont{Allevi,
  Jedrkiewicz, Haderka, {Pe\v{r}ina~Jr.}, and Bondani}}]{Allevi2015}
\bibinfo{author}{\bibfnamefont{A.}~\bibnamefont{Allevi}},
  \bibinfo{author}{\bibfnamefont{O.}~\bibnamefont{Jedrkiewicz}},
  \bibinfo{author}{\bibfnamefont{O.}~\bibnamefont{Haderka}},
  \bibinfo{author}{\bibfnamefont{J.}~\bibnamefont{{Pe\v{r}ina~Jr.}}},
  \bibnamefont{and} \bibinfo{author}{\bibfnamefont{M.}~\bibnamefont{Bondani}},
  in \emph{\bibinfo{booktitle}{Proc. of SPIE 9505}}, edited by
  \bibinfo{editor}{\bibfnamefont{K.}~\bibnamefont{Banaszek}} \bibnamefont{and}
  \bibinfo{editor}{\bibfnamefont{C.}~\bibnamefont{Silberhorn}}
  (\bibinfo{publisher}{SPIE}, \bibinfo{address}{Bellingham},
  \bibinfo{year}{2015}{\natexlab{a}}), p. \bibinfo{pages}{95050S}.

\bibitem[{\citenamefont{Allevi et~al.}(2015{\natexlab{b}})\citenamefont{Allevi,
  Lamperti, Machulka, Jedrkiewicz, Brambilla, Gatti, {Pe\v{r}ina~Jr.}, Haderka,
  and Bondani}}]{Allevi2015a}
\bibinfo{author}{\bibfnamefont{A.}~\bibnamefont{Allevi}},
  \bibinfo{author}{\bibfnamefont{M.}~\bibnamefont{Lamperti}},
  \bibinfo{author}{\bibfnamefont{R.}~\bibnamefont{Machulka}},
  \bibinfo{author}{\bibfnamefont{O.}~\bibnamefont{Jedrkiewicz}},
  \bibinfo{author}{\bibfnamefont{E.}~\bibnamefont{Brambilla}},
  \bibinfo{author}{\bibfnamefont{A.}~\bibnamefont{Gatti}},
  \bibinfo{author}{\bibfnamefont{J.}~\bibnamefont{{Pe\v{r}ina~Jr.}}},
  \bibinfo{author}{\bibfnamefont{O.}~\bibnamefont{Haderka}}, \bibnamefont{and}
  \bibinfo{author}{\bibfnamefont{M.}~\bibnamefont{Bondani}}, in
  \emph{\bibinfo{booktitle}{Proc. of SPIE 9505}}, edited by
  \bibinfo{editor}{\bibfnamefont{K.}~\bibnamefont{Banaszek}} \bibnamefont{and}
  \bibinfo{editor}{\bibfnamefont{C.}~\bibnamefont{Silberhorn}}
  (\bibinfo{publisher}{SPIE}, \bibinfo{address}{Bellingham},
  \bibinfo{year}{2015}{\natexlab{b}}), p. \bibinfo{pages}{950508}.

\bibitem[{\citenamefont{Spasibko et~al.}(2012)\citenamefont{Spasibko, Iskhakov,
  and Chekhova}}]{Spasibko2012}
\bibinfo{author}{\bibfnamefont{K.~Y.} \bibnamefont{Spasibko}},
  \bibinfo{author}{\bibfnamefont{T.~S.} \bibnamefont{Iskhakov}},
  \bibnamefont{and} \bibinfo{author}{\bibfnamefont{M.~V.}
  \bibnamefont{Chekhova}}, \bibinfo{journal}{Opt. Express}
  \textbf{\bibinfo{volume}{20}}, \bibinfo{pages}{7507} (\bibinfo{year}{2012}).

\bibitem[{\citenamefont{{P\'{e}rez} et~al.}(2014)\citenamefont{{P\'{e}rez},
  Iskhakov, Sharapova, Lemieux, Tikhonova, Chekhova, and Leuchs}}]{Perez2014}
\bibinfo{author}{\bibfnamefont{A.~M.} \bibnamefont{{P\'{e}rez}}},
  \bibinfo{author}{\bibfnamefont{T.~S.} \bibnamefont{Iskhakov}},
  \bibinfo{author}{\bibfnamefont{P.}~\bibnamefont{Sharapova}},
  \bibinfo{author}{\bibfnamefont{S.}~\bibnamefont{Lemieux}},
  \bibinfo{author}{\bibfnamefont{O.~V.} \bibnamefont{Tikhonova}},
  \bibinfo{author}{\bibfnamefont{M.~V.} \bibnamefont{Chekhova}},
  \bibnamefont{and} \bibinfo{author}{\bibfnamefont{G.}~\bibnamefont{Leuchs}},
  \bibinfo{journal}{Opt. Lett.} \textbf{\bibinfo{volume}{39}},
  \bibinfo{pages}{2403} (\bibinfo{year}{2014}).

\bibitem[{\citenamefont{Pe\v{r}ina}(1991)}]{Perina1991}
\bibinfo{author}{\bibfnamefont{J.}~\bibnamefont{Pe\v{r}ina}},
  \emph{\bibinfo{title}{Quantum Statistics of Linear and Nonlinear Optical
  Phenomena}} (\bibinfo{publisher}{Kluwer, Dordrecht}, \bibinfo{year}{1991}).

\bibitem[{\citenamefont{{Pe\v{r}ina~Jr.} and Pe\v{r}ina}(2000)}]{PerinaJr2000}
\bibinfo{author}{\bibfnamefont{J.}~\bibnamefont{{Pe\v{r}ina~Jr.}}}
  \bibnamefont{and}
  \bibinfo{author}{\bibfnamefont{J.}~\bibnamefont{Pe\v{r}ina}}, in
  \emph{\bibinfo{booktitle}{Progress in Optics, Vol. 41}}, edited by
  \bibinfo{editor}{\bibfnamefont{E.}~\bibnamefont{Wolf}}
  (\bibinfo{publisher}{Elsevier, Amsterdam}, \bibinfo{year}{2000}), pp.
  \bibinfo{pages}{361---419}.

\bibitem[{\citenamefont{Huttner et~al.}(1990)\citenamefont{Huttner, Serulnik,
  and Ben-Aryeh}}]{Huttner1990}
\bibinfo{author}{\bibfnamefont{B.}~\bibnamefont{Huttner}},
  \bibinfo{author}{\bibfnamefont{S.}~\bibnamefont{Serulnik}}, \bibnamefont{and}
  \bibinfo{author}{\bibfnamefont{Y.}~\bibnamefont{Ben-Aryeh}},
  \bibinfo{journal}{Phys. Rev. A} \textbf{\bibinfo{volume}{42}},
  \bibinfo{pages}{5594} (\bibinfo{year}{1990}).

\bibitem[{\citenamefont{Vogel et~al.}(2001)\citenamefont{Vogel, Welsch, and
  Walentowicz}}]{Vogel2001}
\bibinfo{author}{\bibfnamefont{W.}~\bibnamefont{Vogel}},
  \bibinfo{author}{\bibfnamefont{D.~G.} \bibnamefont{Welsch}},
  \bibnamefont{and}
  \bibinfo{author}{\bibfnamefont{S.}~\bibnamefont{Walentowicz}},
  \emph{\bibinfo{title}{Quantum Optics}} (\bibinfo{publisher}{Wiley-VCH,
  Weinheim}, \bibinfo{year}{2001}).

\bibitem[{\citenamefont{Fedorov et~al.}(2008)\citenamefont{Fedorov, Efremov,
  Volkov, Moreva, Straupe, and Kulik}}]{Fedorov2008}
\bibinfo{author}{\bibfnamefont{M.~V.} \bibnamefont{Fedorov}},
  \bibinfo{author}{\bibfnamefont{M.~A.} \bibnamefont{Efremov}},
  \bibinfo{author}{\bibfnamefont{P.~A.} \bibnamefont{Volkov}},
  \bibinfo{author}{\bibfnamefont{E.~V.} \bibnamefont{Moreva}},
  \bibinfo{author}{\bibfnamefont{S.~S.} \bibnamefont{Straupe}},
  \bibnamefont{and} \bibinfo{author}{\bibfnamefont{S.~P.} \bibnamefont{Kulik}},
  \bibinfo{journal}{Phys. Rev. A} \textbf{\bibinfo{volume}{77}},
  \bibinfo{pages}{032336} (\bibinfo{year}{2008}).

\bibitem[{\citenamefont{Fedorov et~al.}(2007)\citenamefont{Fedorov, Efremov,
  Volkov, Moreva, Straupe, and Kulik}}]{Fedorov2007}
\bibinfo{author}{\bibfnamefont{M.~V.} \bibnamefont{Fedorov}},
  \bibinfo{author}{\bibfnamefont{M.~A.} \bibnamefont{Efremov}},
  \bibinfo{author}{\bibfnamefont{P.~A.} \bibnamefont{Volkov}},
  \bibinfo{author}{\bibfnamefont{E.~V.} \bibnamefont{Moreva}},
  \bibinfo{author}{\bibfnamefont{S.~S.} \bibnamefont{Straupe}},
  \bibnamefont{and} \bibinfo{author}{\bibfnamefont{S.~P.} \bibnamefont{Kulik}},
  \bibinfo{journal}{Phys. Rev. Lett.} \textbf{\bibinfo{volume}{99}},
  \bibinfo{pages}{063901} (\bibinfo{year}{2007}).

\bibitem[{\citenamefont{{Pe\v{r}ina~Jr.}}(2015)}]{PerinaJr2015}
\bibinfo{author}{\bibfnamefont{J.}~\bibnamefont{{Pe\v{r}ina~Jr.}}},
  \bibinfo{journal}{Phys. Scr.} p. \bibinfo{pages}{in print}
  (\bibinfo{year}{2015}).

\bibitem[{\citenamefont{Fedorov and Miklin}(2014)}]{Fedorov2014}
\bibinfo{author}{\bibfnamefont{M.~V.} \bibnamefont{Fedorov}} \bibnamefont{and}
  \bibinfo{author}{\bibfnamefont{M.~I.} \bibnamefont{Miklin}},
  \bibinfo{journal}{Contemporary Phys.} \textbf{\bibinfo{volume}{55}},
  \bibinfo{pages}{94} (\bibinfo{year}{2014}).

\bibitem[{\citenamefont{Eltschka and Siewert}(2013)}]{Eltschka2013}
\bibinfo{author}{\bibfnamefont{C.}~\bibnamefont{Eltschka}} \bibnamefont{and}
  \bibinfo{author}{\bibfnamefont{J.}~\bibnamefont{Siewert}},
  \bibinfo{journal}{Phys. Rev. Lett.} \textbf{\bibinfo{volume}{111}},
  \bibinfo{pages}{100503} (\bibinfo{year}{2013}).

\bibitem[{\citenamefont{Gatti et~al.}(2012)\citenamefont{Gatti, Corti,
  Brambilla, and Horoshko}}]{Gatti2012}
\bibinfo{author}{\bibfnamefont{A.}~\bibnamefont{Gatti}},
  \bibinfo{author}{\bibfnamefont{T.}~\bibnamefont{Corti}},
  \bibinfo{author}{\bibfnamefont{E.}~\bibnamefont{Brambilla}},
  \bibnamefont{and} \bibinfo{author}{\bibfnamefont{D.~B.}
  \bibnamefont{Horoshko}}, \bibinfo{journal}{Phys. Rev. A}
  \textbf{\bibinfo{volume}{86}}, \bibinfo{eid}{053803} (\bibinfo{year}{2012}).

\bibitem[{\citenamefont{Horoshko et~al.}(2012)\citenamefont{Horoshko, Patera,
  Gatti, and Kolobov}}]{Horoshko2012}
\bibinfo{author}{\bibfnamefont{D.~B.} \bibnamefont{Horoshko}},
  \bibinfo{author}{\bibfnamefont{G.}~\bibnamefont{Patera}},
  \bibinfo{author}{\bibfnamefont{A.}~\bibnamefont{Gatti}}, \bibnamefont{and}
  \bibinfo{author}{\bibfnamefont{M.~I.} \bibnamefont{Kolobov}},
  \bibinfo{journal}{Eur. Phys. J. D} \textbf{\bibinfo{volume}{66}},
  \bibinfo{eid}{239} (\bibinfo{year}{2012}).

\bibitem[{\citenamefont{{Di~Lorenzo~Pires}
  et~al.}(2009)\citenamefont{{Di~Lorenzo~Pires}, Monken, and
  {van~Exter}}}]{DiLorenzoPires2009}
\bibinfo{author}{\bibfnamefont{H.}~\bibnamefont{{Di~Lorenzo~Pires}}},
  \bibinfo{author}{\bibfnamefont{C.~H.} \bibnamefont{Monken}},
  \bibnamefont{and} \bibinfo{author}{\bibfnamefont{M.~P.}
  \bibnamefont{{van~Exter}}}, \bibinfo{journal}{Phys. Rev. A}
  \textbf{\bibinfo{volume}{80}}, \bibinfo{eid}{022307} (\bibinfo{year}{2009}).

\bibitem[{\citenamefont{Fedorov et~al.}(2005)\citenamefont{Fedorov, Efremov,
  Kazakov, Chan, Law, and Eberly}}]{Fedorov2005}
\bibinfo{author}{\bibfnamefont{M.~V.} \bibnamefont{Fedorov}},
  \bibinfo{author}{\bibfnamefont{M.~A.} \bibnamefont{Efremov}},
  \bibinfo{author}{\bibfnamefont{A.~E.} \bibnamefont{Kazakov}},
  \bibinfo{author}{\bibfnamefont{K.~W.} \bibnamefont{Chan}},
  \bibinfo{author}{\bibfnamefont{C.~K.} \bibnamefont{Law}}, \bibnamefont{and}
  \bibinfo{author}{\bibfnamefont{J.~H.} \bibnamefont{Eberly}},
  \bibinfo{journal}{Phys. Rev. A} \textbf{\bibinfo{volume}{72}},
  \bibinfo{pages}{032110} (\bibinfo{year}{2005}).

\bibitem[{\citenamefont{Chan et~al.}(2007)\citenamefont{Chan, Torres, and
  Eberly}}]{Chan2007}
\bibinfo{author}{\bibfnamefont{K.~W.} \bibnamefont{Chan}},
  \bibinfo{author}{\bibfnamefont{J.~P.} \bibnamefont{Torres}},
  \bibnamefont{and} \bibinfo{author}{\bibfnamefont{J.~H.}
  \bibnamefont{Eberly}}, \bibinfo{journal}{Phys. Rev. A}
  \textbf{\bibinfo{volume}{75}}, \bibinfo{pages}{050101(R)}
  (\bibinfo{year}{2007}).

\bibitem[{\citenamefont{Mikhailova et~al.}(2008)\citenamefont{Mikhailova,
  Volkov, and Fedorov}}]{Mikhailova2008}
\bibinfo{author}{\bibfnamefont{Y.~M.} \bibnamefont{Mikhailova}},
  \bibinfo{author}{\bibfnamefont{P.~A.} \bibnamefont{Volkov}},
  \bibnamefont{and} \bibinfo{author}{\bibfnamefont{M.~V.}
  \bibnamefont{Fedorov}}, \bibinfo{journal}{Phys. Rev. A}
  \textbf{\bibinfo{volume}{78}}, \bibinfo{eid}{062327} (\bibinfo{year}{2008}).

\bibitem[{\citenamefont{Pe\v{r}ina}(1985)}]{Perina1985}
\bibinfo{author}{\bibfnamefont{J.}~\bibnamefont{Pe\v{r}ina}},
  \emph{\bibinfo{title}{Coherence of Light}} (\bibinfo{publisher}{Kluwer,
  Dordrecht}, \bibinfo{year}{1985}).

\bibitem[{\citenamefont{Pe\v{r}ina and K\v{r}epelka}(2005)}]{Perina2005}
\bibinfo{author}{\bibfnamefont{J.}~\bibnamefont{Pe\v{r}ina}} \bibnamefont{and}
  \bibinfo{author}{\bibfnamefont{J.}~\bibnamefont{K\v{r}epelka}},
  \bibinfo{journal}{J. Opt. B: Quant. Semiclass. Opt.}
  \textbf{\bibinfo{volume}{7}}, \bibinfo{pages}{246} (\bibinfo{year}{2005}).

\bibitem[{\citenamefont{{Pe\v{r}ina~Jr.}
  et~al.}(2012)\citenamefont{{Pe\v{r}ina~Jr.}, Haderka, Hamar, and
  Mich\'{a}lek}}]{PerinaJr2012a}
\bibinfo{author}{\bibfnamefont{J.}~\bibnamefont{{Pe\v{r}ina~Jr.}}},
  \bibinfo{author}{\bibfnamefont{O.}~\bibnamefont{Haderka}},
  \bibinfo{author}{\bibfnamefont{M.}~\bibnamefont{Hamar}}, \bibnamefont{and}
  \bibinfo{author}{\bibfnamefont{V.}~\bibnamefont{Mich\'{a}lek}},
  \bibinfo{journal}{Opt. Lett.} \textbf{\bibinfo{volume}{37}},
  \bibinfo{pages}{2475} (\bibinfo{year}{2012}).

\bibitem[{\citenamefont{{Pe\v{r}ina~Jr.}
  et~al.}(2013)\citenamefont{{Pe\v{r}ina~Jr.}, Haderka, Mich\'{a}lek, and
  Hamar}}]{PerinaJr2013a}
\bibinfo{author}{\bibfnamefont{J.}~\bibnamefont{{Pe\v{r}ina~Jr.}}},
  \bibinfo{author}{\bibfnamefont{O.}~\bibnamefont{Haderka}},
  \bibinfo{author}{\bibfnamefont{V.}~\bibnamefont{Mich\'{a}lek}},
  \bibnamefont{and} \bibinfo{author}{\bibfnamefont{M.}~\bibnamefont{Hamar}},
  \bibinfo{journal}{Phys. Rev. A} \textbf{\bibinfo{volume}{87}},
  \bibinfo{pages}{022108} (\bibinfo{year}{2013}).

\bibitem[{\citenamefont{Brida et~al.}(2009{\natexlab{b}})\citenamefont{Brida,
  Meda, Genovese, Predazzi, and {Ruo-Berchera}}}]{Brida2009b}
\bibinfo{author}{\bibfnamefont{G.}~\bibnamefont{Brida}},
  \bibinfo{author}{\bibfnamefont{A.}~\bibnamefont{Meda}},
  \bibinfo{author}{\bibfnamefont{M.}~\bibnamefont{Genovese}},
  \bibinfo{author}{\bibfnamefont{E.}~\bibnamefont{Predazzi}}, \bibnamefont{and}
  \bibinfo{author}{\bibfnamefont{I.}~\bibnamefont{{Ruo-Berchera}}},
  \bibinfo{journal}{J. Mod. Opt.} \textbf{\bibinfo{volume}{56}},
  \bibinfo{pages}{201} (\bibinfo{year}{2009}{\natexlab{b}}).

\end{thebibliography}
\bibliographystyle{apsrev}

\end{document}